\documentclass{jfm}

\usepackage{graphicx}
\usepackage{newtxtext}
\usepackage{newtxmath}
\usepackage{natbib}
\usepackage{hyperref}
\hypersetup{
    colorlinks = true,
    urlcolor   = blue,
    citecolor  = black,
}
\usepackage{booktabs,multirow,siunitx}
\newcommand\Web{\mbox{\textit{We}}}
\newcommand\Knud{\mbox{\textit{Kn}}}
\newcommand\Ohne{\mbox{\textit{Oh}}}

\newcommand\Sto{\mbox{\textit{St}}}

\usepackage{layouts}

\newcommand{\RomanNumeralCaps}[1]

\newcommand{\mb}[1]{\boldsymbol{#1}} %
\newcommand{\mt}[1]{\mathsfbi{#1}}%
\usepackage{mathtools}
\DeclarePairedDelimiter\abs{\lvert}{\rvert}
\newcommand{\ep}{\varepsilon}
\DeclarePairedDelimiterXPP\order[1]{\textit{O}}\lparen\rparen{}{#1}

\usepackage{diffcoeff}
\difdef { f, s } { D }
{ op-symbol = \mathrm{D} }
\usepackage{siunitx}
\DeclareSIUnit{\stoke}{St}

\usepackage{makecell}

\usepackage{subcaption}

\newcommand{\comment}[1]{\textcolor{red}{#1}}

\title{Bouncing or Merging in Drop-Bath Impacts}

\author{
Peter Lewin-Jones\aff{1,4}
\corresp{\email{peter.lewin-jones@bristol.ac.uk}},
   Daniel M. Harris\aff{2}
   \corresp{\email{daniel\_harris3@brown.edu}},
  Radu Cimpeanu \aff{1}
  \corresp{\email{Radu.Cimpeanu@warwick.ac.uk}},
  Duncan A. Lockerby\aff{3}\corresp{\email{duncan.lockerby@warwick.ac.uk}}
 \and James E. Sprittles\aff{1}\corresp{\email{j.e.sprittles@warwick.ac.uk}}
 }

\affiliation{\aff{1}Mathematics Institute, University of Warwick, Coventry, CV4 7AL, UK.
\aff{2} School of Engineering, Brown University, Providence, RI 02912, USA.
\aff{3}School of Engineering, University of Warwick, Coventry, CV4 7AL, UK.
\aff{4}School of Engineering Mathematics and Technology, University of Bristol, Bristol, BS8 1QU, UK.
}

\begin{document}
\maketitle
\begin{abstract}
When a liquid drop impacts a liquid surface, it deforms the surface, trapping a thin layer of air between drop and bath. The drop then either bounces off the bath or, if the gas film ruptures, merges with it. We present a model for drop-bath impacts that can accurately predict whether a drop will bounce or merge, and compare it to theory-driven experiments. The model incorporates a lubrication description of the deforming trapped gas film, which accounts for its significant curvature, within a finite-element framework. Capturing this outcome requires micro- and nano-scale physics: van der Waals-driven disjoining pressure and gas-kinetic effects that depend on the mean free path of the gas. Using only independently measured material parameters, we obtain the first quantitative agreement with experiment for the critical impact speed separating bounce from merge. We explore the behaviour of drop-bath impacts at low ambient air pressure, and then investigate the bounce-merge threshold for impacts onto shallow baths and thin films, finding a bounce-merge-bounce-merge transition as the bath depth decreases.
\end{abstract}


\section{Introduction}

Understanding the behaviour of liquid drops that impact liquid surfaces is important to a wide variety of applications, including the processes of air-sea gas exchange when rain drops hit the ocean \citep{Wanninkhof2009}, inkjet printing \citep{Lohse2022}, and spray cooling of hot electronics components \citep{Liang2017}. Although there are many possible outcomes of such an impact, including coalescence, splashing, or the production of a high-speed jet as first identified by \citet{Worthington1908}, impacts can be divided into two primary outcomes: either the drop makes contact with the impacted surface, or it does not, because there is a thin layer of the surrounding gas that becomes sandwiched between the drop and the surface. In the latter case, the lubricating pressure from the gas film is strong enough to prevent contact so that either the drop rebounds, a phenomenon observed as early as in \citet{Reynolds1881}, or floats for a sustained period before eventually coalescing \citep{Walker1978}. Notably, it is now well known that by vibrating the bath a drop can be made to bounce repeatedly \citep{Couder2005} -- with the resulting interaction between the drop and its induced waves producing a quantum analogue system capable of macroscopically reproducing quantum-like behaviour \citep{Bush2010}.  If the drop does make contact, and begins coalescing, a huge variety of outcomes are possible, with a long history of study \citep{Rein1993,Yarin2006,Kavehpour2015}, including the potential for a daughter drop to be ejected as part of a coalescence cascade \citep{Thoroddsen2000}. 

In this article, we are focused on the thresholds in speed (or relevant dimensionless number) at which the drop-bath contact behaviour changes. This threshold has normally been characterised in terms of the Weber number, $\Web=\rho^d R U^2/\gamma^d$, where $\rho^d$ is the drop's density, $R$ its radius, $U$ the impact speed and $\gamma^d$ its interfacial surface tension coefficient. This quantifies the ratio of kinetic energy to surface energy of the drop, and gives a measure of how `violent' the impact is. As the impact velocity is increased, there is a critical Weber number where the outcome of the impact changes from bouncing to merging. Borrowing the language from the equivalent phenomenon in drop-drop collisions \citep{Pan2008}, we refer to this transition, from bouncing to merging with increasing impact velocity (i.e. increasing $\Web$), as the `hard' transition (referred to as hard because it occurs at a higher kinetic energy (that is, higher $\Web$), and is therefore more `violent', than other transitions that are observed). The critical Weber number of this transition has been identified in experiments to depend on many different things, including the ambient pressure of the gas \citep{Schotland1960}. Interestingly, at lower impacts speeds, there is another transition, with the slowest drops merging and faster drops bouncing, named the `soft' transition (named due its lower kinetic energy than the `hard' transition), which was identified experimentally by \citet{Jayaratne1964}. Recently, \citet{Pan2007} and \citet{Tang2016} identified that the critical Weber number of the hard transition depends on the depth of the bath,  with remarkable transitions from merging-bouncing-merging-bouncing with increasing bath depth.

Following the observations of \citet{Walker1978}, the ability of a drop to float for an extended period of time was observed using high-speed photography by \citet{Thoroddsen2000}, who were able to measure the time that the drop was floating for. \citet{Klyuzhin2010} showed that reducing the air pressure reduced the time that such drops would float for. More recently, \citet{Lo2017} observed drops floating above very thin liquid films, with small changes in the table angle having a dramatic effect on the floating time. As mentioned, \citet{Couder2005} found that by vibrating the bath, drops can be made to float indefinitely.

Critical to the bounce-merge threshold is the dynamics of the gas film that is sandwiched, or `trapped' between the drop and the bath. A key observation is that the gas film can be as thin as tens to hundreds of nanometres before contact occurs \citep{Kavehpour2015}. These nano-scale gas films are not accurately described by the Navier-Stokes equations with no-slip boundary conditions, because the mean free path of the gas is comparable to the thickness of the film. Unconventional models are required, which incorporate gas kinetic effects that go beyond the Navier-Stokes paradigm. The general use of these models in drop-impact problems has been reviewed in \citet{Sprittles2024}. 

In the context of drop-bath impacts, \citet{Bach2004}, building on the work of \citet{Gopinath2001}, used a lubrication model for the gas film based on the Boltzmann equation to show that the soft transition for very small drops was particularly sensitive to the mean free path of the gas. These were pioneering works in the study of drop impacts, but were limited to low impact speeds: their model used potential flow in the liquid domains, neglecting viscosity, and was limited to small deformations of the drop surface, assumptions that we have been able to relax. Lubrication models for the gas film between drop and bath have also been used to study the dynamics of drops within the bouncing regime \citep{Phillips2025}, and in the cushioning of the early stages of high-speed impacts of drops onto smooth solids  \citep{Mandre2009,Hicks2011,Moore2021}. 

The problem of drops impacting liquid baths is closely linked to other drop-impact problems, where drops have been shown to bounce rather than merge with, or wet, the impacted surface due to a trapped gas film. A particularly relevant case is impacts of drops onto atomically smooth solids. These drop-solid impacts are much easier to image than drop-bath or drop-drop impacts, and various techniques, such as interferometry, have been used to accurately measure the trapped gas film \citep{Driscoll2011, Kolinski2012,vanderVeen2012}, as reviewed in \citet{Josserand2016} and \citet{Sprittles2024}. More recently, the gas film in a drop-bath impact has also been measured by \citet{Tang2019}, finding similar dynamics for the evolution of the gas film to those seen for drop on solid impacts, with a ``neck'' at the edge of the gas film where the film is thinnest. 

Notably, using pressure-controlled chambers, it has been observed in drop-bath impacts by \citet{Schotland1960} that critical Weber numbers of bounce-merge transitions depend dramatically on the ambient pressure.  Such effects are now well-recognised for both drop-drop collisions \citep{Qian1997,Zhang2026} and drop-solid ones \citep{Zhang2021}, and represent an important feature which theoretical models should expect to reproduce.

As well as gas kinetic effects, another phenomenon that is missing from conventional models but is critical to understanding how a drop makes contact is the van der Waals force between the drop and the bath \citep{Israelachvili2011}. As the gas layer thins, it is this force, acting as a disjoining pressure, that ultimately drives the liquid surfaces together. This has been discussed as the cause of contact as early as \citet{Jayaratne1964}, and its importance was highlighted in the context of drop-drop collisions by \citet{Pan2008}. 

In this article, we present a model for drop-bath impacts that is able to predict, for the first time, the transitions between bouncing and merging. This uses a lubrication equation to model the gas film, which incorporates gas kinetic effects and a van der Waals-driven disjoining pressure. This is coupled to the free-surface Navier-Stokes equations for the flow of the drop and the bath. Together these are solved with an arbitrary Eulerian-Lagrangian finite-element method. This builds on our previous models for drop-drop collisions \citep{Lewin-Jones2024} and drop-solid impacts \citep{Chubynsky2020}, which in turn built on the work of \citet{Li2016} for drop-drop collisions. 
 
The study of drop-bath impacts, unlike these previous studies, is more challenging as one cannot rely on planar symmetry, applicable for the head-on collisions of equal-sized drops, so that two distinct liquid domains have to be simultaneously resolved. Furthermore, the gas film geometry becomes significantly curved, so that we are forced to reformulate the lubrication equation with respect to a curved coordinate system that moves with the deforming drop surface, and provides the stress that is applied to the surfaces of the drop and the bath. Only with a curved lubrication model that is embedded with nanoscale physics are we able to accurately predict the bounce-merge threshold.

To confirm the accuracy of our new model, we have also performed new experiments, using the methodology and performed alongside those of \citet{Harris2026}, to measure the critical Weber number for the hard transition of drop-bath collisions of silicone oil at a range of viscosities and droplet radii. 

This article is structured as follows. In Section~\S\ref{sec:method}, we introduce our model for drop-bath impact, that requires us to formulate a curved lubrication equation that sits within a finite element framework to model the trapped gas film, incorporating the nano-scale van der Waals and gas kinetic effects. This model is then benchmarked in \S\ref{sec:vof_bounce} against volume-of-fluid simulations for the case in which the drop does not merge. The experimental methodology is described in \S\ref{sec:meth_exp}. In \S\ref{sec:results_Deepbath} we compare simulations of the new model to experiments of impacts onto a deep bath, and then explore beyond the current experimental regime to make predictions of the critical Weber numbers for low speed impacts and understand the behaviour at reduced ambient air pressure. In \S\ref{sec:results_shallow_bath}, we consider impacts onto shallow baths, comparing to the experiments of \citet{Tang2016} and explore the limit of vanishing bath depth. Finally, in \S\ref{sec:conclusions} we conclude and discuss future directions for this methodology and the investigation of drop impacts.

\section{Methodology: Modelling Drop-Bath Impacts}\label{sec:method}

Consider the impact of a liquid drop onto a liquid bath. A thin layer of the surrounding gas is trapped between them, which we model using lubrication theory, exploiting the small ratio between the thickness of the gas layer and the size of the drop. We assume that there is no drop or surface deformation until the lubrication approximation is valid, and from thereon it is only at the region of the lubrication model where the drop and bath directly interact. In \S\ref{sec:meth_macro_models} we introduce the models for the drop and the bath. In \S\ref{sec:meth_gas_model}, we describe the model for the gas film, how the (curved) trapped gas film is described and the lubrication equation that is solved, with a full derivation in appendix \ref{app:deriv}. Next, we describe the additional physics, beyond those of classical fluid dynamics, that are required for accurate modelling: gas kinetic effects in \S\ref{sec:meth_gke} and disjoining pressure in \S\ref{sec:meth_vdw}. We introduce the numerical implementation of our new model, using the finite element method, in \S\ref{sec:meth_numerical_method}. In \S\ref{sec:vof_bounce} we benchmark our new model against an established volume-of-fluid method, and in \S\ref{sec:meth_exp} describe the experimental methodology.

\subsection{Drop and Bath Models}\label{sec:meth_macro_models}

\begin{figure}
    \centering
   \includegraphics{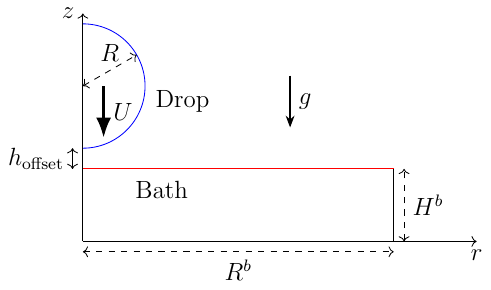}
    \caption[Diagram]{ Initial conditions of our model, with a drop of radius $R$, with base at a height of $h_{\text{offset}}=0.1R$ and initial downward speed $U$ impacting an initially stationary bath of depth $H^b$ and radius $R^b$ (not drawn to scale). }
    \label{fig:initial_conditions}
\end{figure}

We work in a cylindrical coordinate system $(r,z,\theta)$, shown in figure \ref{fig:initial_conditions}, with axisymmetry always assumed, so there is no dependence of quantities on $\theta$ and there is no velocity in the $\theta$-direction. This is a natural setup for such impacts as gravity tends to align drops to fall `down' and `flatten' liquid baths/layers. Unit basis vectors in the $r$- and $z$-directions are denoted by $\mb{e}_r$ and $\mb{e}_z$, respectively, with gravity in the $-\mb{e}_z$ direction. We will use subscripts $r$ and $z$ to denote components of vectors in the $r$ and $z$ directions (for example, $\mb{u}=u_r\mb{e}_r+u_z\mb{e}_z$).

Initially, the drop is a sphere of radius $R$, with uniform velocity $-U\boldsymbol{e}_z$, as shown in figure~\ref{fig:initial_conditions} (this assumes that any oscillations in the drop, from its formation, have had time to settle). We will use superscripts of $d$, $b$ and $g$ to denote quantities associated with the drop, bath and gas respectively. The dynamic viscosity of the liquid drop is $\mu^d$, and its density is $\rho^d$. The base of the upper drop starts from height $h_{\text{offset}}=0.1R$, which is far enough for the gas lubrication pressure to be negligible (see the appendix of \citet{Lewin-Jones2024}, which discusses the equivalent assumptions for drop-drop collisions). The bath, with dynamic viscosity, $\mu^b$, and density, $\rho^b$, has a solid floor and wall, with depth, $H^b$, and radius, $R^b$, and initially is at rest, with a horizontal free surface. Although the model is described in a form that allows different liquid properties in the bath and the drop, all of the results presented in this article use the same liquid for both, leaving such variations \citep{Fudge2021,Wu2022} as an avenue for future work. 

The Navier-Stokes equations \eqref{eq:dim_momcons}--\eqref{eq:dim_masscons} are solved for the velocity $\mb{u}^d$, and pressure, $p^d$, in the liquid drop at time $t$, and are given dimensionally as
\begin{align}
\rho^d \diff.D.{\mb{u}^d}{t}&=-\nabla p^d+\mu^d\nabla^2\mb{u}^d -\rho^d g \mb{e}_z \, ,\label{eq:dim_momcons}\\
0&=\nabla \mb{\cdot}\mb{u}^d\, ,\label{eq:dim_masscons}
\end{align}
where $g$ is the acceleration due to gravity. 

The corresponding Navier-Stokes equations \eqref{eq:dim_momcons_b}--\eqref{eq:dim_masscons_b} are solved for the bath, with superscript $b$ replacing $d$:
\begin{align}
\rho^b \diff.D.{\mb{u}^b}{t}&=-\nabla p^b+\mu^b\nabla^2\mb{u}^b -\rho^b g \mb{e}_z \, ,\label{eq:dim_momcons_b}\\
0&=\nabla \mb{\cdot}\mb{u}^b\, .\label{eq:dim_masscons_b}
\end{align}

On the surface of the drop, where $\mb{n}^d$ is the outer unit normal to the drop, the dynamic boundary condition is
\begin{align}
\mt{T}^{d}\mb{\cdot}\mb{n}^d=\mt{T}^{g}\mb{\cdot}\mb{n}^d-\gamma^d \mb{n}^d\nabla_s\mb{\cdot}\mb{n}^d+\mb{F}^d\, ,\label{eq:ndimdynBC}
\end{align}
where $\mt{T}^d$ is the (drop) stress tensor, $\nabla_s\mb{\cdot}$ is the surface divergence (see \eqref{eq:surf_div} in appendix \S\ref{app:curv_normal}) and $\gamma^d$ is the constant surface tension coefficient of the drop-gas interface. The stress tensor in the drop is given by
 \begin{align}
     \mt{T}^d&=- p^d\mb{I}+\mu^d\left(\nabla\mb{u}^d+(\nabla\mb{u}^d)^\top\right)\, .\label{eq:ndim_stress}
 \end{align}
Similarly, $\mt{T}^g$ is the gas stress tensor. There is also a surface force, $\mb{F}^d$ which models the van der Waals driven disjoining pressure between the drop and the bath, and will be discussed in \S\ref{sec:meth_vdw}. Correspondingly, the dynamic boundary condition on the bath surface is 
\begin{align}
\mt{T}^{b}\mb{\cdot}\mb{n}^b=\mt{T}^{g}\mb{\cdot}\mb{n}^b-\gamma^b\mb{n}^b\nabla_s\mb{\cdot}\mb{n}^b+\mb{F}^b\, ,\label{eq:ndimdynBC_b}
\end{align}
with $\mt{T}^{b}$, $\gamma^b$ and $\mb{F}^b$ being the bath stress tensor, bath-gas surface tension coefficient and bath surface force respectively.

 At $r=0$, the boundary conditions for the drop and the are given by symmetry as $u_r=0$ and $\diffp{u_z}{r}=0$, where here $(u_r,u_z)$ is the velocity of the drop and bath respectively.
The bath initially occupies the region from $z=0$ to $z=-H^b$, and from $r=0$ to $r=R^{b}$. At $z=-H^b$ (the `floor' of the bath), we apply the no-slip boundary condition, $\mb{u}^b=\mb{0}$. At $r=R^b$ (the `wall' of the bath), a no penetration condition, $u^b_r=0$, is applied, as well as a no shear condition, $\mb{e}_r\cdot\mt{T}^b\cdot\mb{e}_z=0$, and a 90$^{\circ}$ contact angle is prescribed at the boundary of the free surface. This choice of boundary condition on the bath wall allows the free-surface to wet/dewet the side, as waves produced by the impact encounter it, without any moving-contact-line issues. In this article $R^b$ will always be chosen so that the edge of the bath is far from the impacting drop, and does not affect whether the drop bounces or merges with the bath; the effect of this will be discussed in \S\ref{sec:soft_transition} and would be an interesting path of future investigation. 

\subsection{Modelling the Gas Film}\label{sec:meth_gas_model}

Previous research for drop impact onto smooth solids \citep{Chubynsky2020}, heated solids in the Leidenfrost regime \citep{Chakraborty2022}, and the head-on collision of two identical drops \citep{Lewin-Jones2024} has successfully used lubrication theory to model the trapped gas film, coupled to the Navier-Stokes equations used to model the dynamics in the drop. Lubrication theory has also been coupled to various reduced-order models for the drop \citep{Bach2004,Mandre2009,Hicks2011,Moore2021,Phillips2025}. In all of these cases the film is horizontal, and the lubrication height was measured vertically, in an Eulerian sense, so that the film height is a function of $r$. 

\begin{figure}
    \centering
   \includegraphics{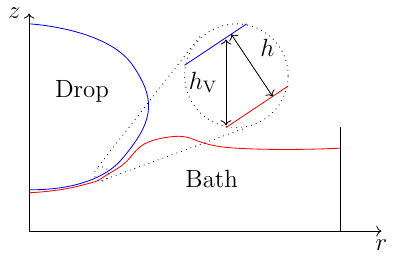}
    \caption[Diagram]{ Two possible definitions for the height of the gas film. Denoted by $h$ is the height measured in the normal direction to the drop, as used by the model in this article. Denoted by $h_\text{V}$ is the vertical height of the film, which is unsuitable for this problem due to the large deformation of the bath. }
    \label{fig:height_def}
\end{figure}

However, for drop-bath impacts, especially at higher impact speeds, the bath is deformed considerably, and we cannot apply the horizontal model, as in \citet{Lewin-Jones2024}, because the vertical distance between the drop and the bath will not be an accurate measurement of the thickness of the gas film. Among other problems, once the bath deforms significantly a lubrication equation based on the vertical height will no longer preserve mass correctly (as this does not correctly measure change in the volume of gas), and the magnitude of the van der Waals force, which depends strongly on height, will be too small. The solution is to replace the horizontal lubrication equation with a model that treats the film thickness as the distance between the drop and the bath in the direction normal to a free surface rather than vertically, and derive a lubrication equation where the governing equations of the gas film are averaged over that `normal' thickness. This is shown in figure~\ref{fig:height_def}, where $h_\mathrm{V}$ shows how the film height would be measured vertically, compared to $h$ which is how the model in this article is developed.

The model we have developed describes a lubricating film between two axisymmetric curved surfaces. The general approach is to stop solving for the gas dynamics in the \emph{fixed} Eulerian frame (e.g. the pressure is solved as a function of $r$) and instead fit a coordinate system to the \emph{moving} surface of the drop. Lubrication theory then provides both the pressure and shear stress terms for the boundary conditions \eqref{eq:ndimdynBC} and \eqref{eq:ndimdynBC_b}. 

We have used the approach of \citet{Howell1994}, originally developed for thin viscous liquid sheets, to consider the governing equations of the gas in a curved, time evolving coordinate system, and applied lubrication theory, integrating across the film in the normal direction to the drop, to obtain a single partial differential equation for the gas pressure. However, the thin film in \citet{Howell1994} is a sheet with free surfaces, rather than our setup consisting of a film trapped between two moving surfaces. To calculate the time evolution the film height, we have adapted the method of \citet{Temizer2016}, who derived a general lubrication equation between two curved surfaces, without any symmetry assumptions.

The choice of film height definition is not unique, as, for example, we could equally work on the bath surface, and define the film height in terms of the normal projection from the bath. It is also possible to formulate the lubrication based on a `lubrication surface' that lies between the two bounding surfaces, with the film height defined as normal projection from either the boundary surfaces or the intermediate lubrication surface. These possibilities are all compared in \citet{Temizer2016}, but as long as the surfaces are approximately locally parallel (as they will be for the lubrication approximation to be valid), the differences between these formulations will be asymptotically small.

\subsection{Curved Lubrication Equation}\label{sec:meth_curved}

A full derivation of the lubrication equation is available in appendix \ref{app:deriv}, which we summarise here. We parametrise the axisymmetric surface of the drop by curvilinear coordinate $s$ and azimuthal coordinate $\theta$. Axisymmetry means that the position vector at $s$ and $\theta$, at time $t$, can be written as $\mb{x}^d(s,\theta,t)=r^d(s,t) \mb{e}_r+z^d(s,t)\mb{e}_z$. Note that in Cartesian coordinates $\mb{e}_r$ depends on $\theta$, but dependencies on $\theta$ will be unstated except where it is important. Correspondingly the bath surface is parametrised by $\xi$ and $\theta$.

\begin{figure}

    \centering
       \includegraphics{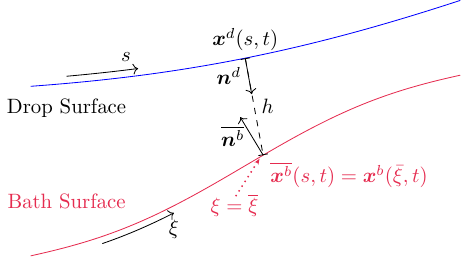}
    \caption[Diagram]{ The definition of the film height $h$ at drop position vector $\mb{x}^d(s,t)$. The height is measured in the direction normal to the drop's free surface. The intersection of the projected normal on the bath surface is denoted by $\overline{\mb{x}^b}$, at coordinate $\overline{\xi}$.}
    \label{fig:Curved_diagram}
\end{figure}

The point of intersection of the drop normal $\mb{n}^d$, with the bath surface is denoted by $\overline{\mb{x}^b}$, and we denote its curvilinear surface coordinate by $\overline{\xi}$, as shown in figure~\ref{fig:Curved_diagram}. Throughout this section, symbols with over-bars denote variables evaluated at the point of projection (this is described in more detail in appendix \S\ref{app:deriv_height_deriv}). Because of the axisymmetry, $\overline{\mb{x}^b}$ is at the same azimuthal coordinate $\theta$ as $\mb{x}^d$.  The film thickness is then denoted by $h=\left(\overline{\mb{x}^b}-\mb{x}^d\right)\mb{\cdot}\mb{n}^d$, so that $h\mb{n}^d=\overline{\mb{x}^b}-\mb{x}^d$, as shown in figure~\ref{fig:Curved_diagram}.

To proceed, we need to change to an alternative, curvilinear coordinate system with respect to the drop surface. The construction of this is shown in appendix \S\ref{app:curv_basis}, based on the approach described in \citet{Howell1994}. This has orthonormal basis $\{\mb{e}_s,\mb{e}_\theta,\mb{n}^d\}$, where $\mb{n}^d$ is the unit normal vector to the drop surface, and $\mb{e}_s$ is the unit vector tangent to the drop surface, as shown in figure~\ref{fig:coord_systems}. Note that $\mb{e}_s$ and $\mb{n}^d$ depend on $s$ and $t$. We denote the components of vectors in the directions of these basis vectors with subscript $s$ and $n$ (for example, $\mb{u}^d=u_n^d\mb{e}_n+u_s^d\mb{e}_s$). Important quantities are the (surface) scaling factors
\begin{align}
a_s&=\abs*{\diffp{\mb{x}^d}{s}}=\sqrt{\diffp{r^d}{s}^2+\diffp{z^d}{s}^2}\, , &
a_\theta&=\abs*{\diffp{\mb{x}^d}{\theta}}=r^d\, .
\end{align}
The parametrised drop surface is moving, and we denote the velocity of this geometric drop surface by $\mb{w}^d=\diffp{\mb{x}^d}{t}$. This is not the same as the material velocity of the drop fluid, as in principle, we could have the geometric surface moving, but the liquid be stationary. However, the kinematic boundary condition does set the normal components of the material fluid velocity and the surface velocity to be equal, $u_n=w_n$. The equivalent velocity of the bath surface is denoted by $\mb{w}^b$.

Applying lubrication theory in this curved coordinate system (see \S \ref{app:deriv_lubrication_scale} in the appendix) gives the lubrication equation \eqref{eq:curved_lub} that must be solved for the gas pressure $p^g$,
\begin{align}
    0&=h\nabla_s\mb{\cdot}\mb{w}^d+\frac{1}{a_sa_\theta}\diffp{}{s}\left(a_\theta\left(\frac{-{ h}^3}{12\mu^ga_s}\diffp{p^g}{s}+\frac{h\left(\overline{u_s^b}+u_s^d-2w_s^d\right)}{2}\right)\right)\nonumber\\ &\quad+ \overline{w^{b}_n}-w^d_n+\frac{1}{a_s}\diffp{h}{s}\left(w_s^d-\overline{w^b_s}\right)\, ,\label{eq:curved_lub}
\end{align}
where $\mu^g$ is the viscosity of the gas. At $r=0$ a symmetry boundary condition is applied, given by $\diffp{p^g}{s}=0$. At the edge of the film, the gas pressure, $p^g$, is set to a reference pressure $p_0$. Within reason, the exact position of this does not affect the results. As described in \citet{Chubynsky2020}, away from the thin gas film, $\diffp{p^g}{s}$ will tend to $0$, and $p^g$ will tend to $p_0$ before reaching the position where the boundary condition is applied. In this article we set this to be where $h$ first exceeds $0.3R$.

Lubrication theory also provides the surface stress on the drop surface due to the gas in the dynamic boundary condition \eqref{eq:ndimdynBC} as 
\begin{align}
    \mt{T}^{g}\mb{\cdot}\mb{n}^d=-p^g\mb{n}^d+\left(\frac{-h}{2a_s}\diffp{p}{s}+\mu^g\frac{\overline{u_s^b}-u_s^d}{ h}\right)\mb{e}_s\, .\label{eq:gas_drop_stress}
\end{align}
Derivations of both of these equations are in appendix \ref{app:deriv}.

For the surface stress due to the gas on the bath surface in the dynamic boundary condition \eqref{eq:ndimdynBC_b}, we instead project a vector (from the bath surface to the drop surface) in the direction normal to the \emph{bath} surface. Now the bars denote evaluation at the point of projection on the drop surface, where the gas pressure $\overline{p^g}$ and the tangential liquid velocity $\overline{u_s^d}$ are evaluated, giving
\begin{align}
     \mt{T}^{g}\mb{\cdot}\mb{n}^b=-\overline{p^g}\mb{n}^b+\left(\frac{h}{2\overline{a_s}}\diffp{\overline{p^g}}{s}+\mu^g\frac{u_s^b-\overline{u_s^d}}{ h}\right)\mb{e}_\xi\, .\label{eq:gas_bath_stress}
     \end{align}
Notably, these forms for the stress vectors (or `tractions') are those which fit most naturally into a finite element formulation.

\subsection{Gas Kinetic Effects}\label{sec:meth_gke}

As we will see in the simulation results in \S\ref{sec:results_Deepbath} below, the trapped gas film can reach thicknesses down to hundreds, or even tens, of nanometers; easily comparable in magnitude to the mean free path, $\lambda$, of the gas, which for air at atmospheric pressure and room temperature is approximately $70$~\si{\nano\m}. The derivation of the lubrication equation started with the incompressible Navier-Stokes equations, and therefore implicitly assumed that the mean free path is much smaller than all length scales of the problem. The Navier-Stokes equations are no longer accurate for flows that involve length scales comparable to the mean free path of the gas, and the linearised BGK-Boltzmann equation is required to accurately simulate these gas flows \citep{Cercignani2006b}. 

The approach we use to account for this effect is to introduce `gas kinetic factors' to the lubrication equation. These factors depend on a local Knudsen number, $\Knud(h)=\lambda/h$. This approach has been used successfully in \citet{Chubynsky2020} for drop-solid impacts and \citet{Lewin-Jones2024} for drop-drop collisions, building on earlier work in free-surface drop collisions \citep{Li2016,Bach2004}, rigid sphere collisions \citep{Davis1972,Hocking1973,Sundararajakumar1996} and lubrication flows in solid-state channels \citep{Fukui1988}.

\begin{figure}
    \centering
    \includegraphics{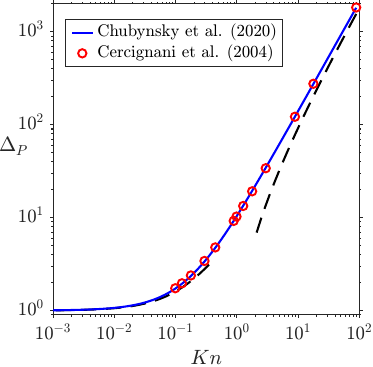}
        \includegraphics{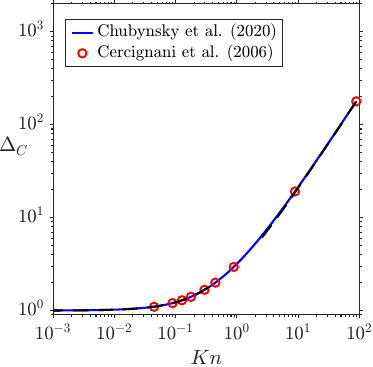}
    \caption[Plots of the gas kinetic factors]{Gas kinetic factors $\Delta_P$ (left) and $\Delta_C$ (right) in \eqref{eq:Delta_P}--\eqref{eq:Delta_C} against Knudsen number. The asymptotic limits for large and small $\Knud$ are shown as black dashed lines, and the red circles show the simulation data from \citet{Cercignani2004} and \citet{Cercignani2006a} used by \cite{Chubynsky2020} to fit them. }
    \label{fig:gke_factors}
\end{figure}

Inspecting the final lubrication equation \eqref{eq:curved_lub} and boundary stress \eqref{eq:gas_drop_stress} we can see that the flow is made up of two components, a pressure gradient driven Poiseuille component given by $\frac{-{ h}^3}{12\mu^ga_s}\diffp{p^g}{s}$ in \eqref{eq:curved_lub}, and a shear driven Couette component given by $\frac{h}{2}\left(\overline{u_s^b}+u_s^d-2w_s^d\right)$ in \eqref{eq:curved_lub}. As derived in \citet{Chubynsky2020} (in the context of drop-solid impact), two factors, $\Delta_P(\Knud)$ for the Poiseuille component and $\Delta_C(\Knud)$ for the Couette component are used to modify the lubrication equation \eqref{eq:curved_lub} and surface stresses \eqref{eq:gas_drop_stress} and \eqref{eq:gas_bath_stress} respectively. The factors that are used here are those derived by \citet{Chubynsky2020}, and assume diffuse scattering of molecules from the surfaces, which is a good approximation for gas-liquid interactions, where the surface is ‘rough’ on a molecular scale. Notably, this formulation relies on the lubrication approximation where the flow is locally akin to a channel flow.

The effect of these factors is to act as an ``effective viscosity'' that depends on the gas film thickness via the mean free path. Specifically, the gas becomes effectively much less viscous as the film thins. At $\Knud=1$ you get approximately ten times more gas flux for a given pressure gradient as compared to the Navier-Stokes equations with no-slip boundary conditions. These factors account for the effect that the Knudsen number has on both the bulk flow of the gas and  the boundary conditions that describe how the gas molecules interact with the surface \citep{Wu2017}. 

The functional forms of the gas kinetic corrections are given in expressions  \eqref{eq:Delta_P}--\eqref{eq:Delta_C}, and plotted in figure~\ref{fig:gke_factors}. Full details of the derivation of these are given in \citet{Chubynsky2020}, which we summarise here.
\begin{align}
     \Delta_P(\Knud)&=1+6.88\Knud +(6\Knud/\pi)\ln\left(1+2.76\Knud+0.127\Knud^2\right)\, ,\label{eq:Delta_P}\\
     \Delta_C(\Knud)&=  1+2\Knud\left(1.1466-0.095\exp\left(\frac{-0.662}{\Knud}\right)-0.0516\exp\left(\frac{-3.3003}{\Knud}\right)\right)\, .\label{eq:Delta_C}
\end{align}
As $\Knud\to 0$, $\Delta_P\sim 1+6.88\Knud$ and $\Delta_C\sim 1+2.2929\Knud$, which are the expressions obtained when a Navier-slip boundary condition is applied, with slip length $1.1466\lambda$ \citep{Albertoni1963}. As $\Knud\to\infty$, they have the asymptotic behaviour approaching free molecular flow, with $\Delta_P\sim (6\Knud/\pi)\ln\left(4\Knud^2/\pi^2\right)$ \citep{Cercignani2004} and $\Delta_C\sim 1+2\Knud$ \citep{Fukui2005}. The combined expressions were then derived in \citet{Chubynsky2020} by asymptotic matching and fitting through simulations of the Boltzmann equation by \citet{Cercignani2004} and \citet{Cercignani2006a} at intermediate $Kn$, as shown in figure~\ref{fig:gke_factors}. 

The factors were originally derived for approximately horizontal gas films, but in our case the film becomes curved, necessitating the formulation of the curved lubrication. However, in the lubrication limit, the flow is locally parallel to the tangent of the drop surface, allowing us to use the kinetic corrections \eqref{eq:Delta_P}--\eqref{eq:Delta_C} developed for horizontal flows, with the film height measured in the normal direction. If the local curvature of the film becomes large enough for this to not be applicable, then the assumptions made in deriving the lubrication equation will have also been broken. In any case, for moderate values of $\Knud$ (up to $\Knud=\order{1}$), the flow rate of a rarefied gas through a curved channel is unaffected by the curvature \citep{Aoki2010}.

With these factors included, the lubrication equation \eqref{eq:curved_lub} and boundary stresses \eqref{eq:gas_drop_stress}--\eqref{eq:gas_bath_stress} become
\begin{align}
    0&=h\nabla_s\mb{\cdot}\mb{w}^d+\frac{1}{a_sa_\theta}\diffp{}{s}\left(a_\theta\left(\frac{-{\Delta_P(\Knud(h)) h}^3}{12\mu^ga_s}\diffp{p^g}{s}+\frac{h\left(\overline{u_s^b}+u_s^d-2w_s^d\right)}{2}\right)\right)\nonumber\\ &\quad+ \overline{w^{b}_n}-w^d_n+\frac{1}{a_s}\diffp{h}{s}\left(w_s^d-\overline{w^b_s}\right)\, ,
    \end{align}
    \begin{align}
    \mt{T}^{g}\mb{\cdot}\mb{n}^d&=-p^g\mb{n}^d+\mu^g\left(\frac{-h}{2a_s}\diffp{p^g}{s}+\frac{\overline{u_s^b}-u_s^d}{\Delta_C(\Knud(h)) h}\right)\mb{e}_s\, ,\\
     \mt{T}^{g}\mb{\cdot}\mb{n}^b&=-\overline{p^g}\mb{n}^b+\mu^g\left(\frac{h}{2\overline{a_s}}\diffp{\overline{p^g}}{s}+\frac{u_s^b-\overline{u_s^d}}{\Delta_C(\Knud(h)) h}\right)\mb{e}_\xi\, ,
     \end{align}
which is the form of the lubrication equation and  stress boundary conditions used in our simulations.

\subsection{Disjoining Pressure}\label{sec:meth_vdw}

As the final part of the model, there is an additional force per unit area acting on the drop in \eqref{eq:ndimdynBC} that takes the form
\begin{align}
    \mb{F}^d=\frac{A}{6\pi {h}^3}\mb{n}^d,
\end{align}
where $A$ is the Hamaker constant for the interaction of the two materials \citep{Israelachvili2011}. This expression arises from integrating the London van der Waals potential over two semi-infinite bodies with a horizontal gap of thickness $h$ between them \citep{Hamaker1937}. Although our surfaces are not perfectly flat, and have finite domains, this is still a very good approximation. The film height is so thin compared with the radius of curvature that treating the drops as infinite from the perspective of this force is reasonable, and \citet{Dai2008} showed that the additional terms that appear for curved films are asymptotically small in the lubrication scaling. Identically, in \eqref{eq:ndimdynBC_b},
\begin{align}
    \mb{F}^b=\frac{A}{6\pi {h}^3}\mb{n}^b\, .
\end{align}

This force acts to pull the drop and the lower surface rapidly together below a critical film height of $\sim 10$~\si{\nano\meter}, which is determined by the stability of perturbations  which grow due to the disjoining pressure and are suppressed by surface tension (see the S.I. of \citet{Chubynsky2020} and similar well-studied effects for thin spinodal rupture of thin liquid films on solids \citep{Oron1997,Craster2009}). The Hamaker constant depends on the liquid considered but can be measured or computed independently of drop-impact experiments (i.e. it isn't `fitted'). For silicone oil we use $A=5.03\times 10^{-20}$~\si{\joule}, as calculated by \citet{Chen2016}.

\subsection{Model Parameters}

For the cases considered in this article, where the drop and bath are the same liquid, there are 7  parameters that are the same for each set of impacts with the same liquid and ambient conditions: the liquid density, $\rho^d$, viscosity, $\mu^d$, surface tension, $\gamma^d$, and Hamaker constant $A$; the gas viscosity, $\mu_g$ and mean free path $\lambda$; and gravitational acceleration, $g$. There are then 4 parameters that specify a particular impact: the drop radius $R$, the impact speed $U$, the bath depth $H^b$ and bath radius $R^b$.

The primary non-dimensional number that characterises a particular impact is the Weber number $\Web=\rho^d R U^2/\gamma^d$, which is proportional to the ratio of the drop's kinetic energy to its surface energy, and has been commonly used to report experimental results for the outcomes of drop impacts since \citet{Schotland1960}. Many non-dimensional numbers other than the Weber number have been used to characterise the outcome of a drop impact, including the Stokes number, $\Sto=\mu^g/(\rho^d UR)$, which is the ratio of the viscous force in  the gas to the inertia of the drop \citep{Josserand2016}, and the two-phase Ohnesorge number, $\Ohne_{g,d
}=\mu^g/\sqrt{\rho^d R \gamma^d}$, characterising the effect of the gas viscosity as compared to the drop’s inertia and surface tension \citep{Huang2021}. 

We choose to consider the critical Weber numbers of the impacts as functions of the dimensional parameters, such as the drop radius and ambient gas pressure, to be consistent with how experimental results have been reported.

\subsection{Numerical Method}\label{sec:meth_numerical_method}

Our model has been implemented in the open-source C++ finite element package \href{https://oomph-lib.github.io/oomph-lib/doc/html/}{oomph-lib} \citep{Heil2022}, with our code available at \url{https://github.com/PLewinJones/drop_bath_impact}. A detailed description is available in \citet{Lewin-Jones2024a}, which we summarise here.

The equations are solved using an Arbitrary-Eulerian formulation, where the mesh is moved at speed $\mb{w}$. With this the material derivative takes the form \citep{Donea2004}
\begin{align}
    \diff.D.{}{t}&=\diffp{}{t}+\left(\mb{u}-\mb{w}\right)\mb{\cdot}\nabla\, .
\end{align}
 We use oomph-lib's pseudo-solid Arbitrary-Lagrangian-Eulerian triangular mesh, where the boundary of the mesh is moved with the normal velocity of the liquid at the drop surface, and the interior elements are moved by treating them as an elastic solid. The time derivatives are adjusted to account for the motion of the mesh \citep{Cairncross2000}. Periodically the domain is remeshed using an adaptive method \citep{Zienkiewicz1987} that refines the size of the mesh elements as needed to keep the inter-element stress jump below a tolerance of $5\times 10^{-5}$. An implicit, adaptive time-stepper is used (the 2\textsuperscript{nd} order backward differentiation formula, BDF2). The time-stepper remains mostly at a maximum dimensionless timestep of $5\times 10^{-4}$, with smaller time-steps only needed during particular moments such as during the initial contact of the drop. Simulations take on average a few hundred hours on a single CPU core, though this varies significantly with contact time (when the simulations end), drop radius and impact speed. 

The bath is solved on a separate triangular mesh to the drop, so that the boundary nodes of the surface elements of the drop and the bath do not necessarily `line up'. Therefore, before each time-step, a loop is performed over the integration points of the drop surface to identify the correct bath surface element it needs to get information from. This is done by solving for the intersection of the normal to the drop with each bath surface element in turn, and checking if the computed bath coordinate actually lies inside that bath surface element. An equivalent loop is performed for the bath surface elements.

\subsection{Benchmarking: Comparison to Volume-of-Fluid Simulations}\label{sec:vof_bounce}

\begin{figure}
\centering
 \begin{subfigure}{0.5\textwidth}
 \centering
\includegraphics[width=0.3\textwidth]{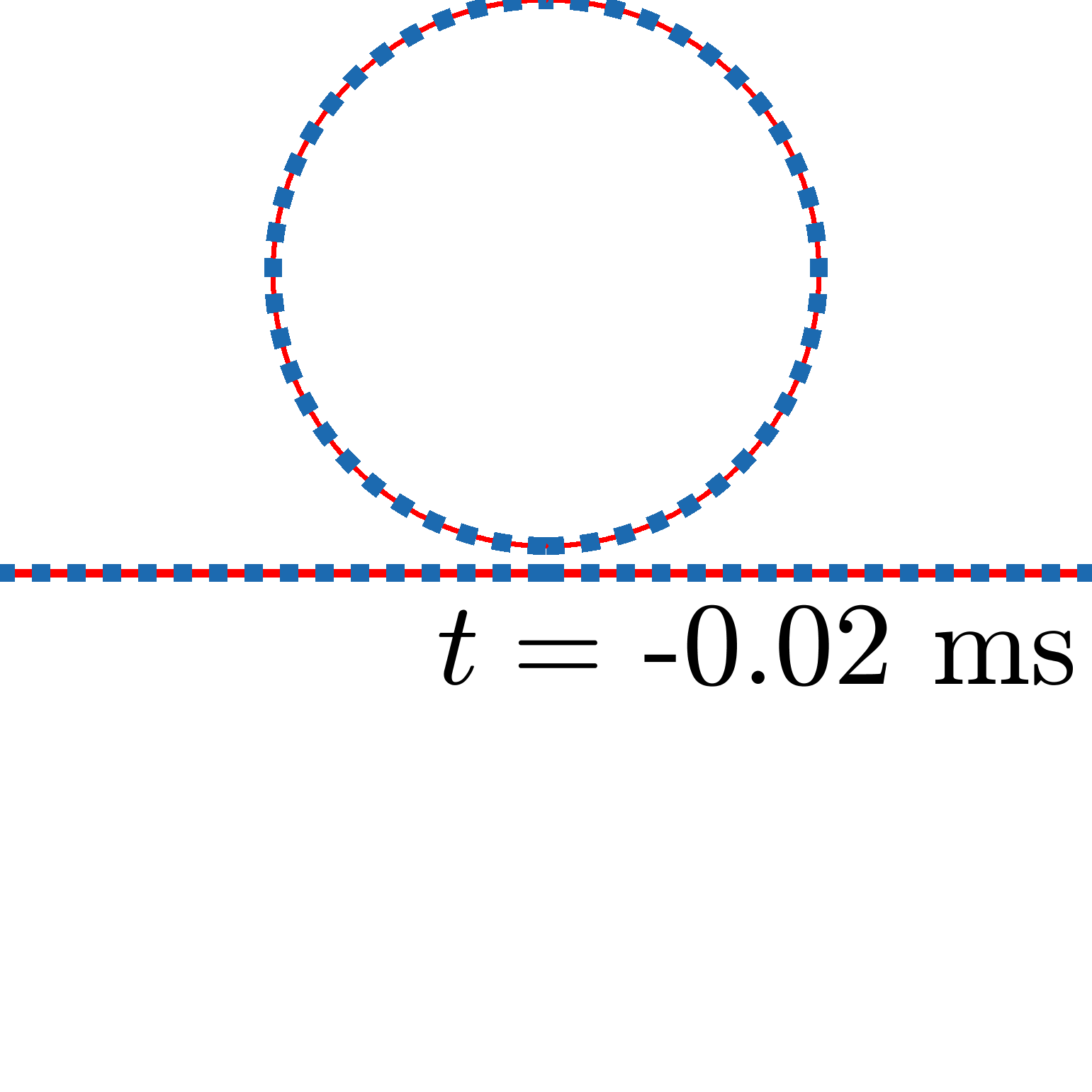}
\includegraphics[width=0.3\textwidth]{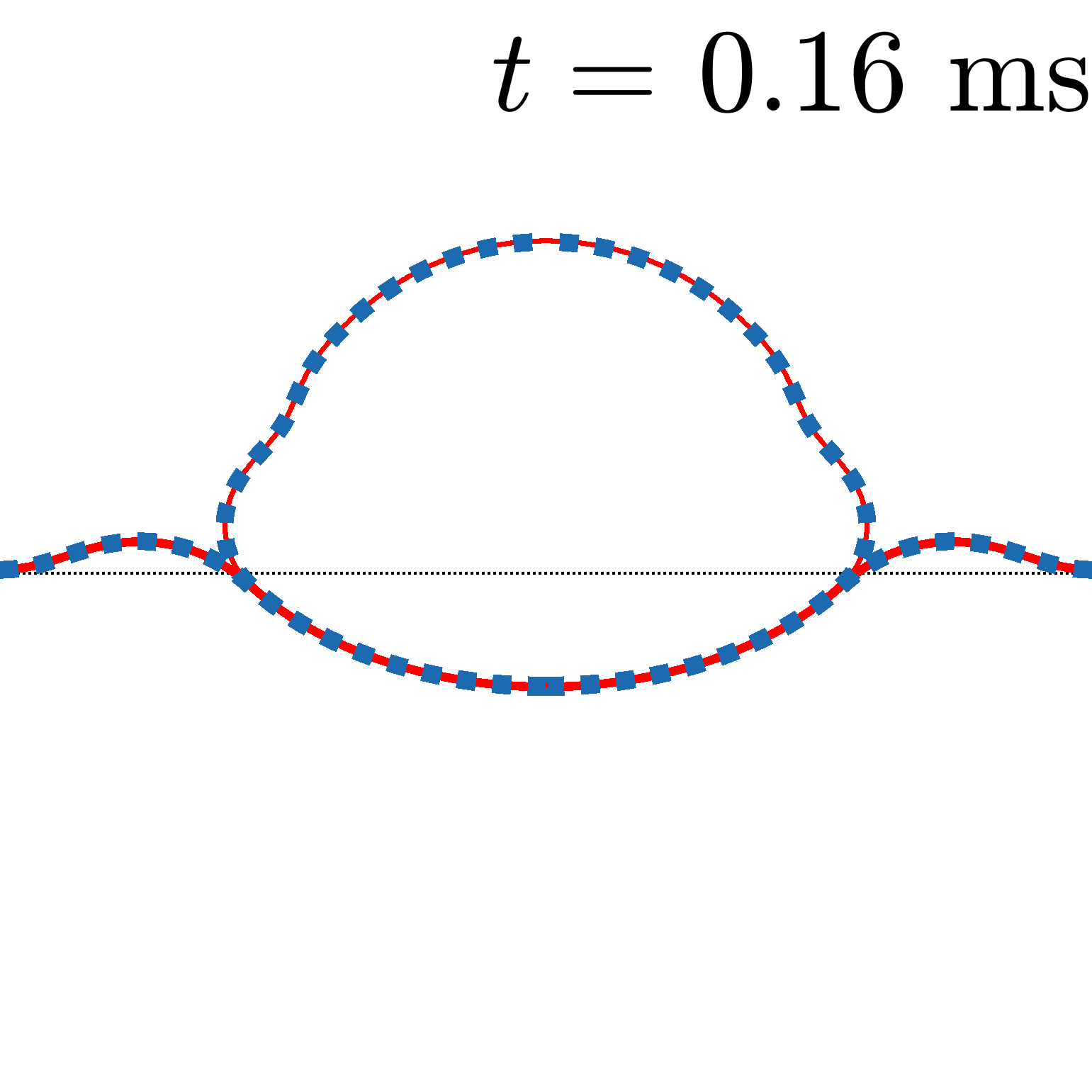}
\includegraphics[width=0.3\textwidth]{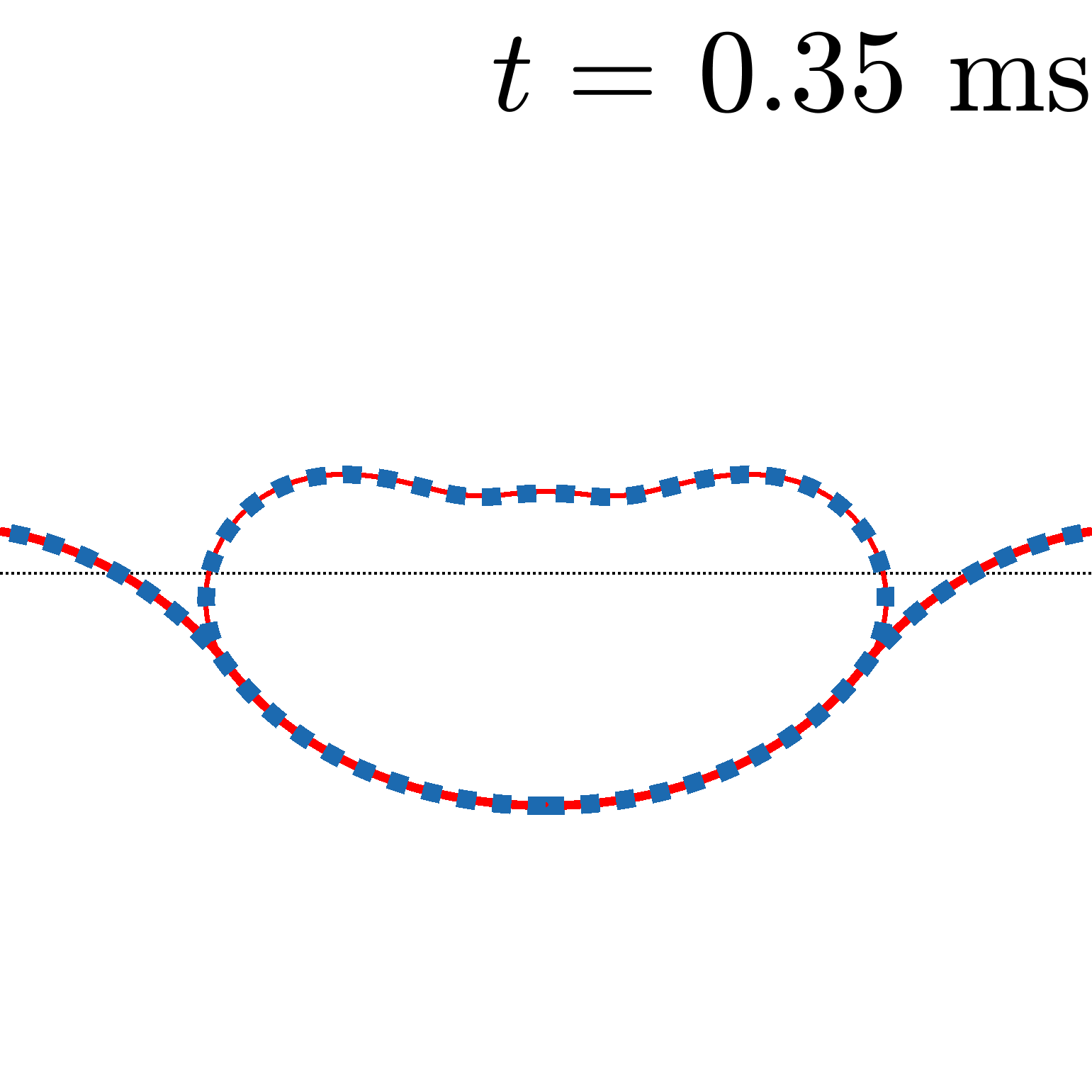}

\includegraphics[width=0.3\textwidth]{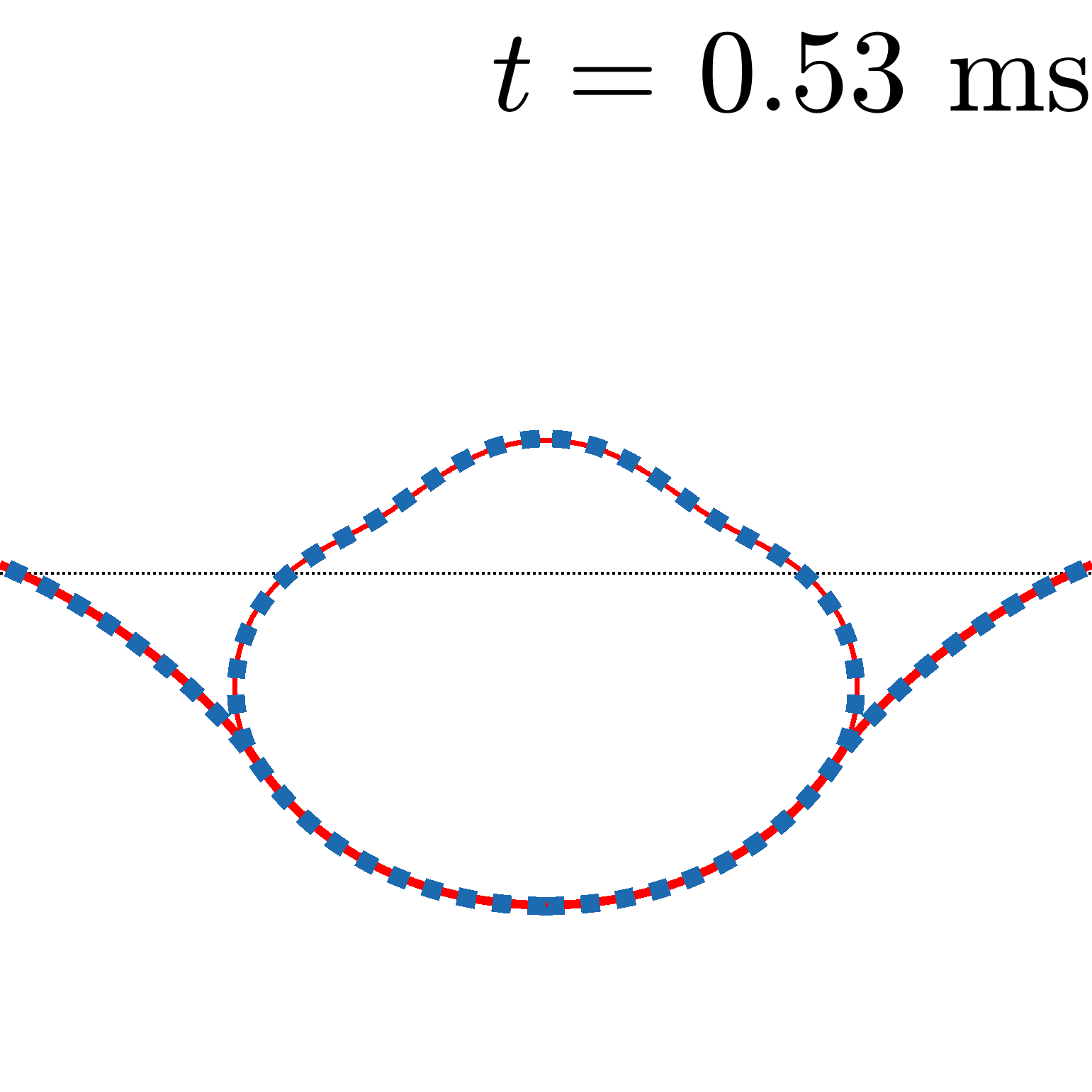}
\includegraphics[width=0.3\textwidth]{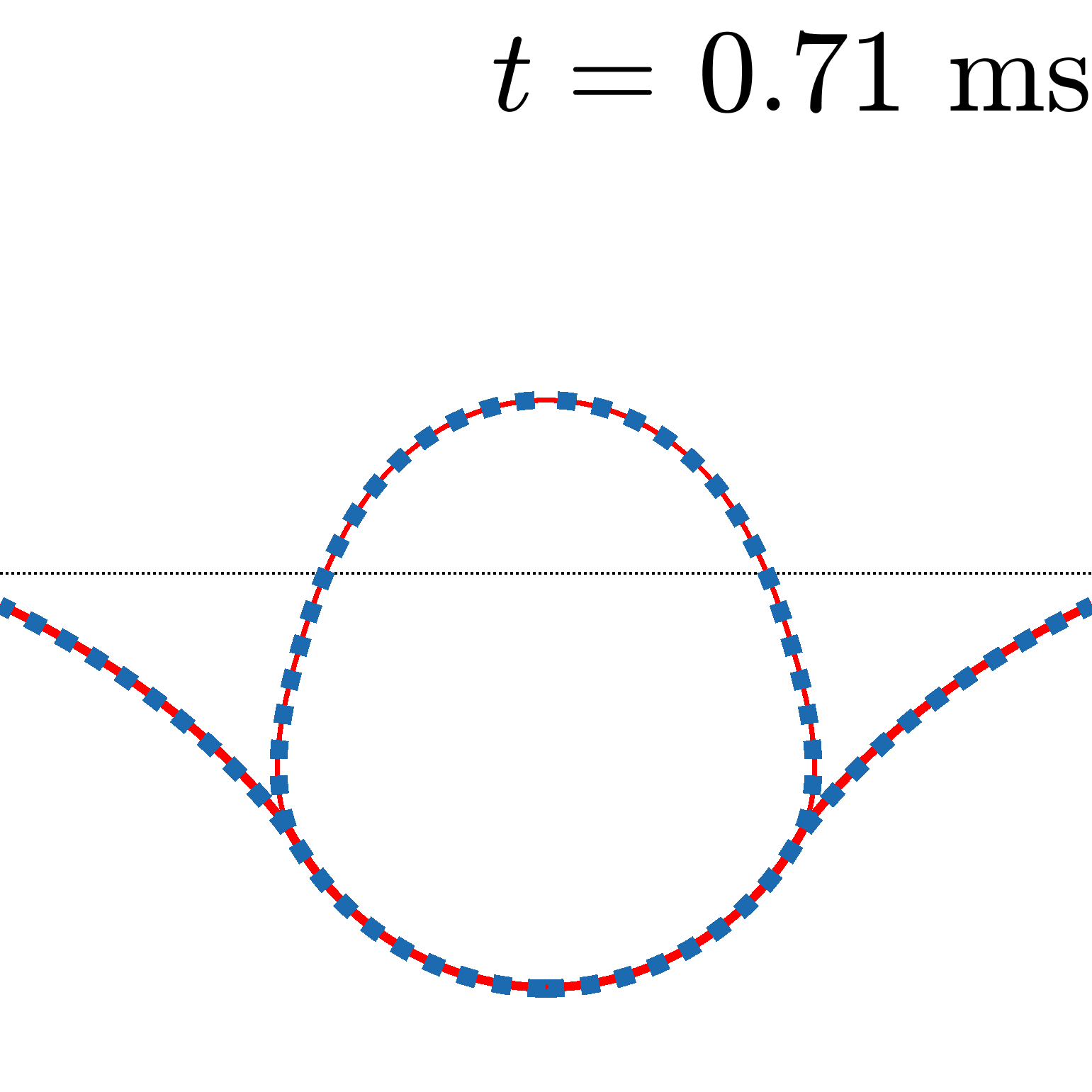}
\includegraphics[width=0.3\textwidth]{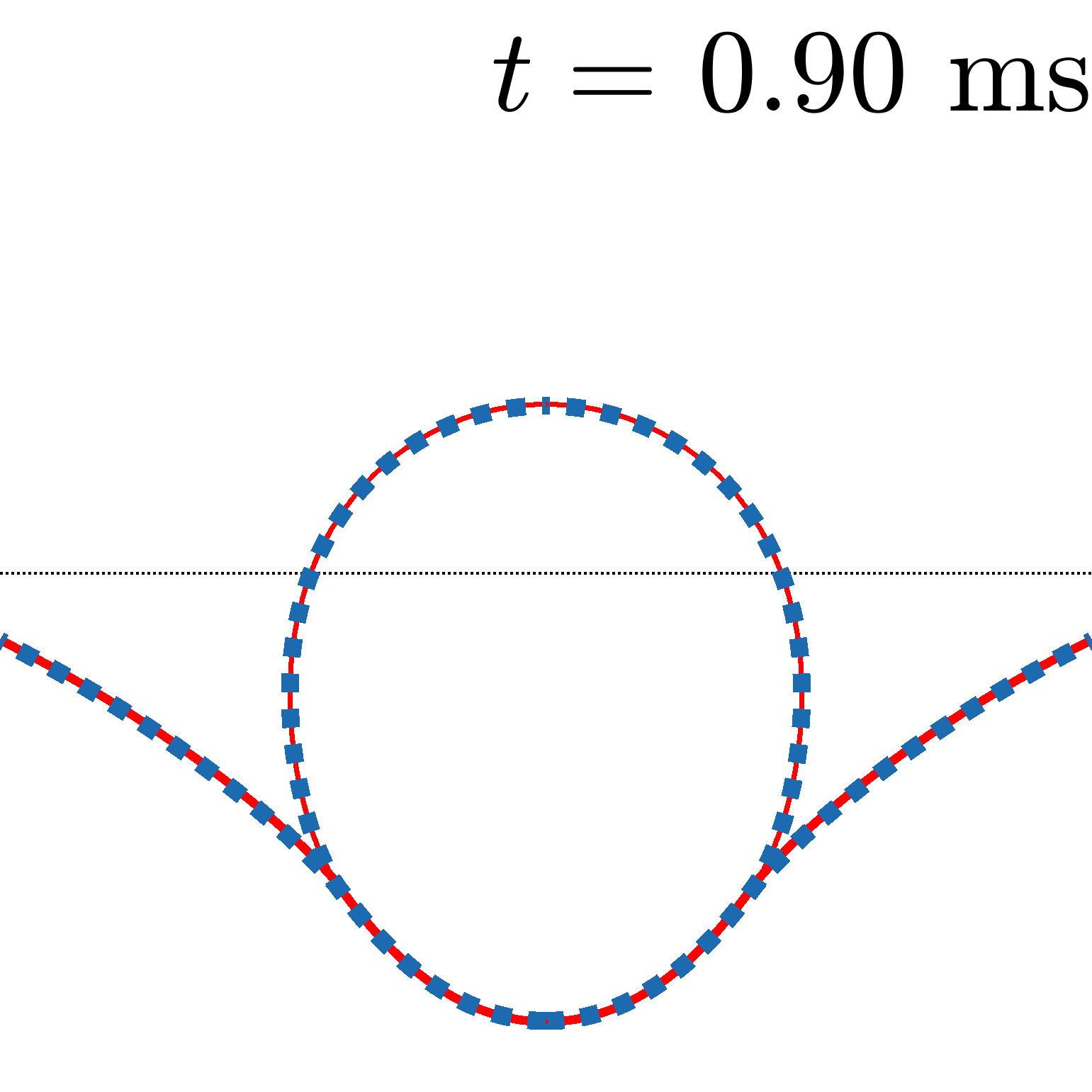}

\includegraphics[width=0.3\textwidth]{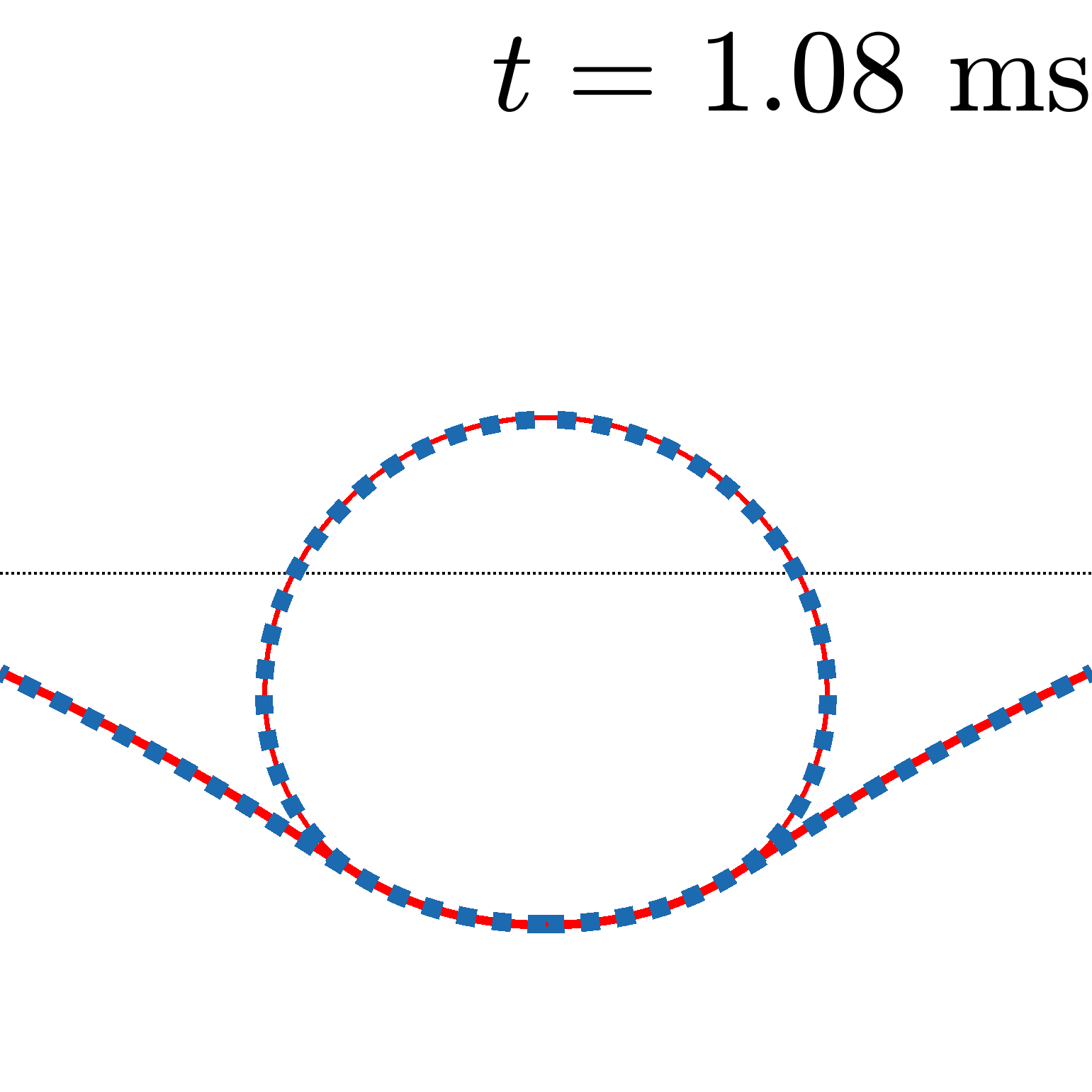}
\includegraphics[width=0.3\textwidth]{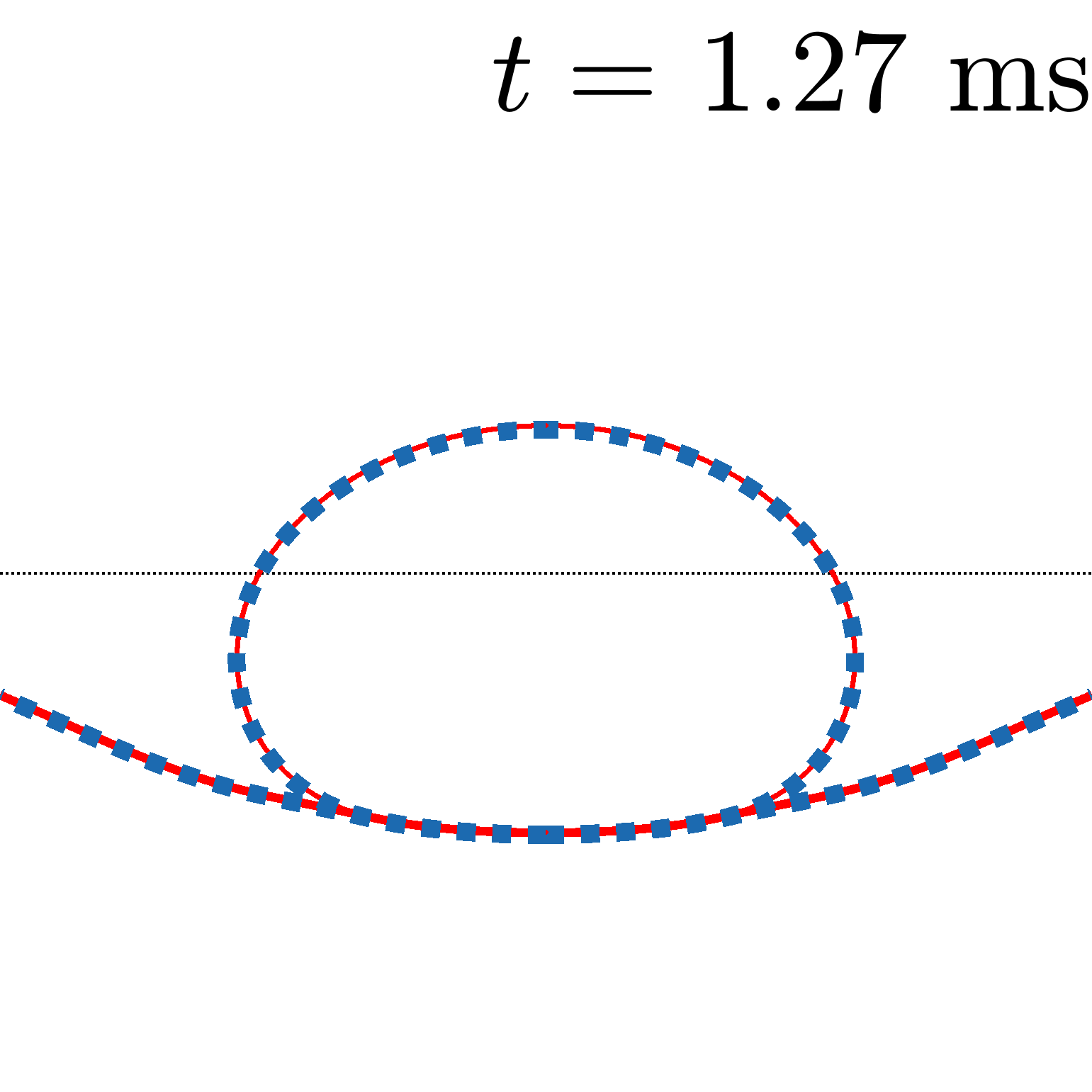}
\includegraphics[width=0.3\textwidth]{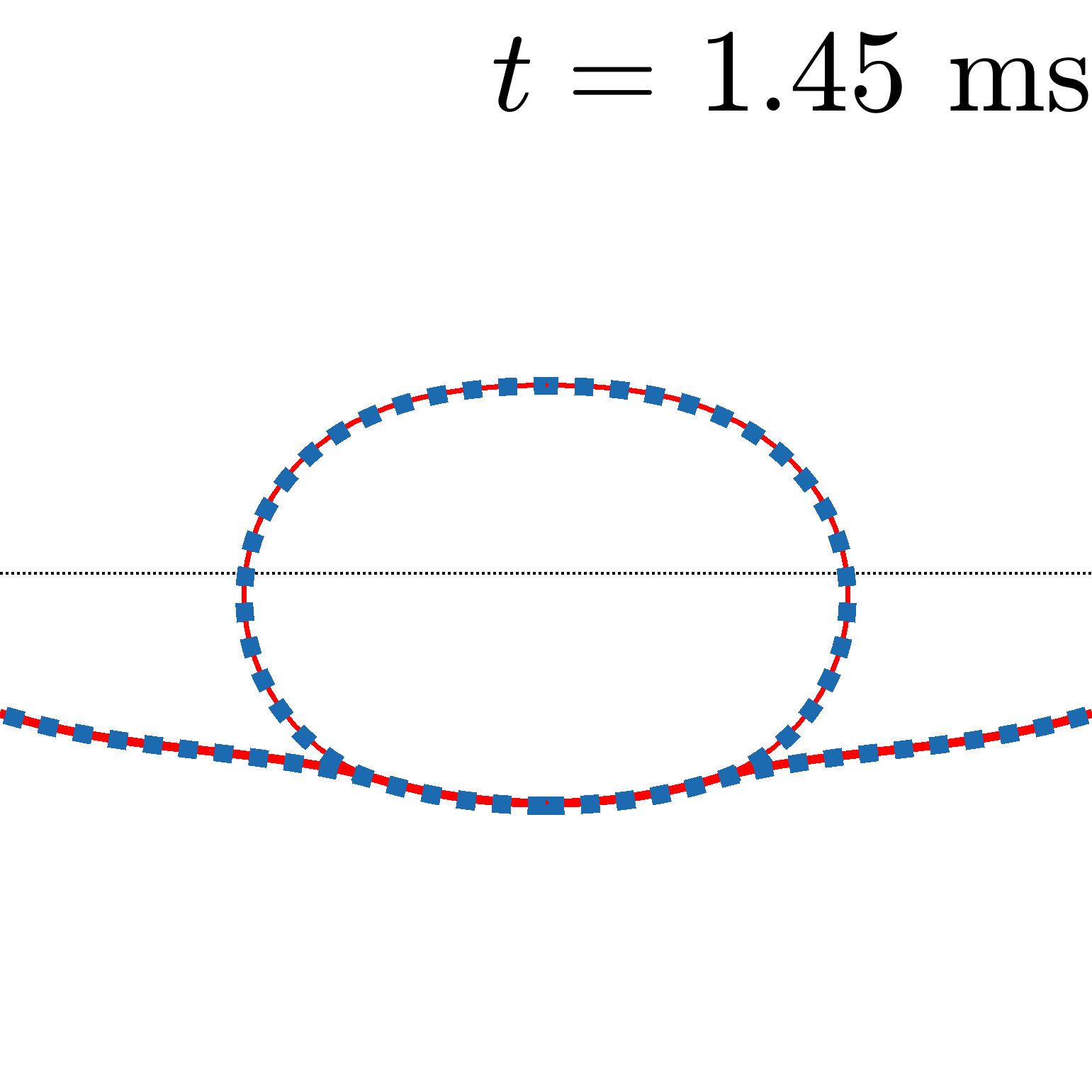}

\end{subfigure}%
~
\begin{subfigure}{0.5\textwidth}
\centering
\includegraphics[width=.9\textwidth]{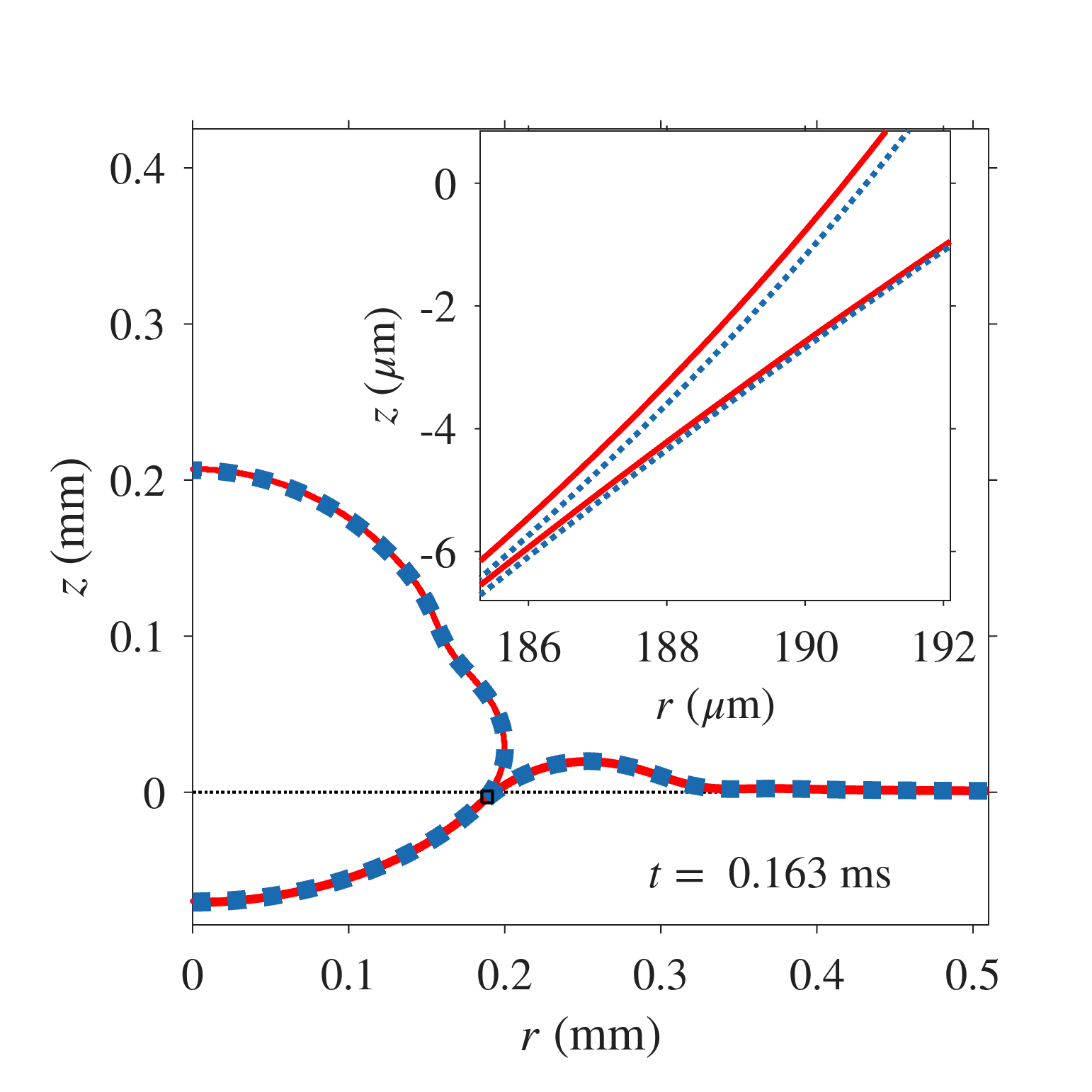}
\end{subfigure}

\caption{Snapshots (time increasing left-to-right, top-to-bottom) of 2~\si{\centi\stoke} silicone oil ($\rho^d=873$~\si{\kg\per\m\cubed}, $\mu^d=1.746$~\si{\milli\Pa\s},  $\gamma^d=18.7$~\si{\milli\N \per\meter}), with radius $R=170$~\si{\micro\meter}, at $\Web=5.5$. The bath depth is $H^b=3R$, and bath radius is $R^b=10R$. We depict the results of the companion forced-bouncing Basilisk VoF simulation in red. Overlaid as a dashed blue line is our finite element simulation. The thin black lines show the undeformed height of the bath. A video available as Movie 1. The right plot shows an enlargement of the second snapshot. The inset shows a zoomed view, showing the small difference between the surface positions of the two methods.}\label{fig:VOF_bounce_snaps}
\end{figure}

For both validation and interest, we take this opportunity to compare our novel finite element based model to results obtained via the well-established Volume-of-Fluid (VoF) method \citep{Hirt1981}.  These simulations were performed using the open-source \textcolor{blue}{\href{http://basilisk.fr/}{Basilisk}} package \citep{Popinet2015}, which has been used successfully for a wide variety of drop impact problems and related free surface flows \citep{Galeano-Rios2021,Alventosa2023,Gabbard2025,Phillips2025,Fudge2023,Sanjay2023,Zhang2022} -- all cases in which rebound is assumed and contact is numerically prevented.

We first compare the results for a case where our experiments (see below in \S\ref{sec:results_Deepbath}) show that the drop bounces from the bath: 2~\si{\centi\stoke} silicone oil with $R=170$~\si{\micro\meter}, at $\Web=5.5$. Here, we make a comparison with a VoF method that prevents coalescence and forces the drop to bounce from the bath for our parametric setup of interest. This method was used for drop-bath bouncing in \citet{Alventosa2023}, based on the method developed by \citet{Ramirez-Soto2020}. It uses two distinct so-called colour functions to define where the boundaries between the liquid and the gas are, one function for the drop and one for the bath. These `colour functions' are equal to 1 in their respective liquid domains (bath and drop) and $0$ in the gas and the other fluid domain.  Cells cut by interfaces take values between $0$ and $1$ and the interfaces are reconstructed using an implicit geometrical algorithm. By construction, this separates the drop and the bath, and computationally prevents them from merging. Code is available at \url{https://github.com/rcsc-group/BouncingDroplets}.

For a case in which our finite element simulation predicts bouncing, we compare to the forced-bouncing VoF simulation in figure~\ref{fig:VOF_bounce_snaps}, with a video available as movie 1. The macroscopic agreement is excellent, validating both methods in cases where the drops are known to bounce. If we `zoom in' on the gas film, as shown in the inset of figure~\ref{fig:VOF_bounce_snaps}, the surfaces do not exactly line up. This is not surprising, as even a perfect VoF simulation does not include gas-kinetic effects and we should not expect agreement on the film thickness. Nevertheless, the overall agreement for the drop shapes is remarkably good and very reassuring.

In appendix \S\ref{app:VoF_postmerge}, we consider a case where the experiments show that the drop merges with the bath, and compare our finite element simulation to a VoF method that allows merging to occur. We find that our finite element method predicts that contact occurs at a much later time than that found by the VoF method, and in \S\ref{app:VoF_postmerge} we discuss reasons for this discrepancy and prospects for improvement for VoF.

\section{Experimental Methodology}\label{sec:meth_exp}

In order to validate our quantitative predictions for the bounce-merge threshold, companion experiments were performed for a range of droplet sizes and viscosities, as summarized in Table \ref{tab:results_db}.  The experimental measurements reported here are a subset of an extended experimental campaign examining the influence of the bounce-merge threshold on motion of the liquid bath \citep{Harris2026}, with the results reported here exclusively for the case where the bath is at rest.  For these experiments, submillimetric droplets of silicone oil are generated using a custom piezoelectric droplet-on-demand generator \citep{harris2015low, ionkin2018note}, with the latest version used for this work documented at \url{https://github.com/harrislab-brown/DropGen}. Droplets are released from a controlled height, and impact normally on a still bath of the same fluid.  The depth of the bath is 4.5 mm, significantly larger than the droplet radii $R$, and thus confidently in the deep-bath regime \citep{Pan2007,Tang2016,tang2018bouncing}.  The dynamics are filmed from the side with a high-speed camera, the recording of which is analysed to compute the droplet size, impact speed, and impact outcome (bounce or merge) as in prior work \citep{Alventosa2023,Harris2026}.  To determine the merge threshold, the generator is incrementally raised by 1 mm until 13 of 15 successive droplets merge.  To determine the bounce threshold, the generator is then lowered by 1 mm until 13 of 15 successive droplets merge.  These transition points are very closely separated, generally by less than 2 cm/s in impact speed.  This process is repeated for a total of nine times, often on different days, and the resulting 18 measurements are averaged for a single prediction of the bounce-merge threshold reported here.  Errors bars include one standard deviation of these measurements and propagated uncertainty on all other parameters, assuming they are uncorrelated.  More extensive details on the experimental protocol and processing are available in the recent publication by \citet{Harris2026}.

\section{Results: Deep Bath Impacts}\label{sec:results_Deepbath}

A natural place to start our investigation is to remove dependencies on bath height and consider the `deep bath limit'.  Later, in \S\ref{sec:results_shallow_bath}, we will look at the height at which this can be assumed to hold. 

\subsection{Experimental Comparison}

\addtolength{\tabcolsep}{-1pt}
\begin{table}
\small %
    \centering

        \begin{tabular}{@{}ccccccc@{}}
             & \multicolumn{3}{c}{2~\si{\centi\stoke} Silicone Oil} 
           & \multicolumn{2}{c}{20~\si{\centi\stoke} Silicone Oil}& 50~\si{\centi\stoke} Silicone Oil  \\
            \cmidrule(lr){2-4}\cmidrule(lr){5-6}\cmidrule(lr){7-7}
        \makecell{Experimental \\   Drop Radius (\si{\micro \m}) }     
           &$168\pm 7$ & $229\pm 5$  & $404\pm 6$ &$205\pm 7$ &   $393\pm 6$ &$439\pm 14$  \\ \midrule%
 \makecell{           Experimental \\Critical Impact \\ Speed (\si{\centi\m\per\s})}
            & $89.9\pm  1.8$ &$80.3\pm 1.3 $&$60.7\pm 1.2 $ &$97.6 \pm 2.3$&  $74.0 \pm 1.7$ & $79.0\pm 1.9$\\ \midrule%
 \makecell{Experimental \\  Critical $\Web$}
            & \makecell{6.34 \\(6.08--6.59) } & \makecell{6.88 \\ (6.65--7.11)} &\makecell{6.95 \\ (6.67--7.23)}  &\makecell{9.00 \\(8.58--9.43)}&\makecell{9.90 \\  (9.44--10.36)}
            &\makecell{12.65 \\  (12.05--13.25)}
            \\ \midrule%
            Simulation $R$ (\si{\micro\m}) &$170$ & $230$  & $405$ &$205$ &   $395$ &$440$  \\\midrule
 \makecell{Simulation \\ Critical $\Web$}
            & \makecell{6.25\\ $\pm0.05$}  & \makecell{6.65\\ $\pm0.05$} &\makecell{6.95\\ $\pm0.05$} &\makecell{8.35\\ $\pm0.05$}& \makecell{ 9.85\\ $\pm0.05$}& \makecell{12.25\\ $\pm0.05$} \\ 
            \bottomrule
        \end{tabular}
         \caption{Comparisons of our simulation results to the experiments for the critical Weber number of the hard transition. The error listed is one standard deviation. Physical parameters used: 2~\si{\centi\stoke} silicone oil: $\rho^d=873$~\si{\kg\per\m\cubed}, $\mu^d=1.746$~\si{\milli\Pa\s},  $\gamma^d=18.7$~\si{\milli\N \per\meter};\\ 20~\si{\centi\stoke} silicone oil: $\rho^d=949$~\si{\kg\per\m\cubed},  $\mu^d=18.98$ ~\si{\milli\Pa\s} $\gamma^d=20.6$~\si{\milli\N \per\meter};\\ 50~\si{\centi\stoke} silicone oil: $\rho^d=960$~\si{\kg\per\m\cubed},  $\mu^d=48.0$ ~\si{\milli\Pa\s} $\gamma^d=20.8$~\si{\milli\N \per\meter}; \\  all simulations: $\mu^g=18.1$~\si{\micro\Pa\s}, $A=5.03\times 10^{-20}$~\si{\joule}, $\lambda=69.0$~\si{\nano\meter}, $g=9.81$~\si{\m \per\second\squared}. 
        \label{tab:results_db}}
   
\end{table}
\addtolength{\tabcolsep}{1pt}

\begin{figure}
    \centering
    
    \includegraphics[width=\linewidth]{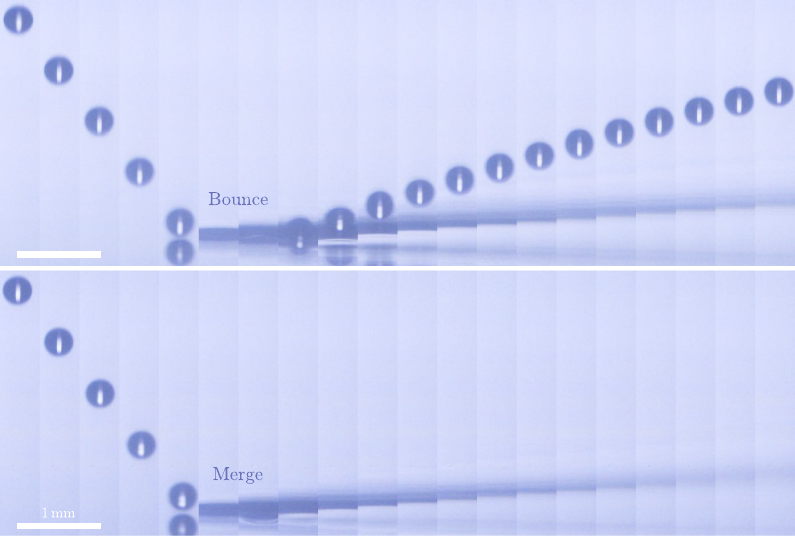}
    \caption{Experimental images straddling the critical We number for 2~\si{\centi\stoke} silicone oil drops with $R=168\pm7$ \si{\micro\m}, showing bouncing at $U=88.6 \pm 0.7$ cm/s (top) and merging at $91.1 \pm 1.0$ cm/s (bottom). The images in each sequence are spaced 1/1500 seconds apart. }
    \label{fig:experimental_images}
\end{figure}

\begin{figure}

 \centering
 \includegraphics[width=0.195\linewidth]{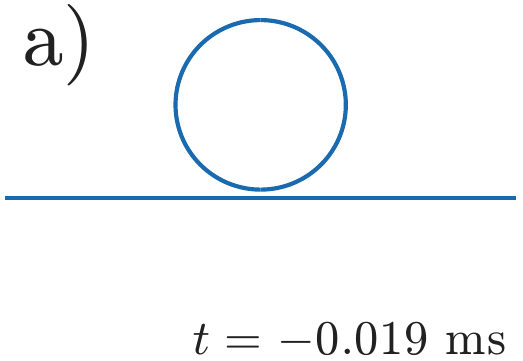}
     \includegraphics[width=0.195\linewidth]{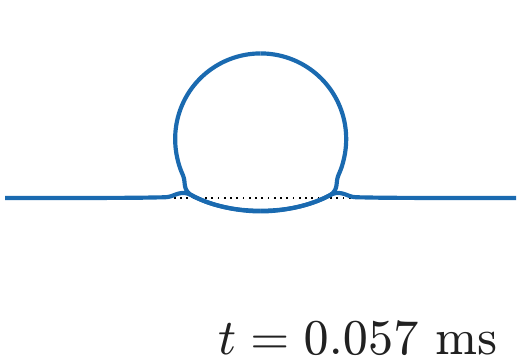}
     \includegraphics[width=0.195\linewidth]{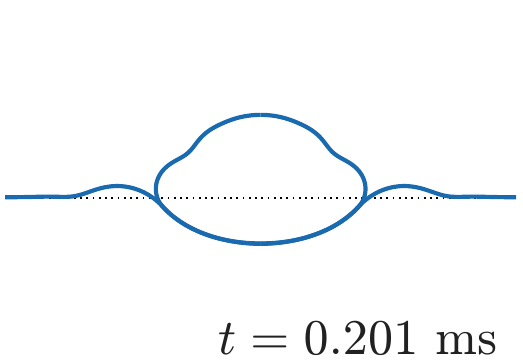}
     \includegraphics[width=0.195\linewidth]{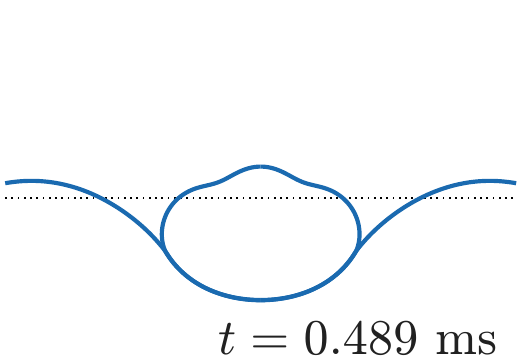}  
     \includegraphics[width=0.195\linewidth]{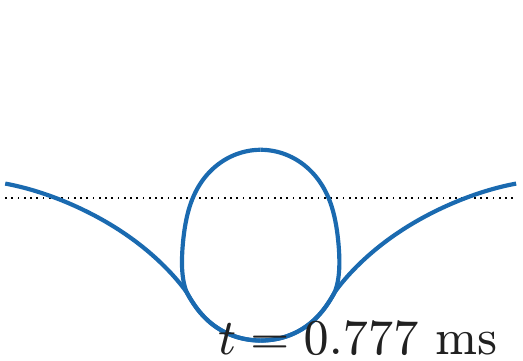} 
     
     \includegraphics[width=0.195\linewidth]{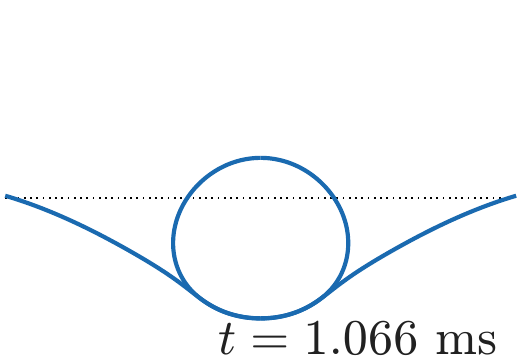}
     \includegraphics[width=0.195\linewidth]{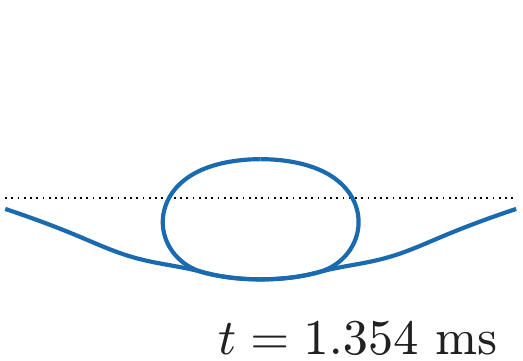}
      \includegraphics[width=0.195\linewidth]{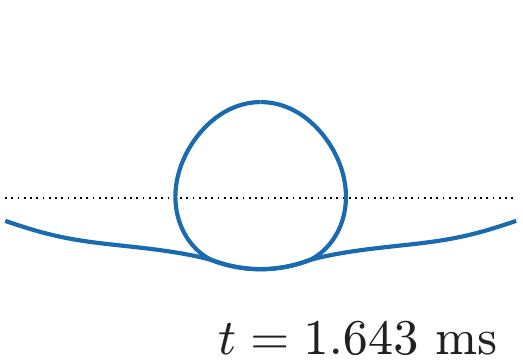}   
     \includegraphics[width=0.195\linewidth]{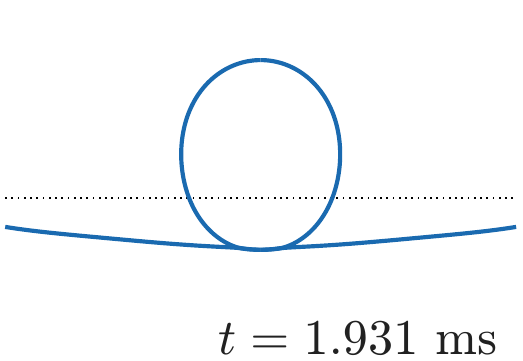}
     \includegraphics[width=0.195\linewidth]{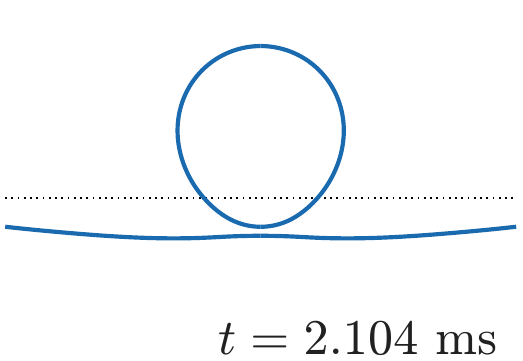}

      \includegraphics[width=.9\linewidth]{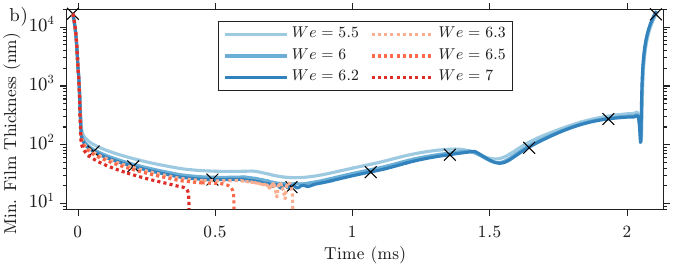}

\includegraphics[width=.9\linewidth]{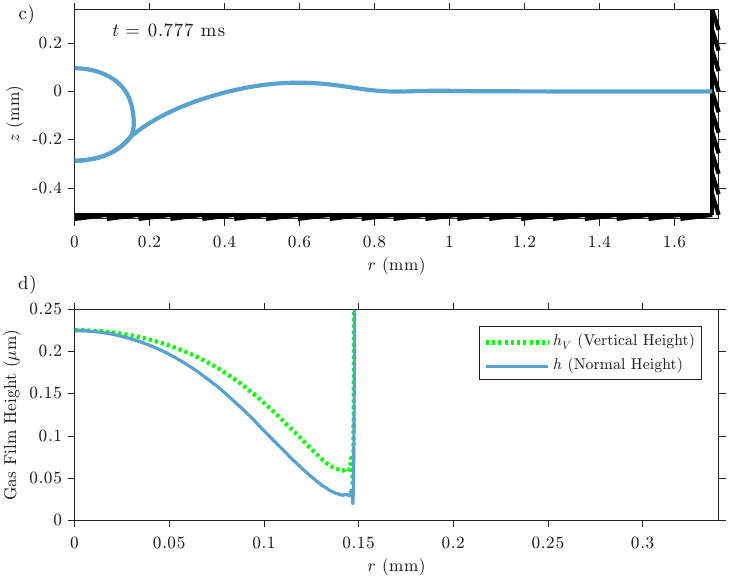}
  
   \caption{a,c) Snapshots of the $\Web=6.2$, $R=170$~\si{\micro\meter}, simulation for 2~\si{\centi\stoke} silicone oil, just below the hard transition (increasing impact speed will lead to merging). The snapshot shown in c) (at $t=0.777$~\si{\milli\s}) shows the full simulation domain at a key moment in the dynamics. In b) the minimum film thickness is plotted against time for a range of Weber numbers either side of the hard transition. The crosses show the times of the snapshots shown in a) and c). Below, in d), is the gas film height at $t=0.777$~\si{\milli\s}, in both the normal and vertical directions.    } \label{fig:min_height_1}
\end{figure}

A quantitative comparison between the results of our finite element simulations and our experiments, for the `hard' transition from bouncing to merging with increasing Weber number, is provided in Table \ref{tab:results_db}. We use silicone oils of three different viscosities, at different drop radii, for a total of six experimental cases. Photographic snapshots of bouncing and merging experiments are shown in figure~\ref{fig:experimental_images}. As can be seen in the table, the agreement for the critical Weber is excellent, with the value found by the simulations lying within the experimental error bar (1 standard deviation), or just outside of it for all the cases considered.

Focusing on a particular experiment, in figure~\ref{fig:min_height_1}a we show the simulation of 2~\si{\centi\stoke} silicone oil at $R=170$~\si{\micro\meter}, which at $\Web=6.2$ is just `below' the transition to merging (i.e. increasing the impact speed further leads to merging). Videos of this case are available as a Supplementary Movie 2. To probe the film dynamics, in Figure~\ref{fig:min_height_1}b one can see the minimum of the film height (in the normal direction, as described in \S\ref{sec:meth_gas_model}) against time, for a range of Weber numbers either side of the bounce-merge transition. For $\Web=6.3$ and above, the drop makes contact with the bath, with the film height rapidly approaching $0$ just before contact, driven by the disjoining pressure.

In figure~\ref{fig:min_height_1}c, we show the full simulation domain (with the shaded boundary showing the edge of the bath) of the $\Web=6.2$ bouncing simulation at $t=0.777$~\si{\milli\s}, which is when contact occurs at $\Web=6.3$. This shows how late through the bouncing process contact can occur, with the gas film persisting for a significant period of time. Below, in figure~\ref{fig:min_height_1}d, is the height of the gas film, with a comparison to the vertically measured height. This shows the significant difference between the two measurements, thus providing further justification for our curved lubrication approach in the film. 

\subsection{Dependence on Drop Radius}

\begin{figure}
\centering
\includegraphics{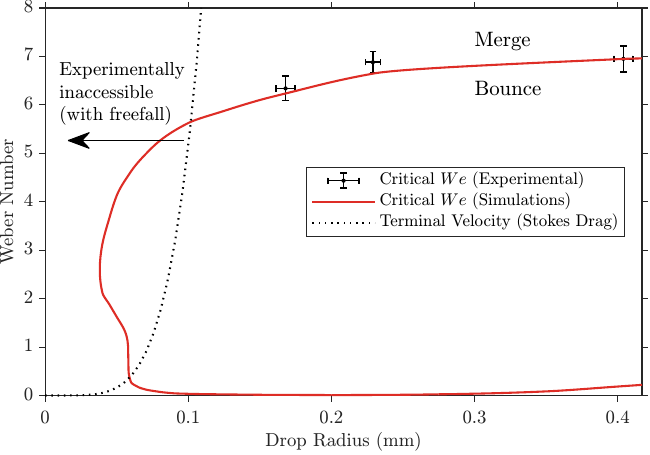}
\caption{Regime diagram for the outcome of the impact of a drop of 2~\si{\centi\stoke} silicone oil onto a deep bath of the same fluid, in air at atmospheric pressure. The red solid curve shows the boundary between bouncing and merging.}\label{fig:phase_1}
\end{figure}

After showing strong agreement against our experiments, further simulations have been performed to produce a regime diagram of Weber number against drop radius, using the same liquid as a basis for parameters, as shown in figure~\ref{fig:phase_1}. One can see that the critical Weber number for the hard transition (from bouncing to merging with increasing Weber number) tends towards a constant Weber number with increasing drop radii (in this case, around $\Web=7$). On the other hand, as the drop radius decreases, the critical Weber number decreases rapidly, until a critical drop radius at around $R=0.04$~\si{\milli\meter} is reached, below which the drop always merges with the bath, a phenomena similar to that seen for head-on drop-drop collisions both experimentally \citep{Huang2021} and in simulations using our methodology \citep{Lewin-Jones2024}. However, this always-merging regime is not accessible using the experimental technique used here, which varies the impact speed via the height that the drop is released from. This means there is a maximum speed (and therefore Weber number) that is accessible with this technique: the terminal velocity of the drop. Estimating the drag on the drop with Stokes drag, the Weber number at the terminal velocity is also plotted in figure~\ref{fig:phase_1}, creating a region of the regime diagram that is inaccessible to these free-fall experiments. \footnote{The assumption of Stokes drag here is not entirely valid, as the Reynolds number reaches $\sim 10$ for the terminal velocity curve shown here, whereas Stokes drag is valid only for $\Rey\ll 1$. However, Stokes drag is an underestimate of the drag, and therefore the inaccessible region is larger than that plotted, and the critical radius definitely lies within the inaccessible region.}  The critical radius below which the drop always merges is within this inaccessible region, and so has not been seen by our current experiments. In future, it may be possible to access this region by using an experimental technique that propels the drop above the terminal speed, or by exploring a different part of the parameter space by changing the properties of the gas, which we will discuss in \S\ref{sec:pressure}.

\subsection{Low Speed Impacts: Merging vs Floating}\label{sec:soft_transition}

\begin{figure}
\centering
\includegraphics{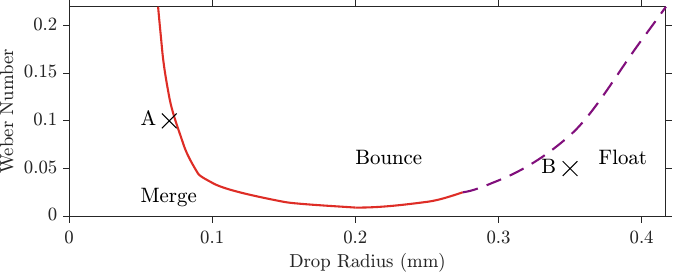}
   \caption{A zoomed view of the lower part of figure~\ref{fig:phase_1}, showing the `soft' transition with increasing Weber number. For small drops, the outcome changes from merging to bouncing with increasing Weber number, and the red solid curve shows the critical Weber number. The simulations at and around the point labelled A is shown in figure~\ref{fig:soft_merge}. For large drops, instead the outcome changes from floating to bouncing with increasing Weber number, and the dashed purple line shows the critical Weber number. The simulations around the point labelled B are shown in figure~\ref{fig:soft_float}.} \label{fig:phase_1_zoom}
\end{figure}

Figure~\ref{fig:phase_1_zoom} shows the lower part of figure~\ref{fig:phase_1}, magnified to show the low speed `soft' transition from merging to bouncing with increasing Weber number. Notably, as the drop radius gets larger (beyond $R\approx0.2$mm), the critical Weber number starts \emph{increasing} with the drop radius. In this region, we find that there is a qualitative change in the behaviour of the drops, labelled `Merge' and `Float' in figure~\ref{fig:phase_1_zoom}.

\begin{figure}
\centering
\includegraphics[width=0.19\textwidth]{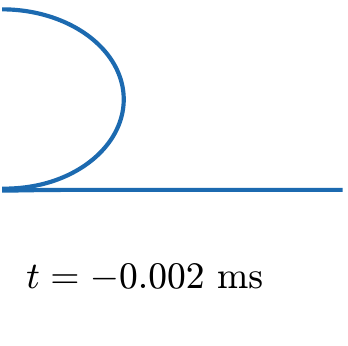}
\includegraphics[width=0.19\textwidth]{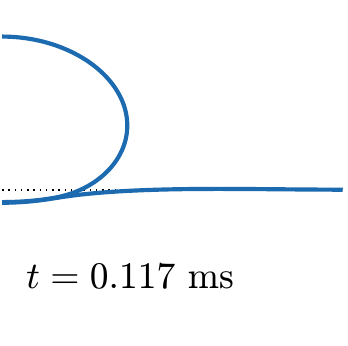}
\includegraphics[width=0.19\textwidth]{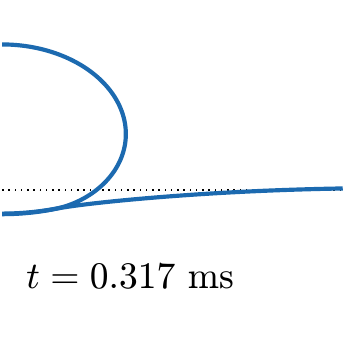}
\includegraphics[width=0.19\textwidth]{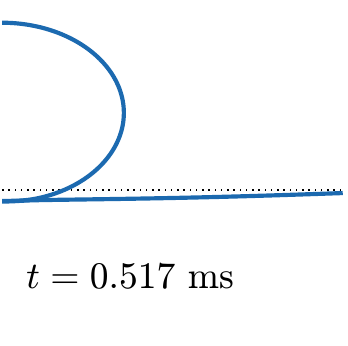}
\includegraphics[width=0.19\textwidth]{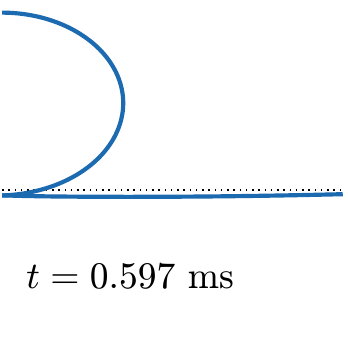}

 \includegraphics[width=\linewidth]{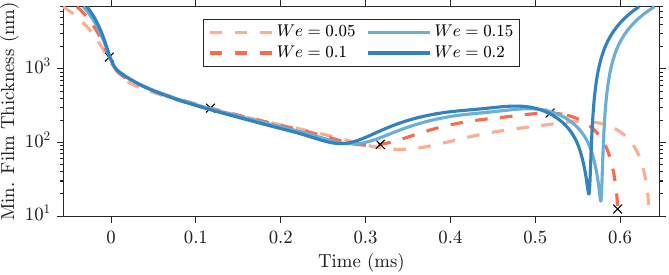}
  \caption{Minimum film heights for simulations either side of the `soft' transition for $R=70$~\si{\milli\m} drops of 2~\si{\centi\stoke} silicone oil, showing the small radius regime labelled `Merge' in figure \ref{fig:phase_1_zoom}. The snapshots above show the $\Web=0.1$ simulation (point A in figure~\ref{fig:phase_1_zoom}) at the points labelled by crosses, with a video as Movie 3.} \label{fig:soft_merge}
\end{figure}

The first regime, for small radii, is labelled `Merge' in figure~\ref{fig:phase_1_zoom}. A minimum height plot at $R=70$\si{\micro\m} is shown in figure~\ref{fig:soft_merge}, with the $\Web=0.1$ simulation corresponding to the point labelled `A' in figure~\ref{fig:phase_1_zoom}, with a video available as supplementary movie 3. In this regime, below the transition the drop merges at the end of the impact, at the moment it would otherwise bounce away from the bath surface. This type of ‘dimple collapse’ contact was seen in simulations of \cite{Bach2004}, and for drop-solid impacts by \citet{Chubynsky2020}.

\begin{figure}
\centering
\includegraphics[width=0.19\textwidth]{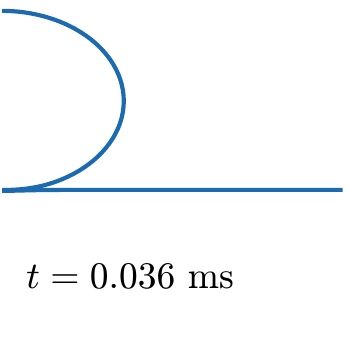}
\includegraphics[width=0.19\textwidth]{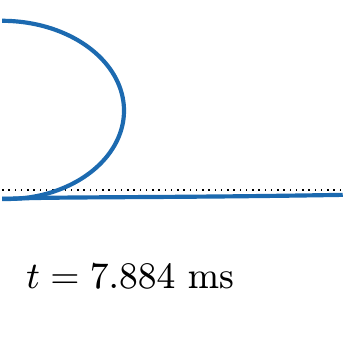}
\includegraphics[width=0.19\textwidth]{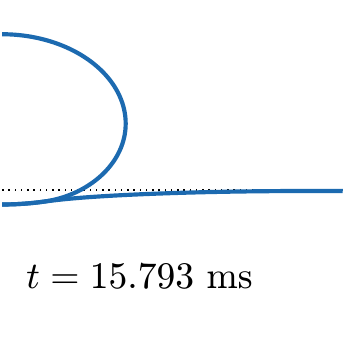}
\includegraphics[width=0.19\textwidth]{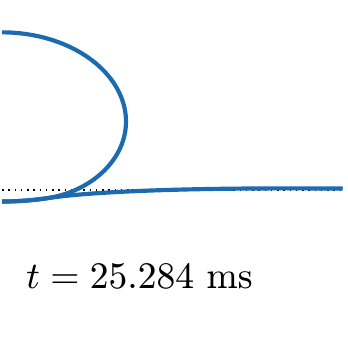}
\includegraphics[width=0.19\textwidth]{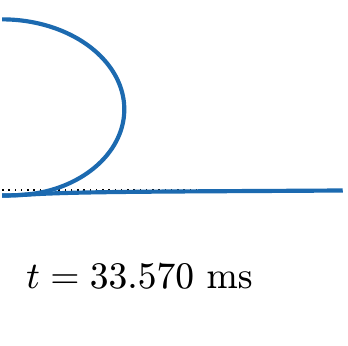}

    \includegraphics[width=\linewidth]{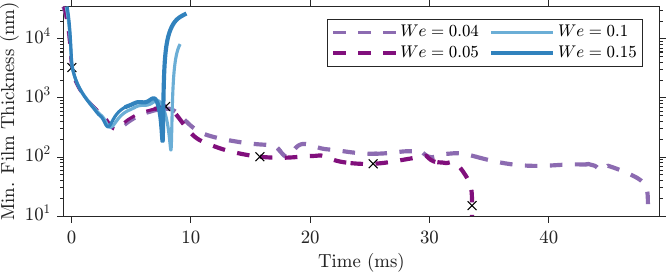}
 \caption{Minimum film heights for simulations of $R=350$~\si{\milli\m} drops of 2~\si{\centi\stoke} silicone oil, either side of the `soft' transition in the large radius bounce-float regime. The snapshots above show the $\Web=0.05$ simulation (point B in figure~\ref{fig:phase_1_zoom}) is the point labelled B in figure \ref{fig:phase_1_zoom} at the points labelled by crosses, with a video available as movie 4.} \label{fig:soft_float}
\end{figure}

For large radii, the drop fails to bounce but does not immediately merge -- we call this the `floating' regime.  A transition between bouncing and floating has previously been identified, both experimentally and in simulations by, among others, \citet{Sanjay2023} and \cite{Alventosa2023}. This is characterised as where the coefficient of restitution tends to zero, and, notably, does not require the micro-scale physics included in our simulations to be identified. An example is shown in figure~\ref{fig:soft_float} for the minimum film height at $R=350$\si{\micro\m} for a range of $\Web$. At $\Web=0.05$, the point labelled `B' in figure~\ref{fig:phase_1_zoom}, the drop floats for over 48~\si{\milli\s}, where as if it was in the `merging' regime it would merge at $\sim 10$\si{\milli\s}. 

\begin{figure}
\centering

\includegraphics[width=.8\linewidth]{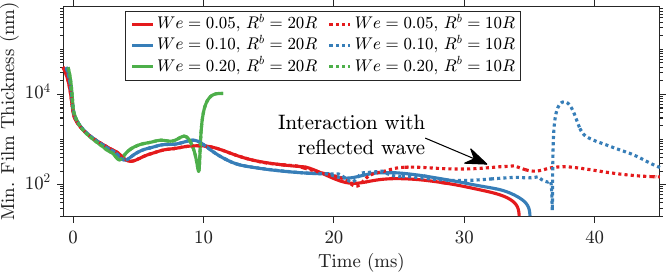}
\caption{Minimum film heights for simulations of $R=350$~\si{\milli\m} drops of 2~\si{\centi\stoke} silicone oil, either side of the bounce-float regime. Two different bath radii are used: $R^b=20R$, and $R^b=10R$  (as used in all other simulations in this article). The floating $R^b=10R$ simulations are affected by reflections of the waves on the bath surface, but only once the drops have started floating. }\label{fig:float_farfield}
\end{figure}

In the floating regime, we do observe some interaction with the far wall of the bath (as shown in figure~\ref{fig:initial_conditions}), as the wave produced by the impact has time to be reflected by the far wall. In figure~\ref{fig:float_farfield} we show the minimum film height for two sets of simulations of the 2~\si{\centi\stoke} silicone oil at $R=400$~\si{\micro\m}: one with the bath radius at its regular value of $R^b=10R$, and the second with the bath radius doubled to $R^b=20R$. We find that increasing the bath radius decreases the time the drop floats for in the floating regime ($\Web=0.1$ and $\Web=0.15$ in figure~\ref{fig:float_farfield}), as the reflected wave no longer has time to interact with the floating drop. Importantly, this does not affect whether the impact in question is in the floating regime or not (as can be seen for the bouncing $\Web=0.2$ case in figure~\ref{fig:float_farfield}), only affecting the dynamics that occur once the drop has started floating. 

\subsection{Effect of the Ambient Air Pressure}\label{sec:pressure}

\begin{figure}
\centering
\includegraphics{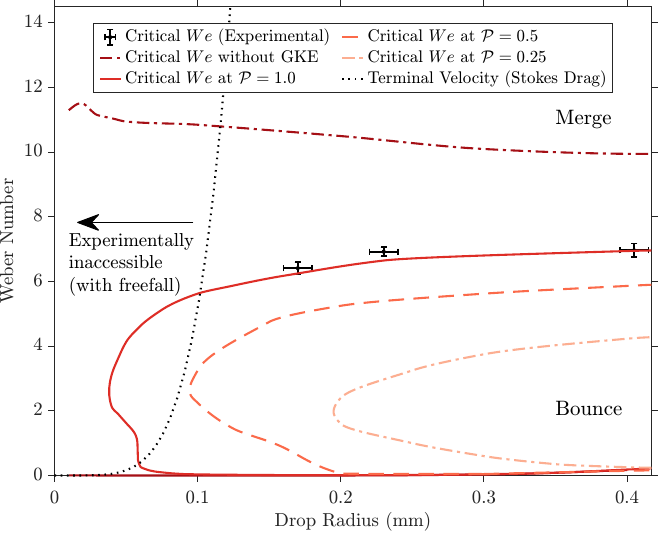}
\caption{Regime diagram for the outcome of the impact of a drop of 2~\si{\centi\stoke} silicone oil onto a deep bath of the same fluid, in air at different ambient pressures, where $\mathcal{P}$ is the ratio of the ambient pressure to atmospheric pressure. Also shown is the outcome when gas kinetic effects are removed, corresponding to $\mathcal{P}\to\infty$.} \label{fig:phase_1_all}
\end{figure}

Thus far, we have considered the gas to be air at room temperature and pressure, where the mean free path is $69$~\si{\nano\meter}. However, varying the mean free path can have a dramatic effect on the outcome of a drop impact and this can be achieved by either changing the gas or the ambient air pressure, which we now consider. 

For fixed temperature, the mean free path path of a gas varies inversely proportionally to the ambient pressure: for an ideal gas, the mean free path is related to the gas pressure $p$ by $\lambda=k_BT/(\sqrt{2}\pi d^2 p)$, where $k_B$ is the Boltzmann constant, $T$ is the temperature and $d^2$ is a collisional cross-sectional area. However, we do not use this expression directly. Instead, using the same approach as our simulations of drop-drop collisions in \citet{Lewin-Jones2024}, we denote the ratio of the ambient pressure relative to a reference pressure by $\mathcal{P}=p_0/p_{\mathrm{atm}}$, and use a known value of the mean free path, $\lambda_{\mathrm{atm}}$, of our chosen gas at our fixed (room) temperature at this reference pressure. Then the mean free path at the ambient pressure $p_0$ is given by $\lambda=\lambda_{\mathrm{atm}}/\mathcal{P}$. In particular, we use atmospheric pressure $p_{\mathrm{atm}}= 101,325$~\si{\pascal}, with $\lambda_{\mathrm{atm}}=69$~\si{\nano\meter}, so that $\mathcal{P}$ can be viewed as the ambient pressure measured in atmospheres.

The bounce-merge transition curves of simulations of 2~\si{\centi\stoke} silicone oil at two additional pressures ($\mathcal{P}=0.5$ and $\mathcal{P}=0.25$) are plotted in figure~\ref{fig:phase_1_all}, alongside the previous simulations at $\mathcal{P}=1$ from figure~\ref{fig:phase_1}. There is a clear trend that as the ambient pressure decreases, the bouncing region of the regime diagram shrinks. In particular, the critical Weber number of the hard transition decreases, and that of the soft transition (at least in the `merge' regime) increases. Because the transition between floating and bouncing (for large drops at low $\Web$, as in figure~\ref{fig:soft_float}) is determined by the macroscopic dynamics of the coefficient of restitution going to zero, this regime is not directly affected by the change of pressure, and for both $\mathcal{P}=1.0$ and $\mathcal{P}=0.5$ the transition has the same critical Weber number is this regime (e.g. at $R=0.35$~\si{\milli\m}). At $\mathcal{P}=0.25$, however, the critical Weber number at $R=0.35$~\si{\milli\m} has increased compared to $\mathcal{P}=0.5$, because at this reduced pressure the drop now merges (rather than floats) below the transition.

An important finding is that the critical radius (below which the drops always merge) also increases at decreased ambient pressures. This has the effect of moving the critical radius to within the the region of the regime diagram that is accessible to falling drop experiments (noting that Stokes drag is independent of gas pressure), suggesting that experiments at reduced pressure should be able to identify this critical radius, which should be a key goal of future experiments into this regime. 

Finally, also plotted in figure~\ref{fig:phase_1_all} are transition curves that are found when the gas kinetic effects are `turned off' (setting $\Delta_P=1$ and $\Delta_C=1$ in equations \eqref{eq:Delta_P}--\eqref{eq:Delta_C}), which corresponds to $\lambda\to0$, or, equivalently, to $\mathcal{P}\to\infty$. The critical Weber number for the hard transition now increases with decreasing drop radius as around twice the value found for $\mathcal{P}=1$. Moreover, there is no critical radius below which the drops always merge. The critical Weber number for the soft transition is also wildly different: for small radii the critical Weber number of the soft transition lies below $\Web=0.001$, at least 10 times smaller than the smallest value for $\mathcal{P}=1$, highlighting the importance of gas kinetic effects to correctly predict these critical Weber numbers.

\section{Results: Shallow Bath Impacts}\label{sec:results_shallow_bath}

Next, we relax the assumption of a deep bath and study the influence of this depth on the impact outcome, which allows us to compare to experiments in the literature.  We refer to such cases as `shallow baths' to distinguish them from the deep baths previously considered.

\subsection{Comparison to Experiments of Tang et. al (2016)} 

\begin{figure}
\centering
    \includegraphics[width=.4\linewidth]{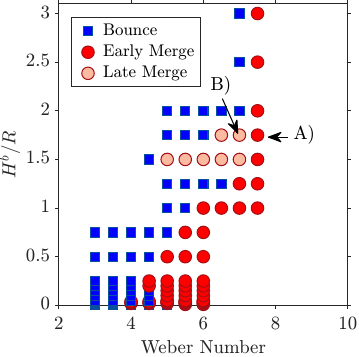}
      \includegraphics[width=.4\linewidth]{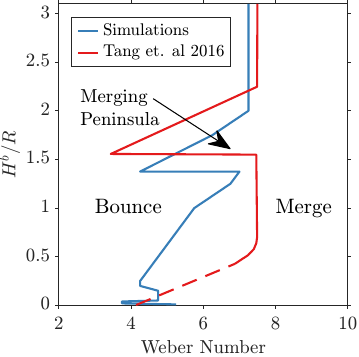}
    \caption{Simulations of $R=800$~\si{\micro\m} tetradecane, comparing to the experiments of  \citet{Tang2016}. In a) the simulation points are plotted, with the merging simulations labelled A) and B) shown in detail in figure \ref{fig:bath_height_merge}. In b) the boundary between bouncing and merging from a) is shown, in comparison to the boundary found experimentally by \citet{Tang2016}. (Note that they define the Weber number with respect to the drop diameter, so we have rescaled their results to match our radius-based definition). The dashed line shows the part of their phase diagram where there is a large amount of uncertainty due to a lack of experimental points.   Physical parameters are $\rho^d=760.0$\si{\kg\per\m\cubed}, $\mu^d=2.3$\si{\milli\pascal\s}, $\gamma^d=27.0$~\si{\milli\newton\per\m}, $A=5.2\times 10^{-20}$~\si{\joule}, $\mu^g=18.27$~\si{\micro\pascal\s}. } \label{fig:bath_height_Tang}
\end{figure}

Of particular interest are the experiments of \citet{Tang2016}, where they found that with decreasing bath depth, there is a remarkable regime where drops bounce, then merge, then bounce again and then merge again, see figure~\ref{fig:bath_height_Tang}b, a phenomenon first found by \citet{Pan2007}. Notably, we can see that our simulations, shown in figure~\ref{fig:bath_height_Tang}, also reproduce this complex effect. The direct comparison of experiments to simulations for the bounce-merge boundary is then shown in figure~\ref{fig:bath_height_Tang}b.

The qualitative agreement is very good, as our simulations have reproduced the bounce-merge-bounce-merge transition observed experimentally. We show strong quantitative agreement for the critical Weber number in the deep bath limit, and  identify the `merging peninsula', where for intermediately deep baths (around $H^b=1.5R$) the drops merge for lower Weber numbers than for both deeper and shallower baths. The position of our merging peninsula is slightly different to that found by \citet{Tang2016}, with ours at a slightly shallower bath, but the agreement is still remarkably good. 

\begin{figure}
\centering
    \includegraphics[width=\linewidth]{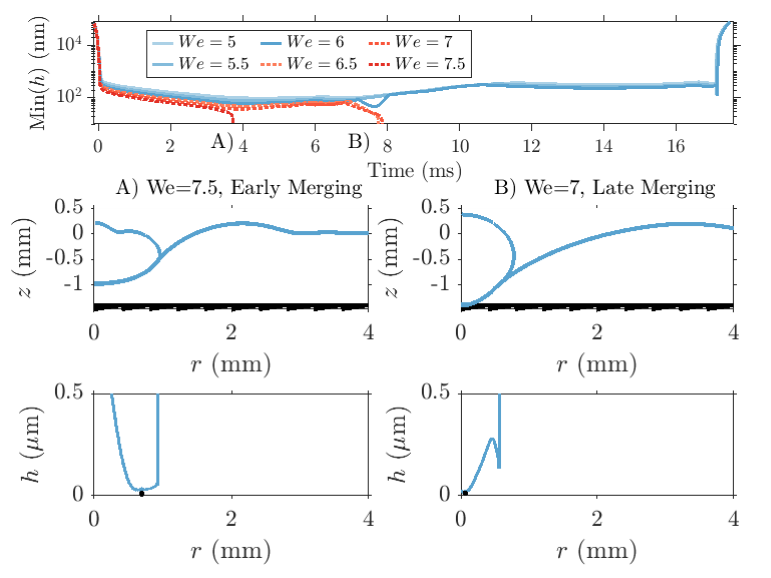}
     
    \caption{Top: the minimum film height against time for $R=800$~\si{\micro\m} tetradecane drops impacting a bath of depth $H^b=1.75$. In the left column, the moment of the contact for the $\Web=7.5$ impact, in the Early Merging regime is shown, which is point A) in figure~\ref{fig:bath_height_Tang}. In the right is the corresponding plots for $\Web=7$, showing Late Merging. Below are the gas film heights for the two simulations at the moment of contact, with the black dots showing the position of contact.} \label{fig:bath_height_merge}
\end{figure}

As observed by \citet{Tang2016}, the contact between the drop and bath occurs differently in the merging peninsula, at a significantly later time as compared to the higher speed contacts that occur for all bath depths. Our simulations also demonstrate this behaviour, with the points where this occurs labelled `Late Merge' in figure~\ref{fig:bath_height_Tang}. A comparison between simulations of the two modes of contact is plotted in figure \ref{fig:bath_height_merge}, showing that the `late merge' mode occurs due to the interaction between the trapped gas film (between the drop and bath), and the smooth bottom of the bath itself, with contact occurring near $r=0$ after the air is `squeezed out', which is consistent with the observations of \citet{Tang2016}. Note that the drop does not contact the bottom of the bath directly, and there is no change in the bath's topology before the merging occurs. For the case shown in figure~\ref{fig:bath_height_merge}, the height of the bath at $r=0$ at the moment of contact is only 9.3~\si{\micro\m}, compared to the initial
bath depth of 1400~\si{\micro\m}. By contrast, in the `early merge' mode shown in figure~\ref{fig:bath_height_merge}, observed at higher Weber numbers, contact happens both well above the bottom of the bath and away from $r=0$.

For even shallower baths, just below the merging peninsula, we find the same behaviour as the experiments, with the critical Weber number suddenly increasing as the bath depth decreases around $\Web=1.5$. As the depth decreases further, however, the quantitative agreement is less good, with our simulations showing merging for Weber numbers where the experiments found bouncing, particularly for bath depths between $ 0.25R$ and $1.0R$. We do not know the cause of this discrepancy, as there are no obvious numerical issues in this regime, and it is clearly worthy of future investigation in which theory and experiments are done in parallel.

\subsection{The Thin Film Limit}

\begin{figure}
\centering
    \includegraphics[width=\linewidth]{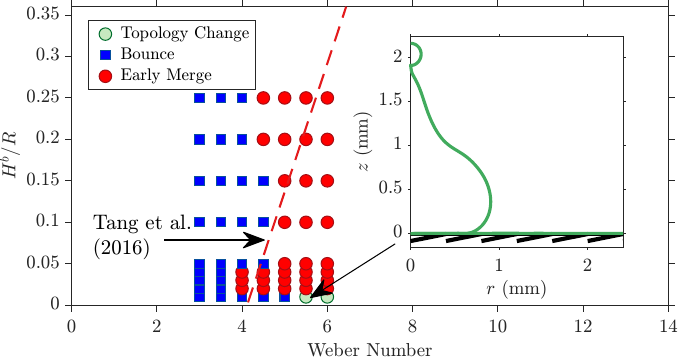}
    \caption{A zoomed version of the phase diagram in figure~\ref{fig:bath_height_Tang}, showing the limit as the bath depth tends to zero. The dashed line shows the bounce-merge boundary from the experiments in \citet{Tang2016}. The inset shows the final moment of the simulation at $H^b=0.01R$, $\Web=5.5$, where a change in topology would occur, ending our simulation. } \label{fig:bath_height_Tang_zoom}
\end{figure}

Finally, we consider the limit as the bath height tends to zero, so the drop now impacts a thin film of liquid. The experiments of \citet{Tang2016} have only a few experimental points in this region, but their boundary is plotted with a dashed line in figure~\ref{fig:bath_height_Tang}b, which suggests a limit $\sim \Web=4$. Our simulations for shallow baths also suggest a limit around this value, however, something surprising happens for our simulations of the shallowest bath we consider, at $H^b=0.01R$: the critical Weber number suddenly increases as compared to the slightly deeper bath of  $H^b=0.02R$. A zoomed view of the phase diagram figure~\ref{fig:bath_height_Tang}a is shown in figure~\ref{fig:bath_height_Tang_zoom}. For $H^b=0.01R$ we see bouncing up to $\Web=5$. 

For $\Web=5.5$, before seeing a bounce or contact, we find that the drop produces a jet, and will release a satellite droplet, as shown in the inset of figure~\ref{fig:bath_height_Tang_zoom}.  Similar behaviour to this was seen (both experimentally and in volume of fluid simulations) by \citet{Zhang2022} and \citet{Sharma2021} for impacts of drops onto superhydrophobic solids. Our current finite-element methodology does not have the functionality to continue beyond this change in topology, and therefore we cannot determine if this drop would bounce or merge with the solid. It is possible for this functionality to be implemented, and this is a clear possibility of future work. Regardless, our results suggest that the bounce-merge boundary as $H^b$ tends to $0$ is \emph{not} the apparent limit suggested by our deeper-bath simulations.

This limit is particularly of interest because thin liquid films have been used in experiments to emulate smooth solids. Typically, these films have been more viscous than the drops, for example in \citet{Pack2017}, \citet{Zhang2021} and \cite{Zhang2022a}. Deep baths of highly viscous liquid
have also been used, such as in \citet{Marcotte2019}. Intriguingly, in \citet{Zhang2021} a discrepancy was found between their drop-on-thin-film experiments and the drop-on-solid simulations from \cite{Chubynsky2020} for the critical Weber number. In other words, there is a suggestion that there is a fundamental difference between drop-on-smooth-solid and drop-on-ultra-thin-liquid-film cases. 

As a first step towards investigating the drop-on-ultra-thin-liquid-film limit, we briefly consider a final impact problem, with a different set of material parameters, chosen so that we do not observe the problem of topology change found in our comparisons to \citet{Tang2016} above.

\begin{figure}
\centering
\includegraphics[width=\linewidth]{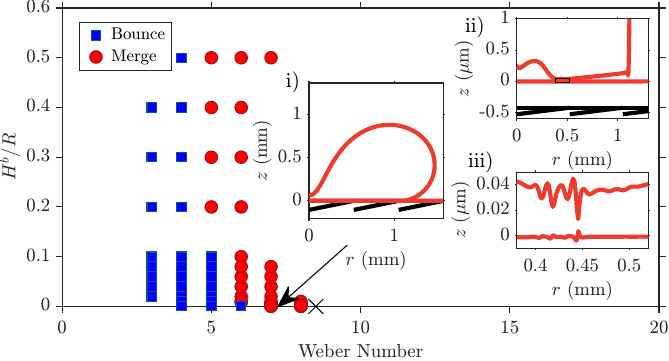}
\caption{Simulations of impact of a drop onto a thin liquid film, with material parameters chosen to not exhibit the topology change seen in figure~\ref{fig:bath_height_Tang_zoom}. The insets show the moment of contact for the simulation at  $\Web=7$, $H^b=0.0004R$: i) shows the full drop, ii) is vertically zoomed to show the film, and iii) is zoomed to show the point of contact.  Physical parameters are $R=1.05$\si{\milli\metre}, $\rho^d=1168.1$\si{\kg\per\m\cubed}, $\mu^d=11.7$\si{\milli\pascal\s}, $\gamma^d=68.0$~\si{\milli\newton\per\m}, $A=6.08\times 10^{-20}$~\si{\joule}, $\mu^g=18.27$~\si{\micro\pascal\s}.} \label{fig:bath_height_2}
\end{figure}

The results of these simulations are shown in figure~\ref{fig:bath_height_2}. Here, we see the same behaviour for very shallow baths that the critical Weber number suddenly increases with decreasing bath height. In inset i) of figure~\ref{fig:bath_height_2} we show the moment of contact for $\Web=7$ at a bath depth of $H^b=0.0004R$, the shallowest bath depth we were able to resolve. In inset ii) we show a zoomed view of the gas film, which is still many times thinner than the bath. Importantly, the drop has not visibly deformed the bath macroscopically. However, in inset iii) we zoom even further around the point of contact, where we discover that the van der Waals force on the bath has deflected the bath surface upwards, and so this is not behaving the same as an impact of a drop onto a \emph{rigid} smooth solid, despite the macroscopic appearance. We also ran simulations  of the identical drop impacting a smooth solid (using the method of \cite{Chubynsky2020}, reimplemented in oomphlib as described in \citet{Lewin-Jones2024}), which finds a critical Weber number of $8.5\pm0.5$, significantly higher than the drop-on-very-thin-film results, and is shown by the X in figure~\ref{fig:bath_height_2}. We have therefore identified a possible reason for the discrepancy between the drop-solid and drop-on-film results found by \citet{Zhang2021}, in that the van der Waals force is able to deform the film and cause contact where a solid impact would bounce. We emphasise however that this is a speculative result, and much more work is needed on this problem before we can be confident of this. In particular, this will involve using our methodology to consider impacts onto films of more viscous liquids, and making direct comparisons to the results of \citet{Zhang2021}.

\section{Conclusions}\label{sec:conclusions}

In this article we have presented a model for the impact of liquid drops onto liquid baths that is able to predict the transitions between merging and bouncing and back to merging with increasing impact speed. The model incorporates micro-scale gas kinetic effects and van der Waals driven disjoining pressure into a lubrication model of the trapped gas film. This model builds on our previous modelling of drop-solid impacts \citep{Chubynsky2020} and drop-drop collisions \citep{Lewin-Jones2024}, and required re-formulating our lubrication model in a curved coordinate system. 

We demonstrated excellent agreement with experimental values for the hard transition between bouncing and merging for drops of silicone oil. Two distinct regimes for the behaviour of impacts below the low speed merge-bounce transition were identified, with small drops merging quickly whilst large drops float on the surface for extended periods of time before merging. We also predicted that there is a critical radius below which these drops would always merge. We then extended this to show how the behaviour depends on the ambient pressure of the gas, and demonstrated how the thresholds are dramatically different if gas kinetic effects are not included in our model, showing that these effects are key to understanding these transitions. 

Impacts onto shallow baths were then considered, successfully comparing to the experiments of \citet{Tang2016} to identify transitions from bouncing to merging to bouncing and back to merging with decreasing bath depth. Finally, we investigated the limit as the bath depth tends to zero, and discussed a possible cause of observed discrepancies between impacts on smooth solids and very thin films.

\section{Future Directions}

In this article we have only considered impacts of drops onto baths of the same liquid. It is very simple to extend to two different immiscible liquids with different physical properies such as density and viscosity, and that will be an important direction of future work using our approach. This includes both impacts onto deep baths of different liquids, such as in the experiments of \citet{Wu2022} and \citet{Fudge2021}, as well as impacts onto highly viscous films, which have been used to emulate a smooth solid \citep{Zhang2021,Pack2017}. 

A major shortcoming of our methodology is our current inability to continue beyond the moment of contact, which would require performing `numerical surgery' on the mesh \citep{Anthony2023}. Other methodologies, such as volume-of-fluid methods, are routinely able to allow changes in topology. However, at the moment they are unable to make quantitive predictions of when merging will occur in drop impact problems in the regimes near the transitions between bouncing and merging. In appendix \ref{app:VoF_postmerge}, we make a comparison between our finite element approach and a VoF method that allows merging, which show the drop merging with the bath substantially earlier than the finite element model predicts.

Our results have suggested several important directions for future experimental investigations. We predict that changing the ambient pressure of the surrounding gas will have a substantial effect on the dynamics. Experiments in a controlled atmosphere would be able to explore the always-merging regime that we predict occurs for drops below a critical radius. Experiments that are able to produce drop impacts at very low $\Web$ would allow the soft transitions, and in particular the boundary between the merge-bounce and merge-float regimes to be explored.

The floating drops seen in our simulations only float for tens of milliseconds, however drops have also been observed to float for much longer, such as by \citet{Lo2017}, in their case for over 1000~\si{\milli\s}. Their floating drops are almost stationary, whereas here the drops are still visibly oscillating, and it would be interesting to investigate in future if this is the same mechanism that allows the drop to float.

Our methodology is currently only valid for head-on impacts, as we assume axisymmetry. To explore impacts at angles, and bounce-merge transitions with changing impact angles, such as those observed by \citet{Jayaratne1964}, the model would need to be extended to a full three-dimensional system. Adapting the curved lubrication equation to 3D is reasonably straightforward, however there are significant challenges to the computational implementation, including creating high-quality tetrahedral meshes, and developing a robust parallel solution of the resulting non-linear equations. Another possibility, for impacts onto very thin liquid films, is using a lubrication model for the liquid film as well as the gas film, as formulated by \citet{Duchemin2020}.

Finally, the curved lubrication formulation could be extended to any impact of liquid objects onto any smooth surface, not just drops onto baths.  These include impacts onto sessile droplets, where the off-set of the impact has been shown to affect the outcome of the impact \citep{Moon2018,Ramirez-Soto2020}, and collisions of unequally sized drops, a process that is critically important to understand raindrop formation \citep{Poydenot2024}. Beyond Newtonian fluids, in the Kaye effect, a jet of viscoelastic fluid bounces off a pool of the same liquid due to a trapped air film, with the bouncing shown to depend on the ambient pressure of the surrounding air \citep{King2019}.

\backsection[Supplementary data]{\label{SupMat}Supplementary material and movies are available at \\https://doi.org/[...]}

\backsection[Funding]{P.L.-J. was supported by a studentship within the UK EPSRC-supported Centre for Doctoral Training in the Modelling of Heterogeneous Systems (HetSys), EP/S022848/1, and also acknowledges support from the Engineering and Physical Sciences Research Council (EPSRC) under Grant No. UKRI093. J.E.S. acknowledges support from the Engineering and Physical Sciences Research Council (EPSRC) under Grant No. UKRI3193. D.M.H. and R.C. acknowledge support from the National Science Foundation (NSF CBET-2123371) and the Engineering and Physical Sciences Research Council (EPSRC EP/W016036/1).}

\backsection[Declaration of interests]{ The authors report no conflict of interest.}

\backsection[Data availability statement]{Code for the finite element simulations is available at \url{https://github.com/PLewinJones/drop_bath_impact}. Code for the VoF simulations is available at \url{https://github.com/rcsc-group/BouncingDroplets}. }

\backsection[Author ORCIDs]{  P. Lewin-Jones, https://orcid.org/0000-0001-5592-1243; J. E. Sprittles, https://orcid.org/0000-0002-4169-6468; D. A. Lockerby, https://orcid.org/0000-0001-5232-7986; R. Cimpeanu https://orcid.org/0000-0002-7934-7909; D. M. Harris https://orcid.org/0000-0003-2615-9178}

\appendix

\section{Comparison with VoF: Post-merging Dynamics}\label{app:VoF_postmerge}

In this appendix, in contrast to the forced bouncing model used in \S\ref{sec:vof_bounce}, we use a more traditional free-surface VoF method, where a single colour function is used that is equal to 1 in the liquid (drop or bath) and equal to 0 in the gas, as used, for example, in \citet{Josserand2016a,Cimpeanu2018a,Sykes2023}. In this case, the drop will merge with the bath when the surfaces lie within the same mesh cell. Simulations were performed with three different levels of mesh refinement. Specifically, the minimum allowed mesh cell size was varied. Basilisk uses an adaptive multi-grid quadtree method in space and associated timestep restriction, and we set the mesh refinement strategy according to pre-specified criteria with stringent tolerances, which include changes in magnitude of velocity components, vorticity as well as location-specific criteria near the time-evolving interfaces between liquid and gas. The side-length of the smallest mesh cell is $2^{-N}\times L_{\mathrm{Box}}$, where $N$ is the `level' of refinement and $L_{\mathrm{Box}}$ is the side-length of the square simulation domain, which in these simulations is $10R$. For the highest refinement level used here, level 14, this corresponds to a side-length of the smallest computational cell of $104$\si{\nano\meter}. This is several times larger than the minimum film thicknesses seen in the bouncing finite element simulations (in e.g. figure~\ref{fig:min_height_1}), but computational resource-wise it is a comparable numerical effort, and represents a significantly finer grid than required in previously investigated forced bouncing scenarios. This is a finer grid than the maximum refinement used by \citet{Sharma2021}, who were able to fully resolve a lower speed drop bouncing from a smooth solid surface with a VoF method that allowed merging.

\begin{figure}
    \centering
     \includegraphics[width=0.23\linewidth]{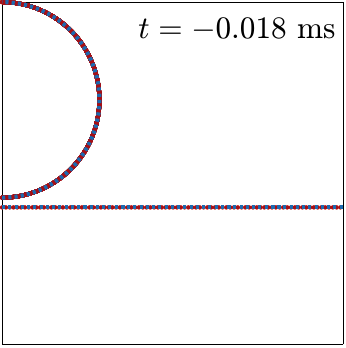}
     \includegraphics[width=0.23\linewidth]{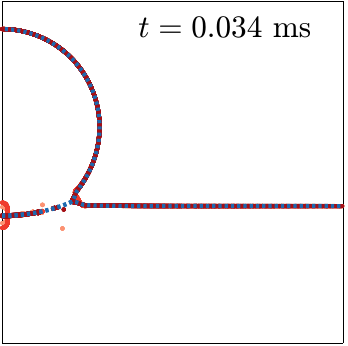}
     \includegraphics[width=0.23\linewidth]{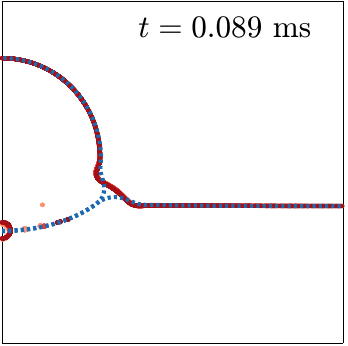}
     \includegraphics[width=0.23\linewidth]{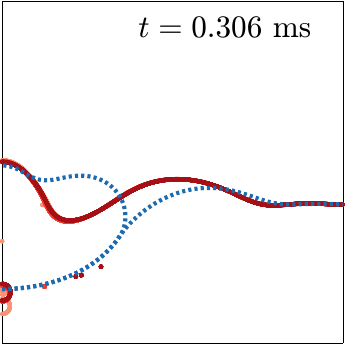}
     
    \includegraphics[width=\linewidth]{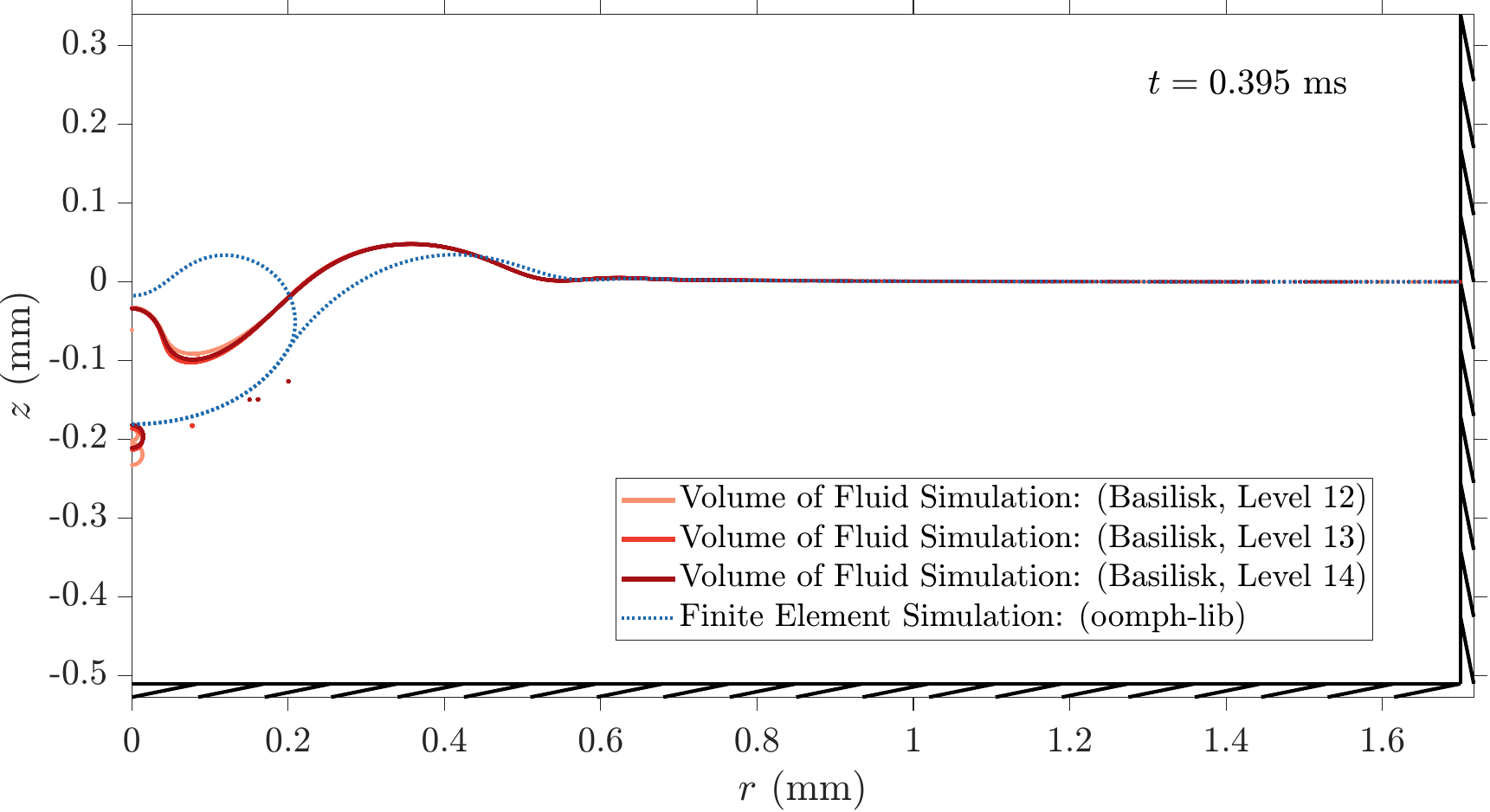}
        \caption{ Impact of a $R=170$~\si{\micro\meter}, 2~\si{\centi\stoke} silicone oil at $\Web=7$, where experiments show merging. Above are snapshots in time, with time increasing left-to-right, bottom-to-top.  Below is the moment of contact. The blue curves show the finite element simulation, and overlaid in orange are three Basilisk VoF simulations, at different levels of mesh refinement, which merge much earlier. A video is available as movie 5. }
    \label{fig:db_VOF_merge}
\end{figure}

The simulations using this method at $\Web=7$ are shown in figure~\ref{fig:db_VOF_merge}, overlaid on our finite element simulation. A video is available as \comment{HERE}, and the code at \url{https://github.com/rcsc-group/BouncingDroplets}. In all cases, the drop merged with the bath much earlier than in our finite element simulations. The dots and bubbles in the VoF simulations in figure~\ref{fig:db_VOF_merge} show where air has been trapped within the bath after the drop merged. At the highest resolution, contact occurs at $t=0.029$~\si{\milli\second}, (i.e. almost straight-away) compared to $t=0.4$~\si{\milli\second} in our simulation (as shown in figure~\ref{fig:min_height_1}). The state of all the simulations at $t=0.4$~\si{\milli\second} is shown in the lower panel of figure~\ref{fig:db_VOF_merge}, demonstrating that the premature merging in VoF causes substantially different (inaccurate) dynamics, and that at this level of (still-significant) refinement, VoF methods are unable to predict when a drop will merge with a bath. 

Interestingly, if the additional physical effects were included within the VoF model, they would actually further promote merging. Recently, these effects have been incorporated into a VoF framework in the context of drop-drop collisions \citep{Wang2025,wang2026}, creating mesh-independent capabilities for the VoF approach, but so far they have required artificially large Hamaker constants to find accurate bounce-merge transitions. 

In summary, with current capabilities a VoF approach that allows merging is feasible in scenarios when merging is known to happen, especially when that merging happens very early in the dynamics, but otherwise is unsuitable in transitional sections of the parameter space, and, currently, carries no predictive capabilities in terms of identifying the merging threshold.  This is slightly disappointing, as the VoF approach carries the major advantage of being able to `go through' the change in topology in a computationally simple way.  There are a few potential routes to progress: (i) to increase the resolution further, and resolve nanometric film scales, (ii) to try to incorporate a subgrid model in VoF that captures the film's dynamics or (iii) to use the FEM calculation as an initial condition for a post-contact VoF simulation.

\section{Derivation of the Curved Lubrication Equation}\label{app:deriv}

In this appendix, we derive the curved lubrication equation \eqref{eq:curved_lub} for the gas film. This appendix will be self-contained, so there will be some repetition from the main text. In \S\ref{app:deriv_geometry}, we describe how the (curved) trapped gas film is described, and a surface coordinate system constructed. In \S\ref{app:deriv_height_deriv}, we derive an expression for the time derivative of the film height. Then we introduce the equations for the gas flow in \S\ref{app:deriv_gas_equations}, and apply lubrication theory to these in \S\ref{app:deriv_lubrication_scale}.
\subsection{Geometry of the Gas Film}\label{app:deriv_geometry}
\begin{figure}

    \centering
       \includegraphics{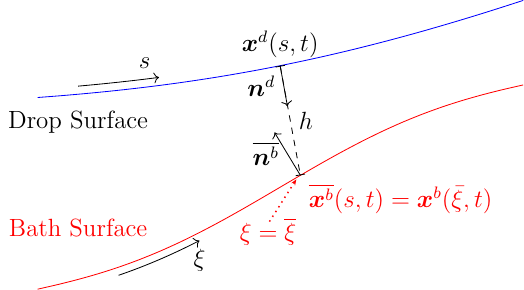}
    \caption[Diagram]{ The definition of the film height $h$ at drop surface coordinate $\mb{x}^d(s,t)$. It is measured in the normal direction to the drop surface, with unit normal vector $\mb{n}^d$. The position vector of the intersection of the projected normal on the bath surface is denoted by $\overline{\mb{x}^b}$.}
    \label{fig:Curved_diagram_app}
\end{figure}

The axisymmetric surface of the drop is parameterised by curvilinear coordinate $s$ and azimuthal coordinate $\theta$. Axisymmetry means that the position vector at $s$ and $\theta$, at time $t$, can be written as $\mb{x}^d(s,\theta,t)=x^d_r(s) \mb{e}_r+x^d_z(s)\mb{e}_z$. Note that in Cartesian coordinates $\mb{e}_r$ depends on $\theta$, but dependencies on $\theta$ will be unstated except where it is important. Correspondingly the bath surface is parametrised by $\xi$ and $\theta$.

The intersection of the drop normal, $\mb{n}^d$, at $\mb{x}^d$ with the bath surface is denoted by $\overline{\mb{x}^b}$, at curvilinear surface coordinate $\overline{\xi}$ as shown in figure \ref{fig:Curved_diagram_app}. Throughout this section, symbols with over-bars denote variables evaluated at the point of projection (this is described in more detail below in \S\ref{app:deriv_height_deriv}). Because of the axisymmetry, $\overline{\mb{x}^b}$ is at the same azimuthal coordinate $\theta$ as $\mb{x}^d$.  The film thickness is then defined as $h=\left(\overline{\mb{x}^b}-\mb{x}^d\right)\mb{\cdot}\mb{n}^d$, so that $h\mb{n}^d=\overline{\mb{x}^b}-\mb{x}^d$, as shown in figure \ref{fig:Curved_diagram_app}.

\begin{figure}
    \centering
   \includegraphics{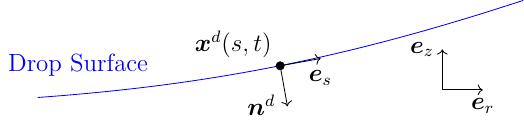}
    \caption[Diagram]{ The curvilinear coordinate system, with basis vectors $\mb{e}_s$ and $\mb{n}^d$ in the directions along the drop surface and in the normal direction. These are compared to the Eulerian basis vectors $\mb{e}_r$ and $\mb{e}_z$. The third basis vector is the azimuthal direction $\mb{e}_\theta$, pointing `into the page'.}
    \label{fig:coord_systems}
\end{figure}

Our aim is to use this definition of height to formulate a lubrication equation that is solved on the drop surface. To proceed, we need to change from $\{\mb{e}_r,\mb{e}_\theta,\mb{e}_z\}$ to an alternative, curvilinear coordinate system with respect to the drop surface. The construction of this is shown in appendix \S\ref{app:curv_basis}, based on the approach described in \citet{Howell1994}. This has orthonormal basis $\{\mb{e}_s,\mb{e}_\theta,\mb{n}^d\}$, where $\mb{n}^d$ is the unit normal vector to the drop surface, and $\mb{e}_s$ is the unit vector tangent to the drop surface, as shown in figure \ref{fig:coord_systems}. Note that $\mb{e}_s$ and $\mb{n}^d$ depend on $s$ and $t$. We denote the components of vectors in the directions of these basis vectors with subscript $s$ and $n$ (for example, $\mb{u}^d=u_n^d\mb{e}_n+u_s^d\mb{e}_s$). Important quantities are the (surface) scaling factors
\begin{align}
a_s&=\abs*{\diffp{\mb{x}^d}{s}}=\sqrt{\diffp{r^d}{s}^2+\diffp{z^d}{s}^2}\, , &
a_\theta&=\abs*{\diffp{\mb{x}^d}{\theta}}=r^d\, ,
\end{align}
and the principle curvatures 
\begin{align}
\kappa_s&=\frac{1}{a_s^3}\left(\diffp{r^d}{s}\diffp[2]{z^d}{s}-\diffp{z^d}{s}\diffp[2]{r^d}{s}\right)\, , &
\kappa_\theta&=\frac{1}{r^da_s}\diffp{z^d}{s}\, .
\end{align}
Full details of this construction are shown in appendix \S\ref{app:curv_basis}. We will also denote the velocity of the drop surface by $\mb{w}^d=\diffp{\mb{x}^d}{t}$ and the bath surface by  $\mb{w}^b=\diffp{\mb{x}^b}{t}$. The normal to the bath is denoted $\mb{n}^b$, and the unit tangent vector to the bath at curvilinear coordinate $\xi$ by $\mb{e}_\xi$, where $\mb{n}^b$ is oriented anticlockwise to $\mb{e}_\xi$ in the $(r,z)$ plane.

\subsection{Time Derivative of the Film Height}\label{app:deriv_height_deriv}
The bath surface is parametrised by $\xi$ and $\theta$ (again dropping the $\theta$ dependence), with the position vector denoted by $\mb{x}^b(\xi,t)$. We denote the point on the surface of the bath where the normal to the drop surface intersects the bath surface (the `point of projection') by $\overline{\mb{x}^b}$, where bar denotes a quantity evaluated at the point of projection. The point of evaluation is a function of $s$, $\theta$ and $t$, so that $\overline{\xi}(s,t)$, defined so that $\overline{\mb{x}^b}(s,\theta,t)=\mb{x}^b(\overline{\xi}(s,t),\theta, t)$ (dropping $\theta$ dependence from here onwards).

For the lubrication equation, the derivative of $h$ with respect to time at coordinate $s$ is needed, in terms of quantities that we know via our simulations of the drop and bath. Specifically, we need to know $\diffp{h}{t}$ as a function of the position vector and velocity of the drop surface ($\mb{x}^d$ and $\mb{w}^d$), the tangent and normal vectors to the drop surface ($\mb{e}_s$ and $\mb{n}^d$), as well as the corresponding quantities at the point of projection on the bath surface: the position vector and velocity of the bath surface ($\overline{\mb{x}^b}$ and $\overline{\mb{w}^b}$), and the tangent and normal vectors to the bath surface ($\overline{\mb{e}_\xi}$ and $\overline{\mb{n}^b}$).

The film height can change due to the motion of the drop, from its normal motion as well as from change in shape, and due to the motion of the bath, as we will now see. The approach derived below is based on the approach in \citet{Temizer2016} for general surfaces, which we have simplified to be specific to the axisymmetric case.

Differentiating the definition of the film height,
\begin{align}
\diffp{h}{t}&=\left(\diffp{}{t}\left(\overline{\mb{x}^b}\right)-\diffp{\mb{x}^d}{t}\right)\mb{\cdot}\mb{n}^d+\left(\overline{\mb{x}^b}-\mb{x}^d\right)\mb{\cdot}\diffp{\mb{n}^d}{t}\, ,\label{eq:dhdt_1}
\end{align}
and we can see that this last term vanishes:
\begin{align}
\left(\overline{\mb{x}^b}-\mb{x}^d\right)\mb{\cdot}\diffp{\mb{n}^d}{t}=h\mb{n}^d\mb{\cdot}\diffp{\mb{n}^d}{t}=\frac{h}{2}\diffp{(\mb{n}^d\mb{\cdot}\mb{n}^d)}{t}=\frac{h}{2}\diffp{(1)}{t}=0\, .
\end{align}
So what remains is to calculate the velocity of the projection point, $\diffp{}{t}\left(\overline{\mb{x}^b}\right)$. The projection point can change due to the motion of both the drop surface \emph{and} the bath surface. 

Writing out the dependence in full,
\begin{align}
\diffp{}{t}\left(\overline{\mb{x}^b}\right)&=\diffp{}{t}\mb{x}^b(\overline{\xi}(s,t), t)\, ,\\
&=\diffp{\overline{\xi}(s,t)}{t}\diffp{\mb{x}^b}{\xi}(\xi, t)\vert_{\xi=\overline{\xi}(s,t)}+\diffp{\mb{x}^b}{t}(\xi, t)\vert_{\xi=\overline{\xi}(s,t)}\, ,\\
&=\diffp{\overline{\xi}}{t}\overline{\diffp{\mb{x}^b}{\xi}}+\overline{\diffp{\mb{x}^b}{t}}
=\diffp{\overline{\xi}}{t}\overline{a_\xi}\ \overline{\mb{e}_\xi}+\overline{\mb{w}^b}\, ,\label{eq:dxbdt}
\end{align}
noting the careful use of the overline for evaluation at the point of projection, and introducing the bath surface velocity $\mb{w}^b=\diffp{\mb{x}^b}{t}$ and surface scaling factor and basis vector via $\overline{\diffp{\mb{x}^b}{\xi}}=\overline{a_\xi}\ \overline{\mb{e}_\xi}$.
Substituting \eqref{eq:dxbdt} into \eqref{eq:dhdt_1} gives
\begin{align}
\diffp{h}{t}
&=\left(\overline{\mb{w}^b}-\mb{w}^d+\overline{a_\xi}\ \overline{\mb{e}_\xi}\diffp{\overline{\xi}}{t}\right)\mb{\cdot}\mb{n}^d\, .\label{eq:dhdt_mid}
\end{align}
 Similarly
\begin{align}
\diffp{}{s}\left(\overline{\mb{x}^b}\right)
&=\diffp{\overline{\xi}}{s}\overline{\diffp{\mb{x}^b}{\xi}}=\overline{a_\xi}\ \overline{\mb{e}_\xi}\diffp{\overline{\xi}}{s}\, .
\end{align}

Our aim is now to calculate $\diffp{\overline{\xi}}{t}$, which we do by differentiating the identity $0=h\mb{n}^d\mb{\cdot}\mb{e}_s$. We have
\begin{align}
0&=\diffp{}{t}\left((\overline{\mb{x}^b}-\mb{x}^d)\mb{\cdot}\mb{e}_s\right)\, ,\\
0&=\left(\diffp{\overline{\mb{x}^b}}{t}-\diffp{\mb{x}^d}{t}\right)\mb{\cdot}\mb{e}_s+\left(\overline{\mb{x}^b}-\mb{x}^d\right)\mb{\cdot}\diffp{\mb{e}_s}{t}\, ,\\
0&=\left(\overline{\mb{w}^b}-\mb{w}^d+\overline{a_\xi}\ \overline{\mb{e}_\xi}\diffp{\overline{\xi}}{t}\right)\mb{\cdot}\mb{e}_s+h\mb{n}^d\mb{\cdot}\frac{1}{a_s}\left(\diffp{w_n^d}{s}+a_s\kappa_sw_s^d\right)\mb{n}^d\, ,\\
\diffp{\overline{\xi}}{t}&=\frac{-1}{\overline{a_\xi}\ \overline{\mb{e}_\xi}\mb{\cdot}\mb{e}_s}\left(\left(\overline{\mb{w}^b}-\mb{w}^d\right)\mb{\cdot}\mb{e}_s+\frac{h}{a_s}\left(\diffp{w_n^d}{s}+a_s\kappa_sw_s^d\right)\right)\, .\label{eq:dxidt}
\end{align}
Where we have used that 
\begin{align}
    \diffp{\mb{e}_s}{t}&=\frac{1}{a_s}\left(\diffp{w_n^d}{s}+a_s\kappa_s w_s^d\right)\mb{n}^d\, , 
\end{align}
which is derived in \ref{app:time_derivs}.
 Note it is safe to divide by $\overline{\mb{e}_\xi}\mb{\cdot}\mb{e}_s$ because if this is $0$ the two surfaces are perpendicular and we definitely aren't in a lubrication regime! Substituting \eqref{eq:dxidt} into \eqref{eq:dhdt_mid} we obtain
\begin{align}
\diffp{h}{t}
&=\left(\overline{w^b_n}-w_n^d\right)-\frac{\overline{\mb{e}}_\xi\mb{\cdot}\mb{n}^d}{\overline{\mb{e}}_\xi\mb{\cdot}\mb{e}_s}\left(\left(\overline{w^b_s}-w_s^d\right)+\frac{h}{a_s}\left(\diffp{w_n^d}{s}+a_s\kappa_sw_s^d\right)\right)\label{eq:dhdt_mid2}\, .
\end{align}

Finally, to understand the term $\frac{\overline{\mb{e}}_\xi\mb{\cdot}\mb{n}^d}{\overline{\mb{e}}_\xi\mb{\cdot}\mb{e}_s}$ we see that 
\begin{align}
\diffp{h}{s}&=\left(\diffp{\overline{\mb{x}^b}}{s}-\diffp{\mb{x}^d}{s}\right)\mb{\cdot}\mb{n}^d+h\mb{n}^d\mb{\cdot}\diffp{\mb{n}^d}{s}\, ,\\
&=\left(\overline{\diffp{\mb{x}^b}{\xi}}\diffp{\overline{\xi}}{s}-a_s\mb{e}_s\right)\mb{\cdot}\mb{n}^d
=\overline{a_\xi} \diffp{\overline{\xi}}{s} \overline{\mb{e}_\xi}\mb{\cdot}\mb{n}^d\label{eq:dhds_mid}\, ,
\end{align}
then to compute $\diffp{\overline{\xi}}{s}$, we differentiate the identity $0=h\mb{n}^d\mb{\cdot}\mb{e}_s$ to see that
\begin{align}
0&=\diffp{}{s}\left((\overline{\mb{x}^b}-\mb{x}^d)\mb{\cdot}\mb{e}_s\right)\, ,\\
0&=\left(\diffp{\overline{\mb{x}^b}}{s}-\diffp{\mb{x}^d}{s}\right)\mb{\cdot}\mb{e}_s+(\overline{\mb{x}^b}-\mb{x}^d)\mb{\cdot}\diffp{\mb{e}_s}{s}\, ,\\
0&=\left(\overline{a}_\xi\overline{\mb{e}}_\xi\diffp{\overline{\xi}}{s}-a_s\mb{e}_s\right)\mb{\cdot}\mb{e}_s+h\mb{n}^d\mb{\cdot} a_s\kappa_s\mb{n}^d\, ,\\
\diffp{\overline{\xi}}{s}&=\frac{1}{\overline{a}_\xi\overline{\mb{e}}_\xi\mb{\cdot}\mb{e}_s}a_s\left(1-h\kappa_s\right)\label{eq:dbarxids}\, .
\end{align}

So substituting \eqref{eq:dbarxids} into \eqref{eq:dhds_mid} we find that
\begin{align}
\diffp{h}{s}&=\frac{\overline{\mb{e}}_\xi\mb{\cdot}\mb{n}^d}{\overline{\mb{e}}_\xi\mb{\cdot}\mb{e}_s}a_s\left(1-h\kappa_s\right)\, ,\label{eq:dhds}
\end{align}
and substituting into \eqref{eq:dhdt_mid2} we finally find that
\begin{align}
\diffp{h}{t}
&=\left(\overline{w^b_n}-w_n^d\right)-\frac{1}{a_s\left(1-h\kappa_s\right)}\diffp{h}{s}\left(\left(\overline{w^b_s}-w_s^d\right)+\frac{h}{a_s}\left(\diffp{w_n^d}{s}+a_s\kappa_sw_s^d\right)\right)\, .\label{eq:dhdt_final}
\end{align}

\subsection{Equations of the Gas Film}\label{app:deriv_gas_equations}

To proceed, we need to write down the equations of the gas flow with respect to our new surface coordinate system. To do this, we generalise to position vectors away from the surface in the normal direction by introducing the coordinate $n$, which measures the distance from the drop surface, so that the bath surface is at $n=h$. The position vector within the gas is denoted by $\mb{x}=\mb{x}^d+n\mb{n}^d$. The scaling factors are defined as $l_s=a_s(1-\kappa_s n)$ and $l_\theta=a_\theta(1-\kappa_\theta n)$, with full details in appendix \S\ref{app:curv_normal}.

Under this coordinate system, the vector derivative operators take the form
\begin{align}
\nabla f=\frac{1}{l_s}\diffp{f}{s}\mb{e}_s+\diffp{f}{n}\mb{e}_n\, ,\label{eq:grad_def}
\end{align}
\begin{align}
\nabla\mb{\cdot}\mb{f}=\frac{1}{l_sl_\theta}\left(\diffp{}{s}(l_\theta f_s)+\diffp{}{n}(l_sl_\theta f_n)\right)\, .\label{eq:div_def}
\end{align}

We denote the gas fluid velocity at $\mb{x}=\mb{x}^d+n\mb{n}^d$ by $\mb{u}=u_s\mb{e}_s+u_n\mb{n}^d$.

We assume the gas is incompressible, and that inertia and gravitational force can be neglected, so that it is modelled by the Stokes equations. A discussion of these assumptions can be found in \citet{Lewin-Jones2024}. Mass conservation, using \eqref{eq:div_def}, is given by
\begin{align}
\nabla\mb{\cdot}\mb{u}=\frac{1}{l_sl_\theta}\left(\diffp{}{s}(l_\theta u_s)+\diffp{}{n}(l_sl_\theta u_n)\right)=0\, ,\label{eq:divu}
\end{align}
and the $s$ and $n$ components of the Stokes equations are given by
\begin{align}
0&=-\frac{1}{l_s}\diffp{p}{s}+\frac{\mu^g}{l_\theta}\diffp{}{n}\left( \frac{l_\theta}{l_s}\left(\diffp{}{n}(l_su_s)-\diffp{}{s}(u_n)\right)\right)\, ,\label{eq:stokes1}\\
0&=-\diffp{p}{n}-\frac{\mu^g}{l_sl_\theta}\diffp{}{s}\left( \frac{l_\theta}{l_s}\left(\diffp{}{n}(l_su_s)-\diffp{}{s}(u_n)\right)\right)\, ,\label{eq:stokes2}
\end{align}
using the definition $\nabla^2\mb{f}=\nabla(\nabla\mb{\cdot}\mb{f})-\nabla \times(\nabla\times \mb{f})$ for the vector Laplacian and noting that $\nabla\mb{\cdot}\mb{u}=0$, so we have $\nabla^2\mb{u}=-\nabla \times(\nabla\times \mb{u})$. Expressions for this are given in \S\ref{app:curv_normal}.

The first boundary condition is continuity of velocity, given by 
\begin{align}
    \mb{u}&=\mb{u}^d\text{ at } n=0\, , &  \mb{u}&=\overline{\mb{u}^b}\text{ at } n=h\, .\label{eq:vel_cont}
\end{align}

The kinematic boundary condition is given by 
\begin{align}
    u_n&=w_n^d \text{ at } n=0\, , &  \diff.D.{(h-n)}{t}=0 \text{ at } n=h\, .\label{eq:kinematic_base}
\end{align}

The material derivative of a scalar function $f$ in a moving frame is given by \citep{Donea2004}:
\begin{align}
\diff.D.{f}{t}&=\diffp{f}{t}+\left(\mb{u}-\diffp{\mb{x}}{t}\right)\mb{\cdot}\nabla f\, ,\\
&=\diffp{f}{t}+\left(\mb{u}-\diffp{\mb{x}^d}{t}-n\diffp{\mb{n}^d}{t}\right)\mb{\cdot}\nabla f\, ,\\
&=\diffp{f}{t}+\left(u_s-w_s^d+n\left(\frac{1}{ a_s}\diffp{w_n^d}{s}+\kappa_sw_s^d\right)\right)\frac{1}{l_s}\diffp{f}{s}+\left(u_n-w_n^d\right)\diffp{f}{n}\, ,
\end{align}
using \eqref{eq:grad_def}. The kinematic boundary condition at $n=h$ (the bath surface) is then
\begin{align}
0=\diff.D.{(h-n)}{t}&=\diffp{h}{t}+\left(u_s-w_s^d+h\left(\frac{1}{ a_s}\diffp{w_n^d}{s}+\kappa_sw_s^d\right)\right)\frac{1}{a_s(1-\kappa_s h)}\diffp{h}{s}-\left(u_n-w_n^d\right)\, .
\end{align}

Substituting in our expression \eqref{eq:dhdt_final} for $\diffp{h}{t}$, we get
\begin{align}
0&=\left(\overline{w^b_n}-w_n^d\right)-\frac{1}{a_s\left(1-h\kappa_s\right)}\diffp{h}{s}\left(\left(\overline{w^b_s}-w_s^d\right)+\frac{h}{a_s}\left(\diffp{w_n^d}{s}+a_s\kappa_sw_s\right)\right)\nonumber\\&\quad +\left(u_s-w_s^d+h\left(\frac{1}{ a_s}\diffp{w_n^d}{s}+\kappa_sw_s\right)\right)\frac{1}{a_s(1-\kappa_s h)}\diffp{h}{s}-\left(u_n-w_n^d\right)\, ,\\
0&=\left(\overline{w^b_n}-u_n\vert_{n=h}\right)-\frac{1}{a_s\left(1-h\kappa_s\right)}\diffp{h}{s}\left(\overline{w^b_s}-u_s\vert_{n=h}\right)\, .\label{eq:curved_kinematic}
\end{align}

\subsection{Lubrication Scaling}\label{app:deriv_lubrication_scale}

We are now ready to apply lubrication theory, where we assume that the scale in the $n$ direction is asymptotically small to that along the film. To do this, we will non-dimensionalise, denoting non-dimensional quantities by hats. Formally, let $L$ be a reference length scale along the drop surface, defining $\mb{x}^d=L\hat{\mb{x}}^{d}$, and $H$ be a length scale of the normal direction, defining $n=H\hat{n}$ and $h=H\hat{h}$. We assume that $H/L=\ep$, where $\ep$ is a small parameter. We define the velocity scale $U$, the impact speed, with $\mb{u}=U\hat{\mb{u}}$, $\mb{w}^d=U\hat{\mb{w}}^d$ and $\mb{w}^b=U\hat{\mb{w}}^b$. As is standard in lubrication theory, we also rescale the pressure, with $p=P\hat{p}$, for a convenient pressure scale $P$, that depends on $\ep$, which we will determine later.

We correspondingly define $a_s=L\hat{a}$, $a_\theta=L\hat{a}_\theta$, $\kappa_s=\hat{\kappa}_s/L$, $\kappa_\theta=\hat{\kappa}_\theta/L$, $l_s=L\hat{l}_s=L\hat{a}(1-\ep \hat{n} \hat{\kappa}_s)$ and $l_\theta=L\hat{l}_\theta=\hat{a}_\theta(1-\ep \hat{n} \hat{\kappa}_\theta)$.

To proceed, we perform an asymptotic expansion in powers of $\ep$, expanding $\hat{\mb{u}}=\sum_{i=0}^\infty\ep^i\hat{\mb{u}}^{i}$. The boundary conditions are applied to the $\ep^0$ term in the expansion, so that at $\hat{n}=0$, $\hat{u}_n^0=\hat{w}^d_n$ and $\hat{u}_s^0=\hat{u}_s^d$; and $\hat{u}_n^i=\hat{u}_s^i=0$ for $i>0$. Similarly $\hat{u}_s^0=\hat{u}_s^b$ at $\hat{n}=\hat{h}$, and $\hat{u}_s^i=0$ for $i>0$.

In the mass conservation equation \eqref{eq:divu}, this expansion gives
\begin{align}
0&=\frac{U}{L\hat{a}_s\hat{a}_\theta(1-\ep \hat{n} \hat{\kappa}_s)(1-\ep \hat{n} \hat{\kappa}_\theta)}\Bigg(\diffp{}{s}\left(\hat{a}_\theta(1-\ep \hat{n} \hat{\kappa}_\theta)\left(\hat{u}_s^0+\ep \hat{u}_s^1+\order*{\ep^2}\right)\right)\nonumber\\ &\quad+\frac{1}{\ep}\diffp{}{\hat{n}}\left(\hat{a}_s\hat{a}_\theta(1-\ep \hat{n} \hat{\kappa}_s)(1-\ep \hat{n} \hat{\kappa}_\theta) \left(\hat{u}_n^0+\ep \hat{u}_n^1+\order*{\ep^2}\right)\right)\Bigg)\, .\label{eq:divu_scaled}
\end{align}

 Then at $\order*{\ep^{-1}}$, and dividing by $U/L$, we compute that \eqref{eq:divu_scaled} is
\begin{align}
0&=\frac{1}{\hat{a}_s\hat{a}_\theta}\diffp{}{\hat{n}}(\hat{a}_s\hat{a}_\theta \hat{u}_n^{0})\, ,
\end{align}
so that $\hat{u}_n^{0}$ is independent of $\hat{n}$, and by the boundary condition $\hat{u}_n^0=\hat{w}^d_n$ across the whole width of the film. 

At $\order*{\ep^0}$ we find
\begin{align}
0&=\frac{1}{\hat{a}_s\hat{a}_\theta}\left(\diffp{}{s}\left(\hat{a}_\theta \hat{u}_s^0 \right)-\hat{a}_s \hat{a}_\theta( \hat{\kappa}_s+ \hat{\kappa}_\theta) \hat{w}_n^d +\hat{a}_s \hat{a}_\theta \diffp{}{\hat{n}}( \hat{u}_n^{1})\right)\, .\label{eq:divuep0}
\end{align}

As is standard in lubrication theory, integrating \eqref{eq:divuep0} between $\hat{n}=0$ and $\hat{n}=\hat{h}$ across the film gives
\begin{align}
0&=\int_0^{\hat{h}}\left(\frac{1}{\hat{a}_s\hat{a}_\theta}\diffp{}{s}\left(\hat{a}_\theta \hat{u}_s^0\right)-( \hat{\kappa}_s+ \hat{\kappa}_\theta) \hat{w}_n^d+\diffp{ \hat{u}_n^{1}}{\hat{n}}\right)\dl{\hat{n}}\, ,\\
0&=-\hat{h}( \hat{\kappa}_s+ \hat{\kappa}_\theta) \hat{w}_n^d+\frac{1}{\hat{a}_s\hat{a}_\theta}\int_0^{\hat{h}}\diffp{}{s}\left(\hat{a}_\theta \hat{u}_s\right)\dl{\hat{n}}+ \left[\hat{u}_n^{1}\right]_0^{\hat{h}}\, ,\\
0&=-\hat{h}( \hat{\kappa}_s+ \hat{\kappa}_\theta) \hat{w}_n^d+\frac{1}{\hat{a}_s \hat{a}_\theta }\diffp{}{s}\left(\hat{a}_\theta \int_0^{\hat{h}}\hat{u}_s^0\dl{\hat{n}}\right)-\frac{1}{\hat{a}_s}\diffp{\hat{h}}{s}(\hat{u}_s^0)\vert_{\hat{n}=\hat{h}}+ \hat{u}_n^{1}\vert_{\hat{h}}\, .\label{eq:mass_cons1}
\end{align}

To remove the explicit appearance of curvatures, we note that the surface divergence \eqref{eq:surf_div} of the drop surface velocity is 
\begin{align}
\hat{\nabla}_s\mb{\cdot}\hat{\mb{w}}^d&=\frac{1}{\hat{a}_s\hat{a}_\theta}\left(\hat{a}_\theta\diffp{\hat{w}_s^d}{s} +\diffp{\hat{a}_\theta}{s}\hat{w}_s^d\right)-\left(\hat{\kappa}_s +\hat{\kappa}_\theta  \right)\hat{w}_n^d\, ,
\end{align}
so that
\begin{align}
\frac{1}{\hat{a}_s\hat{a}_\theta}\diffp{}{s}\left(\hat{a}_\theta \hat{h}  \hat{w}_s^d\right)&= \hat{h} \hat{\nabla}_s\mb{\cdot}\hat{\mb{w}}^d+\hat{h}(\hat{\kappa}_s +\hat{\kappa}_\theta) \hat{w}_n^d+ \frac{1}{\hat{a}_s}\diffp{\hat{h} }{s} \hat{w}_s^d\, .\label{eq:divw_term}
\end{align}

The kinematic boundary condition at the bath surface \eqref{eq:curved_kinematic} becomes
\begin{align}
    0&=\left(\overline{\hat{w}^b_n}-\hat{w}_n^d-\ep \hat{u}_n^1\vert_{\hat{h}}+\order*{\ep^2}\right)-\frac{\ep}{\hat{a}_s\left(1-\ep \hat{h}\hat{\kappa}_s\right)}\diffp{\hat{h}}{s}\left(\overline{\hat{w}^b_s}-\hat{u}_s^0\vert_{\hat{h}}+\order*{\ep}\right)\, .
\end{align}
so to $\order*{\ep^0}$, $\overline{\hat{w}^b_n}-\hat{w}_n^d=0$. We define an order $\ep$ correction given by $\ep \overline{\hat{w}^{b1}_n}=\overline{\hat{w}^b_n}-\hat{w}_n^d$ , so that at $\order*{\ep}$,
\begin{align}
    0&=\left( \overline{\hat{w}^{b1}_n}- \hat{u}_n^1\vert_{n=h}\right)-\frac{1}{\hat{a}_s}\diffp{\hat{h}}{s}\left(\overline{\hat{w}^b_s}-\hat{u}_s^b\right)\, .
\end{align}
which substituted into \eqref{eq:mass_cons1} gets to 
\begin{align}
    0&= -\frac{1}{\hat{a}_s\hat{a}_\theta}\diffp{}{s}\left(\hat{a}_\theta \hat{h}  \hat{w}_s^d\right)+\hat{h} \hat{\nabla}_s\mb{\cdot}\hat{\mb{w}}^d+ \frac{1}{\hat{a}_s}\diffp{\hat{h} }{s} \hat{w}_s^d+\frac{1}{\hat{a}_s \hat{a}_\theta }\diffp{}{s}\left(\hat{a}_\theta \int_0^{\hat{h}}\hat{u}_s^0\dl{\hat{n}}\right)\nonumber \\ &\quad-\frac{1}{\hat{a}_s}\diffp{\hat{h}}{s}\hat{u}_s^b+ \overline{\hat{w}^{b1}_n}-\frac{1}{\hat{a}_s}\diffp{\hat{h}}{s}\left(\overline{\hat{w}^b_s}-\hat{u}_s^b\right)\, ,\\
     0&= -\frac{1}{\hat{a}_s\hat{a}_\theta}\diffp{}{s}\left(\hat{a}_\theta \hat{h}  \hat{w}_s^d\right)+\hat{h} \hat{\nabla}_s\mb{\cdot}\hat{\mb{w}}^d+\frac{1}{\hat{a}_s \hat{a}_\theta }\diffp{}{s}\left(a_\theta \int_0^{\hat{h}}\hat{u}_s^0\dl{\hat{n}}\right)\nonumber
     \\ &\quad+ \overline{\hat{w}^{b1}_n}+ \frac{1}{\hat{a}_s}\diffp{\hat{h} }{s} \left(\hat{w}_s^d-\overline{\hat{w}^b_s}\right)\, .
\end{align}

The Stokes equations \eqref{eq:stokes1}--\eqref{eq:stokes2} become
\begin{align}
0&=-\frac{P}{L\hat{l}_s}\diffp{\hat{p}}{s}
+\frac{\mu^g U}{L^2\hat{l}_\theta}\frac{1}{\ep}\diffp{}{\hat{n}}\left(
\frac{\hat{l}_\theta}{\hat{l}_s}\left(\frac{1}{\ep}\diffp{}{\hat{n}}\left(\hat{l}_s\left(\hat{u}_s^0+\order*{\ep}\right)\right)
-\diffp{}{s}\left(\hat{w}_n^d+\ep \hat{u}_n^{1}+\order*{\ep^2}\right)\right)
\right)\, ,\label{eq:stokes1_scale}\\
0&=-\frac{P}{ H}\diffp{\hat{p}}{\hat{n}}-\frac{\mu^g U}{L^2\hat{l}_s\hat{l}_\theta}\diffp{}{s}\left( \frac{\hat{l}_\theta}{\hat{l}_s}\left(\frac{1}{\ep}\diffp{}{\hat{n}}\left(\hat{l}_s\left(\hat{u}_s^0+\order*{\ep}\right)\right)-\diffp{}{s}\left(\hat{w}_n+\ep \hat{u}_n^{1}+\order*{\ep^2}\right)\right)\right)\, .\label{eq:stokes2_scale}
\end{align}

Retaining the highest order pressure term, at $\order*{\ep^{-2}}$ the horizontal momentum equation \eqref{eq:stokes1_scale} is
\begin{align}
0&=-\frac{LP}{\mu^g U}\frac{1}{\hat{a}_s}\diffp{\hat{p}}{s}+\frac{1}{L^2\hat{a}_\theta}\frac{1}{\ep^2}\diffp{}{\hat{n}}\left( \frac{\hat{a}_\theta}{\hat{a}_s}\left(\diffp{}{\hat{n}}(\hat{a}_s\hat{u}_s^0)\right)\right)\, ,\\
&=-\frac{LP}{\mu^g U}\frac{1}{a_s}\diffp{\hat{p}}{s}+\frac{1}{L^2\ep^2}\diffp[2]{}{\hat{n}}\hat{u}_s^0\, ,\label{eq:stokes1_scale2}
\end{align}
and so to retain the pressure balance we choose, following standard lubrication theory, $P=\ep^{-2}\mu^g U/L$.  Then at $\order*{\ep^{-3}}$ the vertical momentum equation \eqref{eq:stokes2_scale} is
\begin{align}
0&=-\frac{1}{\ep^3 }\diffp{\hat{p}}{\hat{n}}\, ,
\end{align}
so $\hat{p}$ is independent of $\hat{n}$. Then integrating \eqref{eq:stokes1_scale2} gives
\begin{align}
    \hat{u}_s^0&=\frac{{\hat{n}}^2-\hat{h}\hat{n}}{2\hat{a}_s}\diffp{\hat{p}}{s}+\frac{\hat{u}_s^b\hat{n}+\hat{u}_s^d(\hat{h}-\hat{n})}{\hat{h}}\, ,\label{eq:u_s}\\
\int_0^{\hat{h}}\hat{u}_s^0\dl{\hat{n}}
&=\frac{-{\hat{h}}^3}{12\hat{a}_s}\diffp{\hat{p}}{s}+\frac{\hat{h}(\hat{u}_s^b+\hat{u}_s^d)}{2}\, .\label{eq:u_sint}
\end{align}
Where we have applied the continuity of velocity boundary conditions.

We also note that in the lubrication scaling that to order $\order*{\ep}$, the spatial derivative of the height \eqref{eq:dhds} becomes
\begin{align}
    \ep \diffp{\hat{h}}{s}&=\frac{\overline{\mb{e}}_\xi\mb{\cdot}\mb{n}^d}{\overline{\mb{e}}_\xi\mb{\cdot}\mb{e}_s}\hat{a}_s\, ,\label{eq:dhds_scale}
\end{align}
noting that $\frac{\overline{\mb{e}}_\xi\mb{\cdot}\mb{n}^d}{\overline{\mb{e}}_\xi\mb{\cdot}\mb{e}_s}$ is of size $\ep$. 

Substituting \eqref{eq:divw_term}, \eqref{eq:u_sint} and \eqref{eq:dhds_scale} into \eqref{eq:mass_cons1} we obtain the curved lubrication equation
\begin{align}
0&=\hat{h} \hat{\nabla}_s\mb{\cdot}\hat{\mb{w}}^d+\frac{1}{\hat{a}_s\hat{a}_\theta}\diffp{}{s}\left(\hat{a}_\theta\left(\frac{-{\hat{h}}^3}{12\hat{a}_s}\diffp{\hat{p}}{s}+\frac{\hat{h}(\overline{\hat{u}_s^b}+\hat{u}_s^d-2\hat{w}_s^d)}{2}\right)\right)+ \overline{\hat{w}^{b}_n}-\hat{w}_n^d\nonumber \\ &\quad +\frac{\overline{\mb{e}}_\xi\mb{\cdot}\mb{n}^d}{\overline{\mb{e}}_\xi\mb{\cdot}\mb{e}_s}\hat{a}_s\left(\hat{w}_s^d-\overline{\hat{w}^b_s}\right)\, .\label{eq:final_lubeq}
\end{align}
Finally, after reverting to the original length scales we obtain equation \eqref{eq:curved_lub} used in the main text.

\subsection{The Stress Tensor}

As well as the lubrication equation itself, we also need an expression for the stress tensor in the gas, as this is used in the dynamic boundary conditions \eqref{eq:ndimdynBC} and \eqref{eq:ndimdynBC_b} on the drop and bath surfaces.

Using the expressions in \S\ref{app:curv_normal}, the stress tensor
\begin{align}
T_{ij}^g=-p\delta_{ij}+\mu^g\left((\nabla \mb{u})_{ij}+(\nabla \mb{u})_{ji}\right)\, ,
\end{align}
has normal components
\begin{align}
T_{sn}^g&=\frac{\mu^g}{l_s}\left(\diffp{u_n}{s}-u_s\diffp{l_s}{n}+l_s\diffp{u_s}{n}\right)\, , &
T_{\theta n}^g&=0\, , &
T_{nn}^g&=-p+2\mu^g\diffp{u_n}{n}\, .
\end{align}
which in the lubrication scaling become
\begin{align}
T_{sn}^g&=\frac{\mu^gU}{L}\frac{1}{\hat{l}_s}\left(\diffp{\hat{u}_n}{s}-\frac{1}{\ep}\hat{u}_s\diffp{\hat{l}_s}{\hat{n}}+\hat{l}_s\frac{1}{\ep }\diffp{\hat{u}_s}{\hat{n}}\right)\, , &
T_{\theta n}^g&=0\, , &
T_{nn}^g&=\frac{\mu^gU}{L}\left(-\frac{1}{\ep^2}\hat{p}+2\frac{1}{\ep }\diffp{\hat{u}_n}{\hat{n}}\right)\, .
\end{align}

And to leading order in $\ep$, we retain the usual contributions for diagonal and off-diagonal components of the non-dimensional stress tensor, which scales with $\frac{\mu^gU}{L}$:
\begin{align}
\hat{T}_{nn}^g&=-\frac{1}{\ep^2}\hat{p}\, ,&
\hat{T}_{sn}^g&=\frac{1}{\ep }\left(\diffp{\hat{u}_s}{\hat{n}}\right)\, .
\end{align}

From \eqref{eq:u_s} we have that
\begin{align}
\diffp{\hat{u}_s}{\hat{n}}&=\frac{2{\hat{n}}-\hat{h}}{2\hat{a}_s}\diffp{\hat{p}}{s}+\frac{\hat{u}_s^b-\hat{u}_s^d}{\hat{h}}\, ,
\end{align}
so that, reverting back to the original length scales,
\begin{align}
T_{nn}^g&=-p\, ,&
T_{sn}^g&=\frac{2{n}-h}{2a_s}\diffp{p}{s}+\frac{u_s^b-u_s^d}{h}\, .\label{eq:stress_expressions}
\end{align}

For the drop-surface stress boundary condition \eqref{eq:ndimdynBC}, we directly calculate that on the drop surface
\begin{align}
    \mt{T}^{g}\mb{\cdot}\mb{n}^d=-p\mb{n}^d+\left(\frac{-h}{2a_s}\diffp{p}{s}+\frac{\overline{u_s^b}-u_s^d}{h}\right)\mb{e}_s\, .\label{eq:stress_gd_final}
\end{align}

For the stress due to the gas on on the bath surface in equation  \eqref{eq:ndimdynBC_b} we have to apply this condition slightly differently. At coordinate $\xi$ on the bath surface, we project in the direction normal to the bath and find the intersection point $\overline{\mb{x}^d}$, and measure the film height. We then use the pressure calculated by \eqref{eq:final_lubeq} at $\overline{\mb{x}^d}$, and the tangential liquid velocity $\overline{u_s^d}$, to calculate
\begin{align}
    \mt{T}^{g}\mb{\cdot}\mb{n}^b=-\overline{p}\mb{n}^b+\left(\frac{h}{2\overline{a_s}}\diffp{\overline{p}}{s}+\frac{u_s^b-\overline{u_s^d}}{h}\right)\mb{e}_\xi\, . \label{eq:stress_gb_final}
\end{align}

We now have the lubrication equation \eqref{eq:final_lubeq} that is solved for the pressure on the drop surface, and the gas stress on the liquid surfaces \eqref{eq:stress_gd_final} and \eqref{eq:stress_gb_final}, and we have closed our system of equations for the evolution of the drop and bath, with the gas acting via these surfaces stresses. The complete system is summarised in the main text in \S\ref{sec:meth_curved}.

\section{Technical Details of the Lubrication Model}
\subsection{Construction of the Surface Coordinate System}\label{app:curv_basis}

A general axisymmetric surface can be parametrised by two curvilinear coordinates, $s$ and $\theta$, where $\theta$ is the same as the azimuthal coordinate of the cylindrical system. Then the position vector can be denoted as $\mb{x}^d(s,\theta)=r^d(s)\mb{e}_r+z^d(s)\mb{e}_z$, with the $\theta$ dependence only appearing via the unit vectors. In addition, the surface is allowed to move in time, so that $\mb{x}^d(s,\theta,t)=r^d(s,t)\mb{e}_r+z^d(s,t)\mb{e}_z$. We will suppress the explicit dependence on $s$ and $t$ unless it is necessary.

The spatial derivatives of the position vector, and the unit normal vector to the surface, are
\begin{align}
\diffp{\mb{x}^d}{s}&=\diffp{r^d}{s}\mb{e}_r+\diffp{z^d}{s}\mb{e}_z\, ,&
\diffp{\mb{x}^d}{\theta}&=r^d\mb{e}_\theta\, ,&
\mb{n}^d
&=\frac{\diffp{r^d}{s}\mb{e}_z-\diffp{z^d}{s}\mb{e}_r}{\sqrt{\diffp{r^d}{s}^2+\diffp{z^d}{s}^2}}\, .\end{align}

We define the scaling factors $a_s$ and $a_\theta$ as
\begin{align}
a_s&=\abs*{\diffp{\mb{x}^d}{s}}=\sqrt{\diffp{r^d}{s}^2+\diffp{z^d}{s}^2}\, , &
a_\theta&=\abs*{\diffp{\mb{x}^d}{\theta}}=r^d\end{align}
and note the derivative of these are
\begin{align}
\diffp{a_s}{s}&=\frac{1}{a_s}\left(\diffp{r^d}{s}\diffp[2]{r^d}{s}+\diffp{z^d}{s}\diffp[2]{z^d}{s}\right)\, ,&
\diffp{a_s}{\theta}&=0\, , &
\diffp{a_\theta}{s}&=\diffp{r^d}{s}\, , &
\diffp{a_\theta}{\theta}&=0\, .
\end{align}

Then an orthonormal basis $\{\mb{e}_s,\mb{e}_\theta,\mb{n}^d\}$ is constructed with
\begin{align}
\mb{e}_s&=\frac{1}{a_s}\diffp{\mb{x}^d}{s}=\frac{1}{a_s}\left(\diffp{r^d}{s}\mb{e}_r+\diffp{z^d}{s}\mb{e}_z\right)\, , & \mb{e}_\theta&=\frac{1}{a_\theta}\diffp{\mb{x}^d}{\theta}\, , &\mb{n}^d
&=\frac{1}{a_s}\left(-\diffp{z^d}{s}\mb{e}_r+\diffp{r^d}{s}\mb{e}_z\right)\, ,\label{eq:curved_basis}
\end{align}
noting that  $\mb{e}_\theta$ is the same as in the original axisymmetric coordinates.

The $s$ derivative of $\mb{e}_s$ is then
\begin{align}
\diffp{\mb{e}_s}{s}&=\frac{-1}{a_s^2}\diffp{a_s}{s}\diffp{\mb{x}^d}{s}+\frac{1}{a_s}\diffp[2]{\mb{x}^d}{s}\, ,\\
&=\frac{1}{a_s}\left(\frac{-1}{a_s^2}\left(\diffp{r^d}{s}\diffp[2]{r^d}{s}+\diffp{z^d}{s}\diffp[2]{z^d}{s}\right)\left(\diffp{r^d}{s}\mb{e}_r+\diffp{z^d}{s}\mb{e}_z\right)+\diffp[2]{r^d}{s}\mb{e}_r+\diffp[2]{z^d}{s}\mb{e}_z\right)\, ,\\
&=a_s\frac{1}{a_s^3}\left(\diffp{r^d}{s}\diffp[2]{z^d}{s}-\diffp{z^d}{s}\diffp[2]{r^d}{s}\right)\mb{n}^d=a_s\kappa_s\mb{n}^d\, ,\label{eq:basis_derives1}
\end{align}
where we introduce the first principle curvature
\begin{align}
\kappa_s&=\frac{1}{a_s^3}\left(\diffp{r^d}{s}\diffp[2]{z^d}{s}-\diffp{z^d}{s}\diffp[2]{r^d}{s}\right)\, .
\end{align}
Similarly the other spatial derivatives are found to be
\begin{align}
\diffp{\mb{e}_s}{\theta}&=\frac{1}{a_s}\diffp{r^d}{s}\mb{e}_\theta=\frac{1}{a_s}\diffp{a_\theta}{s}\mb{e}_\theta\, , &
\diffp{\mb{e}_\theta}{s}&=0\, , &
\diffp{\mb{e}_\theta}{\theta}&=-\mb{e}_r\, .
\end{align}

Inverting \eqref{eq:curved_basis} we find that
\begin{align}
\mb{e}_r&=\frac{1}{a_s}\left(\diffp{r^d}{s} \mb{e}_s-\diffp{z^d}{s}\mb{n}^d\right)\, ,&
\mb{e}_z&=\frac{1}{a_s}\left(\diffp{z^d}{s} \mb{e}_s+\diffp{r^d}{s}\mb{n}^d\right)\, , \label{eq:basis_invert}
\end{align}
so that
\begin{align}
\diffp{\mb{e}_\theta}{\theta}&=-\frac{1}{a_s}\diffp{r^d}{s}\mb{e}_s+\frac{1}{a_s}\diffp{z^d}{s}\mb{n}^d=-\frac{1}{a_s}\diffp{a_\theta}{s}\mb{e}_s+a_\theta\kappa_\theta\mb{n}^d\, ,
\end{align}
where we introduce the second principle curvature 
\begin{align}
\kappa_\theta&=\frac{1}{r^da_s}\diffp{z^d}{s}\, .
\end{align}

It can then be computed that 
\begin{align}
\diffp{\mb{n}^d}{s}&=-a_s\kappa_s\mb{e}_s\, , & \diffp{\mb{n}^d}{\theta}&=-a_\theta\kappa_\theta\mb{e}_\theta\, .\label{eq:basis_derives2}
\end{align}

\subsection{Time Derivatives of Basis Vectors}\label{app:time_derivs}
The surface is moving in the $s$ and $n$ directions with velocity $\mb{w}^d$, and we split these into component 
\begin{align}
\diffp{\mb{x}^d}{t}&=\mb{w}^d=w_s^d\mb{e}_s+w_n^d\mb{n}^d\, .
\end{align}
We now want to compute the derivatives $\diffp{\mb{e}_s}{t}$ and $\diffp{\mb{n}^d}{t}$.

Using \eqref{eq:basis_derives1}--\eqref{eq:basis_derives2}, we compute that
\begin{align}
\diffp{\mb{x}^d}{s,t}
&=\left(\diffp{w_s^d}{s}-a_s\kappa_sw_n^d\right)\mb{e}_s+\left(\diffp{w_n^d}{s}+a_s\kappa_sw_s^d\right)\mb{n}^d\, ,
\end{align}
Then using \eqref{eq:basis_invert},
\begin{align}
\diffp{a_s}{t}&=\diffp{}{t}\left(\diffp{r^d}{s}^2+\diffp{z^d}{s}^2\right)^{1/2}\, ,\\
&=\frac{1}{a_s}\left(\diffp{r^d}{s}\diffp{r^d}{s,t}+\diffp{z^d}{s}\diffp{z^d}{s,t}\right)\, ,\\
&=\mb{e}_s\mb{\cdot}\diffp{\mb{x}^d}{s,t}=\left(\diffp{w_s^d}{s}-a_s\kappa_sw_n^d\right)\, .
\end{align}
and so
\begin{align}
\diffp{r^d}{s,t}
=\diffp{\mb{x}^d}{s,t}\mb{\cdot}\mb{e}_r
&=\frac{1}{a_s}\left(\diffp{a_s}{t}\diffp{r^d}{s}-\left(\diffp{w_n^d}{s}+a_s\kappa_sw_s^d\right)\diffp{z^d}{s}\right)\, ,
\end{align}
\begin{align}
\diffp{z^d}{s,t}
&=\frac{1}{a_s}\left(\diffp{a_s}{t}\diffp{z^d}{s}+\left(\diffp{w_n^d}{s}+a_s\kappa_sw_s^d\right)\diffp{r^d}{s}\right)\, ,
\end{align}

We then find that
\begin{align}
\mb{n}^d&=\frac{1}{a_s}\left(\diffp{r^d}{s}\mb{e}_z-\diffp{z^d}{s}\mb{e}_r\right)\, ,\\
\diffp{\mb{n}^d}{t}&=\frac{-1}{a_s}\mb{n}^d\diffp{a_s}{t}+\frac{1}{a_s}\left(\diffp{r^d}{s,t}\mb{e}_z-\diffp{z^d}{s,t}\mb{e}_r\right)\, ,\\
&=\frac{-1}{a_s}\mb{n}^d\diffp{a_s}{t}+
\frac{1}{a_s^2}\Bigg(\diffp{a_s}{t}\diffp{r^d}{s}\mb{e}_z-\left(\diffp{w_n^d}{s}+a_s\kappa_sw_s^d\right)\diffp{z^d}{s}\mb{e}_z\nonumber \\ &\quad-\diffp{a_s}{t}\diffp{z^d}{s}\mb{e}_r-\left(\diffp{w_n^d}{s}+a_s\kappa_sw_s^d\right)\diffp{r^d}{s}\mb{e}_r\Bigg)\, ,\\
&=\frac{-1}{a_s}\mb{n}^d\diffp{a_s}{t}+
\frac{1}{a_s}\diffp{a_s}{t}\mb{n}^d-\frac{1}{a_s^2}\left(\diffp{w_n^d}{s}+a_s\kappa_sw_s^d\right)\mb{e}_s\, ,\\
&=-\frac{1}{a_s}\left(\diffp{w_n^d}{s}+a_s\kappa_sw_s^d\right)\mb{e}_s\, .
\end{align}

And similarly
\begin{align}
\diffp{\mb{e}_s}{t}&=\diffp{}{t}\left(\frac{1}{a_s}\diffp{\mb{x}^d}{s}\right)=\frac{1}{a_s}\left(\diffp{w_n^d}{s}+a_s\kappa_sw_s\right)\mb{n}^d\, .
\end{align}

\subsection{Coordinates Within the Gas Film}\label{app:curv_normal}

So far we have been looking at objects on the surface of the drop. In order to derive the thin film equation in these curvilinear coordinates, we need to generalise to position vector away from the surface. We add a third coordinate $n$, which measures the distance away from the surface in the normal direction, defining the position vector $\mb{x}=\mb{x}^d+n\mb{n}^d$.

Notating $s_1=s$, $s_2=\theta$ and $s_3=n$, we can write a metric tensor given by
\begin{align}
g_{ij}=\diffp{\mb{x}}{s_i} \mb{\cdot} \diffp{\mb{x}}{s_j}=
\begin{pmatrix}
a_s^2(1-\kappa_s n )^2 & 0&0 \\
0& a_\theta^2(1-\kappa_\theta n)^2&0\\
0&0&1
\end{pmatrix}\, .
\end{align}

Our coordinates are a specific version of a general orthogonal, curvilinear coordinate system, where the metric is given by $g_{ij}=l_i^2\delta_{ij}$, where $\delta_{ij}$ is the Kronecker delta. In our case the scaling factors are $l_s=a_s(1-\kappa_s n)$, $l_\theta=a_\theta(1-\kappa_\theta n)$ and $l_n=1$. In the following we shall slightly abuse notation, using $l_i$ and $l_{s_i}$ interchangeably, as well as the corresponding unit basis vectors $\mb{e}_i$ and $\mb{e}_{s_i}$.

In such a coordinate system, the derivative operators of a scalar function $f$, a vector function $\mb{f}=\sum_i f_i\mb{e}_i$ (and $\mb{g}$) take the following forms \citep{Kreyszig2020, Moon1988, Irgens2008}, 
\begin{align}
    \nabla f&=\sum_{i=1}^3\frac{1}{l_i}\diffp{f}{s_i}\mb{e}_i\, ,\\
    \nabla\mb{\cdot}\mb{f}&=\frac{1}{l_1l_2l_3}\sum_i\diffp{}{s_i}\left(l_1l_2l_3\frac{f_i}{l_i}\right)\, ,\\
    \nabla\times\mb{f}&=\frac{1}{l_1l_2l_3}\sum_{ijk}l_i\ep_{ijk}\diffp{}{s_j}(l_kf_k)\mb{e}_i\, ,\\
    (\mb{f}\mb{\cdot}\nabla)\mb{g}&=\sum_j\sum_k\left(\frac{f_k}{l_k}\diffp{g_j}{s_k}+\frac{g_k}{l_kl_j}\left(f_j\diffp{l_j}{s_k}-f_k\diffp{l_k}{s_j}\right)\right)\mb{e}_j\, ,\\
  \text{for}\ i\neq j: \quad  \left[ \nabla \mb{f}\right]_{ij}+\left[ \nabla \mb{f}\right]_{ji}&=\frac{l_i}{l_j}\diffp{}{s_j}\left(\frac{f_i}{l_i}\right)+\frac{l_j}{l_i}\diffp{}{s_i}\left(\frac{f_j}{l_j}\right)\, , \\
  \left[ \nabla \mb{f}\right]_{ii}&=\diffp{}{s_i}\left(\frac{f_i}{l_i}\right)+\sum_{k}\frac{1}{l_il_k}\diffp{l_i}{s_k}f_k\, .
\end{align}
where $\left[\mt{A}\right]_{ij}$ are the components of a tensor of the form $\mt{A}=\sum_{ij}\left[\mt{A}\right]_{ij}\mb{e}_i\otimes\mb{e}_j$ and $\ep_{ijk}$ is the Levi-Civita symbol. The definition $\nabla^2\mb{f}=\nabla(\nabla\mb{\cdot}\mb{f})-\nabla \times(\nabla\times \mb{f})$ is used for the vector Laplacian.

With our axisymmetric coordinate system, where scalar functions and the components of vectors don't depend on $\theta$, and $l_n=1$, these become
\begin{align}
    \nabla f&=\frac{1}{l_s}\diffp{f}{s}\mb{e}_s+\diffp{f}{n}\mb{n}^d\, ,\\
    \nabla\mb{\cdot}\mb{f}&=\frac{1}{l_sl_\theta}\left(\diffp{}{s}\left(l_\theta f_s\right)+\diffp{}{n}\left(l_sl_\theta f_n\right)\right)\, ,\\
    \nabla\times\mb{f}&=\frac{1}{l_s}\left(\diffp{}{n}(l_sf_s)-\diffp{}{s}(f_n)\right)\mb{e}_\theta\, ,\\
    (\mb{f}\mb{\cdot}\nabla)\mb{g}
&=\left(\frac{f_s}{l_s}\diffp{g_s}{s}+f_n\diffp{g_s}{n}+\diffp{l_s}{n}\frac{f_sg_n}{l_s}\right)\mb{e}_s\nonumber\\
&\quad+\left(\frac{f_s}{l_s}\diffp{g_n}{s}+f_n\diffp{g_n}{n}-\diffp{l_s}{n}\frac{f_sg_s}{l_s}\right)\mb{e}_n\, ,
\\
    \nabla \times\nabla\times\mb{f}&=\frac{1}{l_sl_\theta}\diffp{}{s}\left(\frac{l_\theta }{l_s}\left(\diffp{}{n}(l_sf_s)-\diffp{}{s}(f_n)\right)\right)\mb{n}^d\nonumber\\&\quad-\frac{1}{l_\theta}\diffp{}{n}\left( \frac{l_\theta}{l_s}\left(\diffp{}{n}(l_sf_s)-\diffp{}{s}(f_n)\right)\right)\mb{e}_s\, .
\end{align}

and the components of the stress tensor 
$T_{ij}=-p\delta_{ij}+\left((\nabla \mb{u})_{ij}+(\nabla \mb{u})_{ji}\right)$ are
\begin{align}
T_{ss}&=-p+2\left(\frac{1}{l_s}\diffp{\hat{u}_s}{s}+\frac{u_n}{l_s}\diffp{l_s}{n}\right)
\, ,&
T_{\theta\theta}&=-p+\frac{2}{l_\theta}\left(\frac{\hat{u}_s}{l_s}\diffp{l_\theta}{s}+u_n\diffp{l_\theta}{n}\right)
\, , &
T_{s\theta}&=0
\, ,\\
T_{nn}&=-p+2\diffp{u_n}{n}
\, , &
T_{sn}&=\frac{1}{l_s}\left(\diffp{u_n}{s}-\hat{u}_s\diffp{l_s}{n}+l_s\diffp{\hat{u}_s}{n}\right)
\, , &
T_{\theta n}&=0\, .
\end{align}

Finally, for quantities on the drop surface (so that are not functions of $n$), there are the surface operators (i.e. at $n=0$), 
\begin{align}
     \nabla_s f&=\frac{1}{a_s}\diffp{f}{s}\mb{e}_s\, ,\\
    \nabla_s\mb{\cdot}\mb{f}&=\frac{1}{a_sa_\theta}\left(\diffp{}{s}\left(a_\theta f_s\right)+\diffp{}{n}\left(l_sl_\theta \right)f_n\right)\, 
    =\frac{1}{a_sa_\theta}\diffp{}{s}\left(a_\theta f_s\right)-\left(\kappa_s +\kappa_\theta \right)f_n\, .\label{eq:surf_div}
\end{align}

\bibliographystyle{jfm}
\bibliography{DropBathPaper}
\end{document}